\documentclass[aip,numerical,rsi,reprint,superscriptaddress,a4paper]{revtex4-1}

\usepackage{amssymb}
\usepackage{amsmath}
\usepackage{graphicx}

\newcommand{\xiji}{x_{ij}^{(i)}}
\newcommand{\yiji}{y_{ij}^{(i)}}
\newcommand{\dxi}{\delta_x^{(i)}}
\newcommand{\dyi}{\delta_y^{(i)}}

\begin{document}

\title{A novel technique for single-shot energy-resolved 2D X-ray imaging of plasmas relevant for the Inertial Confinement Fusion}

\author{L. Labate}
\email[E-mail: ]{luca.labate@ino.it}
\altaffiliation[Also at ]{Istituto Nazionale di Fisica Nucleare, Sezione di Pisa, Italy}
\affiliation{Intense Laser Irradiation Laboratory, Istituto Nazionale di Ottica, Consiglio Nazionale delle Ricerche, Pisa, Italy}
\author{P. K\"oster}
\affiliation{Intense Laser Irradiation Laboratory, Istituto Nazionale di Ottica, Consiglio Nazionale delle Ricerche, Pisa, Italy}
\author{T. Levato}
\altaffiliation[Also at ]{Dipartimento di Ingegneria Industriale, Universit\`a di Roma ``Tor Vergata'', Italy and Fyzikální ústav AV ČR v.v.i.,  Praha, Czech Republic}
\affiliation{Intense Laser Irradiation Laboratory, Istituto Nazionale di Ottica, Consiglio Nazionale delle Ricerche, Pisa, Italy}
\author{L.A. Gizzi}
\altaffiliation[Also at ]{Istituto Nazionale di Fisica Nucleare, Sezione di Pisa, Italy}
\affiliation{Intense Laser Irradiation Laboratory, Istituto Nazionale di Ottica, Consiglio Nazionale delle Ricerche, Pisa, Italy}

\begin{abstract}
A novel X-ray diagnostic of laser-fusion plasmas is described,
allowing 2D monochromatic images of hot, dense plasmas to be obtained
in any X-ray photon energy range, over a large domain, on a
single-shot basis. 
The device (named Energy-encoded Pinhole Camera - EPiC) is based upon the use of an array of many pinholes
coupled to a large area CCD camera operating in the
single-photon mode.
The available X-ray spectral domain is only limited by the Quantum Efficiency
of scientific-grade X-ray CCD cameras, thus extending from a few keV up to
a few tens of keV.
Spectral 2D images of the emitting plasma can be obtained at any X-ray photon energy provided that a sufficient
number of photons had been collected at the desired energy.
Results from recent ICF related experiments will be reported in order
to detail the new diagnostic. 
\end{abstract}

\maketitle

\section{Introduction}

X-ray imaging and spectroscopy are recognized among the
most useful techniques to measure properties of hot and dense plasmas (see e.g. \cite{Giulietti_RNC1998,Gizzi_SSCP2011} and References therein). This is also true for plasmas  relevant for the Inertial
Confinement Fusion (ICF) \cite{Schneider_RSI2006,Glenn_RSI2010,Schneider_RSI2010,Kimbrough_RSI2010,Hammel_PoP2011}.
In particular, they are an essential tool, complementary to neutron and optical diagnostics,
to infer parameters such as electron and ion temperature, charge state
distribution, density, implosion velocities and fuel-shell plasma
mixing degree in ICF targets (see \cite{Hansen_PoP2012} and references therein).
Moreover, X-ray spectroscopy and imaging have been employed over the past decade to
measure fast electron generation and transport in experiments
relevant for the fast ignition approach \cite{Tabak_PoP1994,Key_PoP2007} to ICF \cite{Norreys_NF2009,Lancaster_PRL2007}.
In these studies, other complementary diagnostic techniques, such as,
e.g., direct measurements of forward escaping electrons
\cite{Wei_PRE2004,Labate_APB2007}, shadowgraphy
\cite{Norreys_PPCF2006,Lancaster_PRL2007}, optical emission from the
rear surface of the target \cite{Manclossi_PRL2006,Santos_PRL2002} and
proton radiography \cite{Perez_PRL2011,Vauzour_PoP2011} are also
employed. However, the analysis of Bremsstrahlung and/or electron-induced X-ray
fluorescence via X-ray imaging and spectroscopy is by far the most adopted strategy to approach issues related to the fast electron production and propagation 
\cite{Martinolli_PRE2006,Theobald_PoP2006,Gizzi_PPCF2007,Chen_PoP2009Brem,Scott_PoP2012,Yasuike_RSI2001,Nilson_PoP2008,Tanimoto_PoP2009,Stephens_PRE2004,Booth_NIMA2011}.\par

%Booth_JPCS2010

From an experimental point of view, measurement techniques based upon Bremsstrahlung emission from fast electrons up to the hundreds of keV region encompass K-edge filters coupled to image plate dosimeter stacks \cite{Chen_RSI2008,Chen_PoP2009Brem} as well as fluorescer filters coupled to photomultipliers \cite{McDonald_RSI2004,Dewald_RSI2010}.
Such a solution has been successfully implemented and used at National Ignition Facility \cite{Dewald_PoP2006}.\par

When imaging capabilities are needed, as in the study of fast electron transport in high density matter, flat or, more often, bent Bragg crystals coupled to X-ray films, image plates or CCD cameras are commonly employed.
In the first case, either a slit \cite{Labate_PoP2005} or a set of pinholes \cite{Tommasini_RSI2006} are used in order to get a simultaneous spatial resolution. 
In the latter case, a number of different configurations are used, allowing either 1D or 2D imaging of the source to be obtained \cite{King_RSI2005,Norreys_PoP2004} with typical spectral resolution down to a few eV.
Recently, the use of bent Laue crystals coupled to CsI/CCD scintillator/detectors for high resolution spectroscopy up to around $80\,\mathrm{keV}$ has been reported \cite{Zhang_RSI2012}.
Furthermore, the use of pinhole camera schemes with a relatively large number of pinholes with different X-ray filters has been discussed \cite{Park_RSI2010}.
\par

It is worth pointing out here that the use of Bragg crystals is typically
limited to low energy ranges ($ \lesssim 10\,\mathrm{keV}$) due to a) a low reflectivity of the available crystals and b) the increasing image aberrations at the small Bragg angles usually needed at high photon energy \cite{Dirksmoller_OC1995}.
The useful energy range limited to a few keV is currently considered as a major limitation of Bragg crystals in the context of fast electron studies.
Indeed, most of the times the study of the fast electron transport is carried out by looking at the K$\alpha$ emission from fluorescent layers buried in the target.
Due to the behavior of the K-shell ionization cross section by fast electrons as a function of the electron energy \cite{Talukder_IJMS2008}, the K$\alpha$ production would be maximized, at the typical electron energy considered in the fast ignition scenario, by using layers with higher Z numbers than the ones currently used.
This would thus push for X-ray spectrometers with imaging capabilities at energy of the order of a few tens of keV, which is beyond the capabilities of currently available crystal based spectrometers.\par

It is also worth mentioning a further issue related to the use of bent crystals when studying fast electron physics.
While, in general, X-ray 1D imaging can provide informations about the laser-to-fast-electrons conversion efficiencies as well as the fast electron energy spectrum, the study of the electron beam angle aperture requires, in principle, a 2D imaging capability (see for instance \cite{Norreys_NF2009} and references therein).
This kind of study is usually performed, as mentioned above, using
buried layers of fluorescent materials \cite{Batani_LPB2002}, each layers  with its own characteristic K$\alpha$ wavelength requiring its own dedicated X-ray imaging diagnostic (typically on a multi-shot basis).
As recently pointed out in \cite{Perez_PoP2010}, it is rather questionable, from an analytical point of view, whether a correct estimate of the fast electron divergence could be inferred by an image of a buried fluorescent layer.
In the same paper the authors discuss a new kind of target suitable for single-shot studies of the fast electron beam divergence, using a curved spectrometer in a 2D imaging configuration.
On the other hand, it is now well known that the effective
collection efficiency of a given curved crystal 2D imaging
configuration can be heavily affected by a change in the temperature
of the emitting region, even dropping of a factor $10^2$ for a
temperature change of a few tens of eV. 
This is due to spectral changes of the observed lines because of changes
occurring in the quantum configuration and screening potential.
More important, beside the overall collection efficiency, this may also affect the shape of the observed image of the emitting region, as shown in \cite{Akli_PoP2007}.\par

The study of the fast electrons plays an important role also in the shock ignition \cite{Betti_PRL2007,Atzeni_PPCF2011} scenario, due to possible nonlocal transport effects \cite{Bell_PPCF2011}.
A couple of experiments have been carried out recently, employing bent
Bragg crystals to diagnose plasma conditions and fast electrons
\cite{Baton_PRL2012,Nejdl_ProcSPIE2011}.\par

Motivated by the above considerations and starting from some earlier studies in which a single-photon diagnostic were used \cite{Gizzi_PRL1996}, we have been working over the
past few years on the development of a new concept X-ray diagnostic
with both 2D imaging and spectroscopic capabilities, tentatively
operating on a broad X-ray photon energy range.
In a previous paper \cite{Labate_RSI2007} we reported on a first
achievement allowing such goals to be obtained on a multi-shot
operation, thus limited to high repetition rate experiments typical,
for instance, of ultrashort X-ray sources development.
In a recent experiment, using such a technique, we were able to detect the contribution to directional X-ray continuum emission from the plasma electrons \cite{Zamponi_PRL2010}.
In this paper, we present a novel, more general scheme, which we
called \textit{Energy-encoded Pinhole Camera} (EPiC), allowing our novel diagnostic to be used on a single laser shot.
This makes our technique particularly useful
in experiments where ICF relevant plasmas have to be studied.\par

In the next Section, a description of the novel scheme will be given
and both the requirements and the issues related to the achievable
spatial and spectral resolution will be discussed.
In Section 3, in order to show the potential of our diagnostic
in the field of ICF physics, we give a few results from recent
experiments related to ICF, followed by a brief discussion of possible perspectives.

\section{Description of the diagnostic}

\begin{figure*}
\begin{center}
\includegraphics[width=0.7\textwidth]{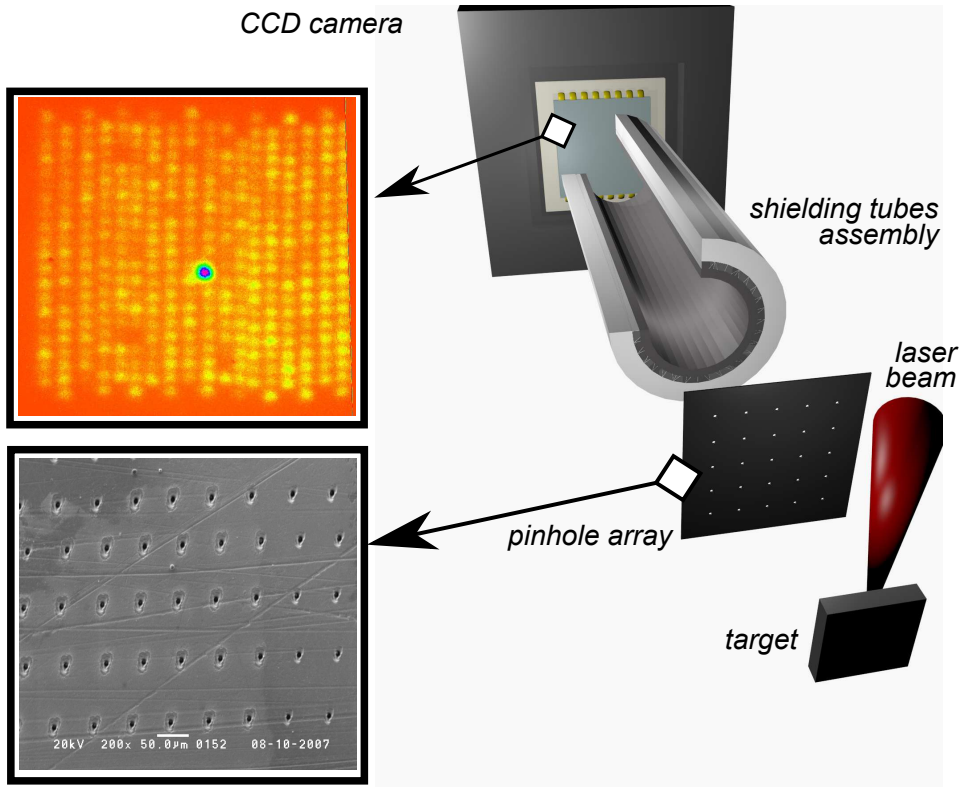}
\caption{(color online) 3D rendering of the EPiC diagnostic. The plasma X-ray source is multiply imaged out onto a large area CCD detector by an array of several pinholes (PHA) (not in scale in the Figure). The CCD detector is forced to operate in the single-photon mode. Typical source-to-PHA distances are of a few centimeters and typical PHA-to-CCD distances range from a few tens of centimeters up to a few meters. The CCD detector can be placed, if needed, into a separate vacuum chamber with respect to the main one, provided that suitable X-ray windows are employed. The lower inset shows a Scanning Electron Microscope image of a homemade pinhole array, drilled onto a $100\,\mathrm{\mu m}$ thick substrate (see text). The upper inset shows a typical acquired image; the ``single-photon'' images from the different pinholes are clearly visible (the brightest spot at the center was produced by a great diameter ($~100\,\mathrm{\mu m}$) pinhole and was used for alignment purposes).
}
\label{figure1}
\end{center}
\end{figure*}

A schematic layout of the EPiC diagnostic is shown in Figure \ref{figure1}.
The scheme basically resembles the one of a conventional X-ray pinhole
camera, where an array of several pinholes is used, instead of a single pinhole, coupled to a CCD camera operating, in this case, in the single-photon mode.
The CCD camera is shielded against visible radiation using suitable filters (not shown in Figure).
In order to prevent high energy electrons, typical of the harsh environments encountered in ICF related experiments \cite{Park_RSI2004}, from directly reaching the pinhole array, a suitable magnet assembly (not visible in Figure) is placed upstream.
Furthermore, a shielding tube, possibly consisting of concentric tubes of different materials (typically plastic+Al), is placed in front of the CCD camera to shield the camera against high energy particles (typically electrons and $\gamma$ photons).

\par

As it is well known, CCD detectors operating in single-photon mode allow the spectrum of the impinging X-ray radiation to be directly retrieved, without any additional dispersive element, from the pulse height distribution, provided that the single-photon response had been characterized using reference X-ray sources \cite{Labate_NIMA2002} and an accurate image processing was carried out (see for instance \cite{Fourment_RSI2009,Maddox_RSI2008Cali,Labate_RSI2007}).
The single-photon operation regime can be obtained, in the typical situation considered in the context of ICF (high flux X-ray sources) either by placing the detector at a large distance from the plasma or by the insertion of a suitable stack of X-ray attenuators.
In the experiments whose results are shown below both strategies were adopted at the same time.\par

Once the CCD camera was forced to operate in the single-photon mode, the presence of a pinhole between the plasma and the detector makes it possible to retrieve for each photon, beside to its energy, the position (in the source plane) where it comes from, just like in a conventional pinhole camera scheme.
Therefore, in the case of a single pinhole, one would get on the CCD detector a collection of photon events whose distribution would peak around the position of the ``true'' (that is, high photon flux) image; for the sake of brevity, we will be calling such a collection a ``single-photon image'' throughout this paper (and we will omit the ``single-photon'' specification when no confusion can be made). 
We already showed in \cite{Labate_RSI2007} how a ``true'' 2D image of the source could be recovered in any given energy range by summing up a large number of single-photon images where only photons belonging to that energy range had been selected.
We postpone to the end of this Section a discussion about the number
of photons/images needed.
In the case considered there, a large number of single-photon images was collected by simply taking a correspondingly large number of laser shots.
This approach is of course not viable when using very high energy, PW-scale, low rep rate laser systems.
In the EPiC scheme a different approach is used, in which an array of a large number of pinholes is employed in order to collect all the needed single-photon images in one laser shot.
This basically hinges on the use of the larger and larger CCD chips now commercially available, which allow images from the different pinholes to be collected onto different regions of the same chip, as shown in the upper inset of Figure \ref{figure1}.
The final image of the source in the EPiC is then recovered by summing up the images provided by each pinhole, retaining only photons in the energy range of interest.
In other words, all the (single-photon) images are collapsed into a single one.
Quantitatively, we can define the final image of the source as
\begin{equation}
I (x_d,y_d) = \sum_{i=0}^{N_x-1}\sum_{j=0}^{N_y-1} D(x_d + \xiji,y_d+\yiji)\label{eq::1}
\end{equation}
where $x_d,y_d$ are the coordinates of a point in the CCD detector plane, $D(x_d,y_d)$ is the signal detected at $(x_d,y_d)$, the sums run over all the $N_x\times N_y$ pinholes and $(\xiji,\yiji)$ is the position of the center of the image through the pinhole $i,j$ in the CCD plane.
To prevent the mixing of images from neighboring pinholes (or, in other words, in order not to count twice the same photons as belonging to different single-photon images), formula (\ref{eq::1}) only makes sense for
\begin{eqnarray}
-\frac{\dxi}{2} < x_d < +\frac{\dxi}{2} \qquad -\frac{\dyi}{2} < y_d < +\frac{\dyi}{2}\label{eq::2}
\end{eqnarray}
where $\dxi$ and $\dyi$ are the distances between neighboring images in the CCD plane in the $x$ and $y$ direction respectively.
It can also be verified that $\dxi= (1+M) \delta_x$, where $M$ is the magnification and $\delta_x$ is the pitch of the pinhole array in the $x$ direction (a similar relation holds in the $y$ direction).
The magnification here is defined as in the usual pinhole camera: $M=q/p$, $p$ being the source-to-pinhole array distance and $q$ the pinhole array-to-CCD distance.
At this stage, it is worth pointing out that the variation of the above quantities from pinhole to pinhole can be safely neglected in a first approximation; indeed, typical pinhole array pitches are in the range $50-200\,\mathrm{\mu m}$ (see lower inset of Figure \ref{figure1}), to be compared with typical $q$ and $p$ values ranging from $\sim 10\,\mathrm{cm}$ up to a few meters.
\par

As for the pinhole array, to our knowledge, no commercial solutions exist with the required number of pinholes (typically a few hundreds). We then made them by laser drilling thick, high Z number substrates.
A detail of a Scanning Electron Microscope image of an array made using this method is shown in the lower inset of Figure \ref{figure1}.
We refer to papers \cite{Levato_NIMA2010,Labate_LPB2009} for details on the fabrication technique and for more specifications of the  final arrays. 
We just mention here that very high aspect ratio, quasi cylindrical shape pinholes were obtained onto $100-200\,\mathrm{\mu m}$ thick W, Pt or Ta foils, with typical diameters of $5-10\,\mathrm{\mu m}$.
Taking into account the typical magnifications used ($\sim 10-20$) and the typical pixel size of available X-ray CCDs ($\sim 10-20\,\mathrm{\mu m}$), this figure basically sets the ultimate limit of the achievable spatial resolution.\par

We notice here that, in the discussion above, we assumed a perfectly shaped pinhole array; in other words, it was assumed that the positions of the different pinholes on the array plane could have been characterized by the two pitches $\delta_x$ and $\delta_y$.
In fact, looking at both the insets of Figure \ref{figure1}, quite clear deviations of the pinhole positions from a regular pattern can be seen, due to the fabrication process.
From a theoretical viewpoint, it is pretty easy to generalize our above (and below) discussion by considering the actual pinhole positions on the pinhole array plane.
From an experimental viewpoint, this means that the actual positions of the image centers through each pinholes have to be  preemptively identified.\par

\begin{figure}
\begin{center}
\includegraphics[width=0.9\columnwidth]{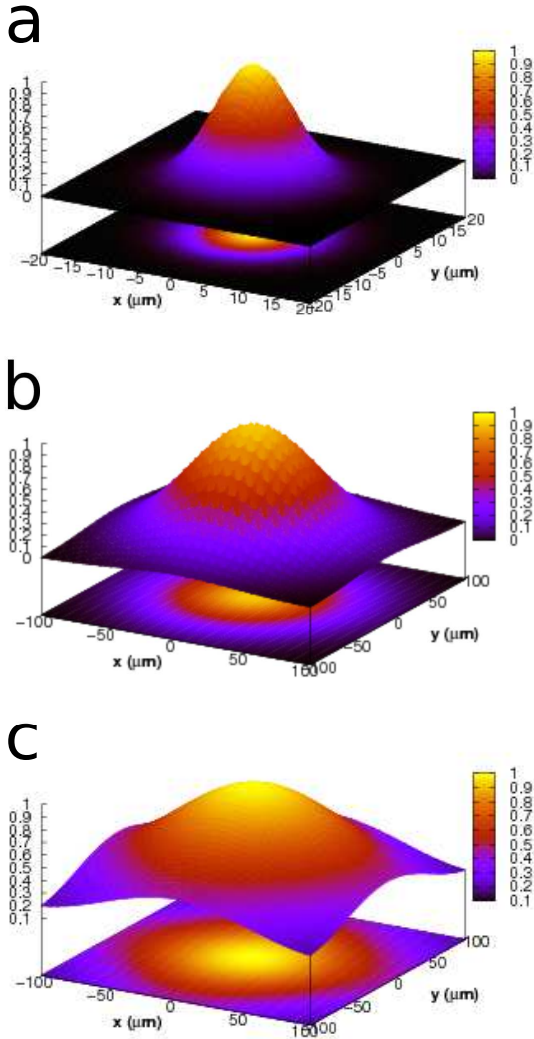}
\caption{Simulated shape of a Gaussian plasma X-ray source imaged out
  at $10\,\mathrm{keV}$ using a $200\,\mathrm{\mu m}$ pitch, $10\times
  10$ pinholes array drilled onto a $100\,\mathrm{\mu m}$ thick Pt
  substrate. Each pinhole is supposed to have a diameter of
  $10\,\mathrm{\mu m}$. 
Three cases corresponding to different source FWHMs are shown, which clarify the effect of a source size comparable to the array pitch: a) $10\,\mathrm{\mu m}$, b) $100\,\mathrm{\mu m}$ and c) $150\,\mathrm{\mu m}$. 
As one goes from a) to c) the useful window (see Eq. (\ref{eq::2})) doesn't accommodate the source image and also overlapping effects start playing a role.}
\label{figure2}
\end{center}
\end{figure}

We will now discuss a few issues related to the reconstruction of the images and the corresponding limitations.
As we said above, care has to be taken in order to avoid spurious effects due to the possible overlap of images from neighboring pinholes.
The conditions given by equation (\ref{eq::2}) just define a rectangular region ($\dxi \times \dyi$ wide) on the CCD plane where the retrieved (collapsed) image makes sense.
However, this is only a necessary condition, not a sufficient one. Indeed, it is basically equivalent to state that each photon had been transmitted through that pinhole which generates the image whose center is the closest to the photon position (on the CCD plane).
This is a pretty artificial assumption, as there is no certainty that the photon had not passed through one of the neighboring pinholes instead.
In order for this possibility to be ruled out (or, better, to reduce
the probability of such a case to occur), the physical parameters of
the pinhole array have to be properly chosen so as to avoid any overlapping between the (single-photon) images through neighboring pinholes.
As we discussed before, the distance between two neighboring images in the CCD plane is given by $\delta_{x,y}^{(i)}=(1+M) \delta_{x,y}$.
Let $\Delta_s$ be the characteristic source size; the image of the source on the detector plane has thus a characteristic size $M \Delta_s$. 
In order for the images from neighboring pinholes not to overlap, this size has to be smaller than the separation between neighboring images: $M \Delta_s \lesssim (1+M) \delta_{x,y}$.
Thus
\begin{equation}
\delta_{x,y} \gtrsim \frac{M}{1+M} \Delta_s\label{eq::3}
\end{equation}
For large values of the magnification, the above formula reduces to $\delta_{x,y}\gtrsim \Delta_s$.
Thus, as a rule of thumb we can state that the pitch of the pinhole array has to be greater than the expected source size.
In order to illustrate this point, we show in Figure \ref{figure2} the
simulated shape of a Gaussian shaped source with different FWHM imaged
using a $10\times10$ pinhole array, with each pinhole having a
diameter  of $10\,\mathrm{\mu m}$. 
The array has pitches $\delta_x = \delta_y = 200\,\mathrm{\mu m}$ and is supposed to be drilled onto a $200\,\mathrm{\mu m}$ thick Pt substrate (this information is used to calculate the transmission through the substrate).
Only photons with $10\,\mathrm{keV}$ energy are considered.
The images are all retrieved in the region defined by Equation (\ref{eq::2}).
It is quite clear the meaning of the condition stated by Equation (\ref{eq::3}).
Furthermore, artifacts on the retrieved image due to the condition ($\ref{eq::3}$) not being fulfilled emerge as the original source size approaches the pinhole array pitch, going from \textit{a)} to \textit{c)}.\par

Finally, it is worth discussing the issue of the Signal-to-Noise
ratio (SNR) of the images obtained using the EPiC scheme.
First of all, considering the typical specifications of currently
available scientific grade X-ray CCD cameras, it can be easily
verified that the main limitation to the SNR stems from the photon
counting statistics (see, for instance, \cite{KnollRadiationDetectionBook_1989}), that is, ultimately, from the number of photons
which can be collected over the energy range considered to retrieve each image.
We point out that this number is limited, in the EPiC scheme, by both
the need to operate in the single-photon regime and the need for the
condition stated by Equation (\ref{eq::3}) to be fulfilled (indeed,
this latter condition ultimately further limits the number of
collected photons
with respect to the case of a CCD used just for photon
counting spectroscopy).
As a matter of fact, the available X-ray flux from ICF relevant
plasmas is by far much higher than the maximum flux required to let
the CCD detector work in the single-photon regime.
For instance, in an experiment carried out at the PALS facility in
Prague (see Section \ref{sec::expres} for more details), we estimated
a spectral flux of $\sim 10^{13}$ Ti K$\alpha$ photons per shot.
Let us consider a distance of a few centimeters between the source and
the pinhole array, pinhole diameters of about $10\,\mathrm{\mu m}$ and
a CCD Quantum Efficiency of a few tens of percent at $\sim
4.5\,\mathrm{keV}$.
By a simple calculation based on the solid angle subtended by each of
the pinholes, it is pretty easy to verify that the number of photons
collected in the CCD chip region available to each pinhole wouldn't
allow the single-photon condition to be fulfilled.
As it was said above, this means that either the CCD has to be put
very far from the source or suitable X-ray attenuators need to be used
(or both).\par

Let $N_{ph}^{(all)}$ be the total number of collected X-ray photons
(integrated over the whole bandwidth).
Clearly $N_{ph}^{(all)}= (N_x N_y) N_{ph}^{(SPI)}$, where
$N_{ph}^{(SPI)}$ is the average number of photons collected in each
single-photon image (for the sake of brevity, SPI from now on). 
$N_x$ and
$N_y$ are, as above, the number of pinholes in the $x$ and $y$ direction respectively.
The upper limit to $N_{ph}^{(SPI)}$ is clearly set  by the total number of pixels available for each
SPI and the single-photon condition. 
Indeed, if no pinhole array was present, the CCD chip would be uniformly
illuminated by X-ray photons.
In other words, all the pixels available to the SPI would have an
equal probability of detecting a photon, so that $N_{ph}^{(SPI)}$ would be simply
$\sim f \, N_{pixel}^{(SPI)}$, where $f<1$ is a factor
accounting for the requirement of well isolated single-photon events as
well as for physical processes occurring after the charge cloud
generation which result in a charge spreading across neighboring
pixels (typically $f\simeq 0.1$) 
\cite{Hopkinson_NIMA1983,Prigozhin_IEEETED2003,Pavlov_NIMA1999}.
In the EPiC case, instead, the value of $N_{ph}^{(SPI)}$ is even lower, since, being  the source actually imaged out through each
pinhole, the probability for each pixel to be hit by a photon is
modified by a function accounting for the source shape in the image
plane. 
As an estimate, we can
write
\begin{eqnarray}
N_{ph}^{(SPI)} \approx f \sum_{p_{ij}\in SPI}{S(p_{ij})}\label{eq::4}
\end{eqnarray}
where the sum has to be performed over all the pixels of the SPI. 
The function $S(p_{ij})$, giving the 2D shape of the source in the image plane, is such that
$\max_{i,j}{S(p_{ij})} =1$, stating that no more than one photon per
pixel has to be collected.
In the typical case, considering a $2k\times 2k$ pixels CCD chip and a
$20\times20$ pinhole array, a chip region of $100\times 100$ pixels is
available for each SPI.
Assuming, for instance, a Gaussian-shaped image with
$\sigma_{x,y}=10\,\mathrm{pixel}$ (this is a pretty conservative
choice as for condition (\ref{eq::3})), one gets from formula (\ref{eq::4})
an estimate of about 60 collected photons  per SPI, resulting in
$N_{ph}^{(all)} \approx 2.4\times 10^4$ photons.
\par

As said above, this is the number of photons collected over the whole CCD bandwidth (typically, from a few up to a few tens of keV).
The SNR of each final retrieved image depends on the fraction of photons belonging to the chosen energy range.
In the typical situation encountered in ICF related experiments, as
the ones discussed in the following Section, a number of photons
$N_{ph}^{(image)}$ of the order of $10^3$ can be collected over an energy range of $\sim 100\,\mathrm{eV}$ centered on the strongest emission lines, such as $K$-shell lines from cold or ionized matter. 
Here, such an energy interval is chosen as it corresponds to the maximum spectral resolution achievable with a CCD in the single-photon counting regime.
On the other hand, one has to consider a much larger spectral range,
typically of a few keV, in order to get the same number of photons
 when the X-ray source has to be imaged out corresponding to continuum emission; examples will be given in the next Section.\par
$N_{ph}^{(image)}$ is the total number of photons typically available for each retrieved image. The detailed distribution of such photons over the image depends, of course, on the source shape.
Of course, the local number of photons is higher at the position where
the maximum of the X-ray emission from the source is imaged out.
As an order of magnitude, we can say that up to about $\sim 100$
photons can be observed at this position, thus resulting, according to
the Poisson statistics, in a SNR of $\sim 10$. At the half maximum
position, the corresponding value for the SNR is about $7$.
As a comparison, a SNR of about 30 was reported in \cite{Howe_RSI2006}
for a crystal based 1D imaging configuration employing a CCD detector,
whereas a much worst SNR ($\sim 3$) was obtained using an X-ray film
as a detector.
\par

As a final remark, we observe that the above figures correspond to the case in which the X-ray attenuators are only used to meet the single-photon condition. In other words, no special spectral behavior is sought for by a suitable combination of attenuators.
However, it is, of course, possible to increase the number of photons collected over a selected spectral region by exploiting the energy dependence of different attenuator materials and CCD Quantum Efficiency \cite{Labate_NIMA2008}.
Furthermore, we notice that a spatial binning may also be applied to increase the SNR. As a matter of fact, in the typical situation encountered (magnification $\sim 10$, pinhole diameter $\sim 5-10\,\mathrm{\mu m}$, pixel size $\sim 10-20\,\mathrm{\mu m}$) the spatial resolution allowed, in principle, by the pixel size would be higher than the theoretical resolution allowed by the pinhole diameter. As a consequence, a spatial binning of $2\times2$ or $3\times 3$ pixels could be safely performed at the expense of the spatial sampling. thus increasing the number of photons locally collected.

\section{Examples of application of the EPiC\label{sec::expres}}

In this Section we show two results from the EPiC technique obtained in recent experiments devoted to the study of fast electron transport in the context of ICF.
Only those aspects strictly related to the diagnostic will be dealt with, leaving to other papers a discussion of the underlying physical issues; a discussion of some of these issues can be found, e.g., in \cite{Koester_PPCF2009,Gizzi_NIMA2010}.\par

\begin{figure*}
\begin{center}
\includegraphics[width=0.9\textwidth]{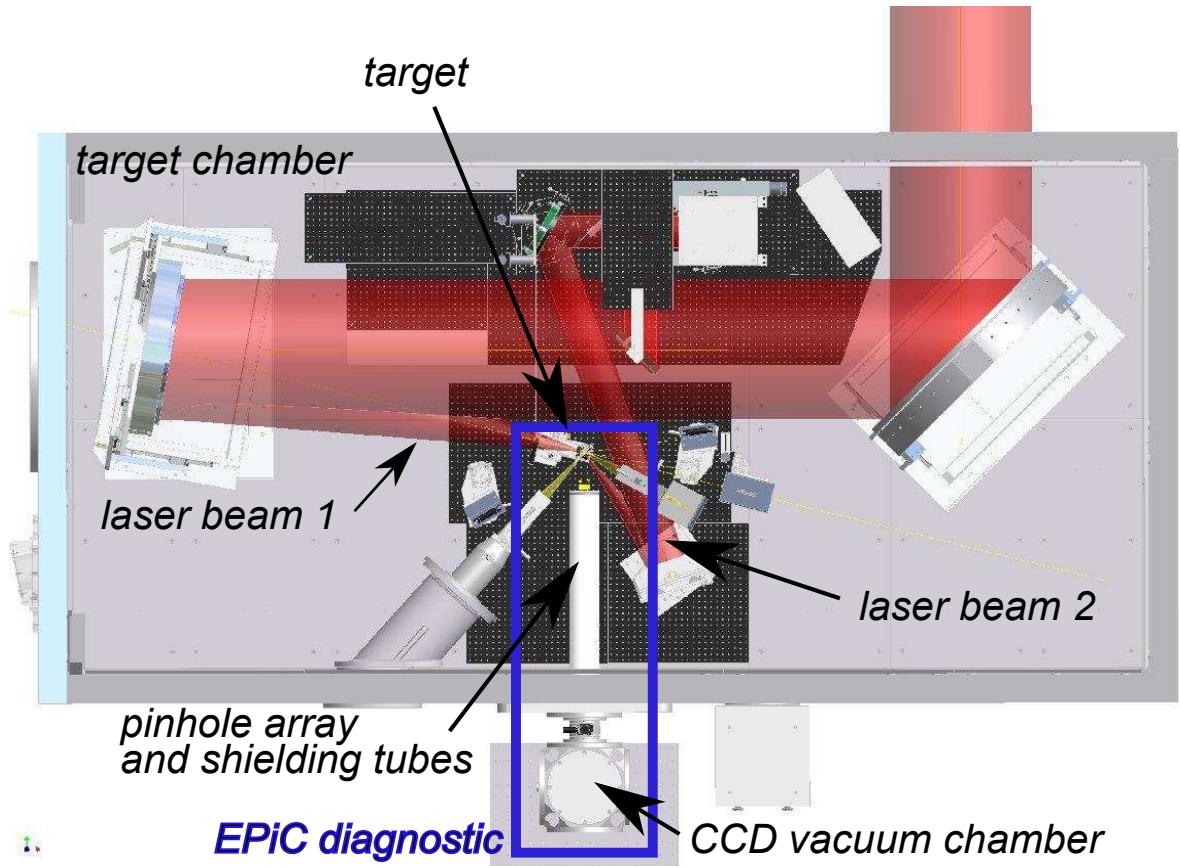}
\caption{(color online) Layout of the experimental setup in a recent experiment carried out at RAL. The target was irradiated by two opposite sides. The overall size of the main vacuum chamber is $2\times 4\,\mathrm{m^2}$. The EPiC is highlighted by the blue box visible at the bottom of the Figure.}
\label{figure3}
\end{center}
\end{figure*}

\begin{figure}
\begin{center}
\includegraphics[width=\columnwidth]{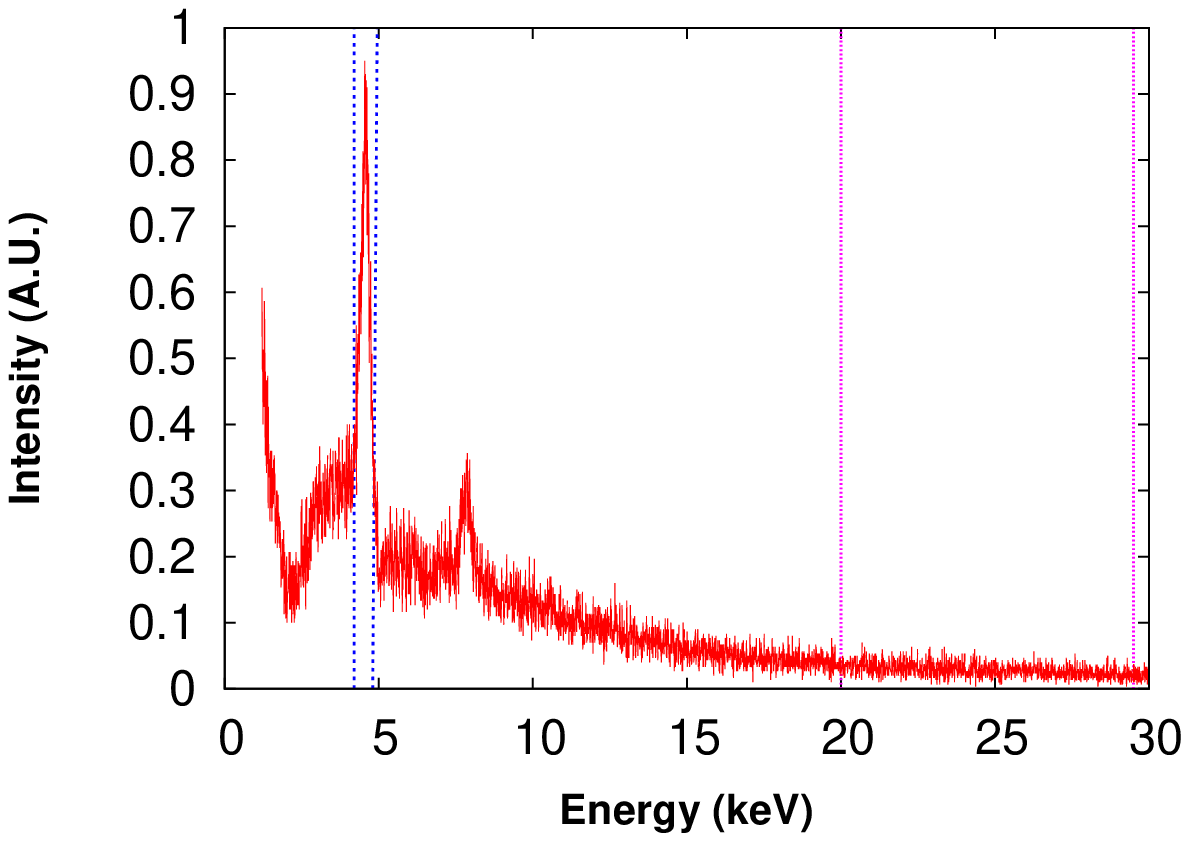}
\caption{(color online) X-ray spectrum retrieved using the EPiC in the RAL experiment. The images of the X-ray source in the two photon energy ranges identified by the vertical lines are shown in Figure \ref{figure5}.}
\label{figure4}
\end{center}
\end{figure}

Figure \ref{figure3} shows a view of the experimental setup in a recent experiment carried out at the Target Area Petawatt (TAP) at the Central Laser Facility of the Rutherford Appleton Laboratory.
The experiment was dedicated to measurement of the heating by fast electrons in the case of two counter-propagating electron beams.
The Vulcan PW laser beam ($700\,\mathrm{fs}$ duration, up to $400\,\mathrm{J}$ energy in that experiment) was split into two beams and focused at an intensity up to about $7\times 10^{18}\,\mathrm{W/cm^2}$ on the opposite sides of solid targets (the two laser beams are shown in the Figure).
A number of different diagnostics were employed (bent crystal spectrometers, X-ray streak-camera, visible diagnostics) which are visible in the Figure; we skip here any further detail on those diagnostics.
The EPiC was used to image out the plasma on a broad energy range.
The diagnostic is highlighted in the Figure by a blue box. The CCD camera was located in a separate vacuum chamber (gray square box at the bottom of the Figure).
The pinhole array was placed at about $15\,\mathrm{cm}$ from the target.
A magnet assembly was placed between the plasma and the array to deflect charged particles and a multi-layer tube (plastic+Al+plastic) was used to reduce background high energy $e^-/\gamma$ particles on the CCD (the tube is clearly visible, entering from the chamber wall at the bottom of the image and going toward the target).
The pinhole array consisted of $20\times 20$ pinholes with $\sim 5 \,\mathrm{\mu m}$ diameter; the pitch was $200\,\mathrm{\mu m}$.
The array substrate was a $200\,\mathrm{\mu m}$ thick Pt foil.
Figure \ref{figure4} shows the emission spectrum, provided by the EPiC, obtained when irradiating a $5\,\mathrm{\mu m}$ thick, $100\times100\,\mathrm{\mu m}^2$ transverse size Ti target.
The main X-ray emission line in the spectrum is the Ti He$\alpha$ line, located  at $\sim 4.75\,\mathrm{keV}$.
The other line in the spectrum was attributed to K-shell emission from Cu, originating from the target stalk which was partially heated by the wings of the laser beams.\par

\begin{figure}
\begin{center}
\includegraphics[width=\columnwidth]{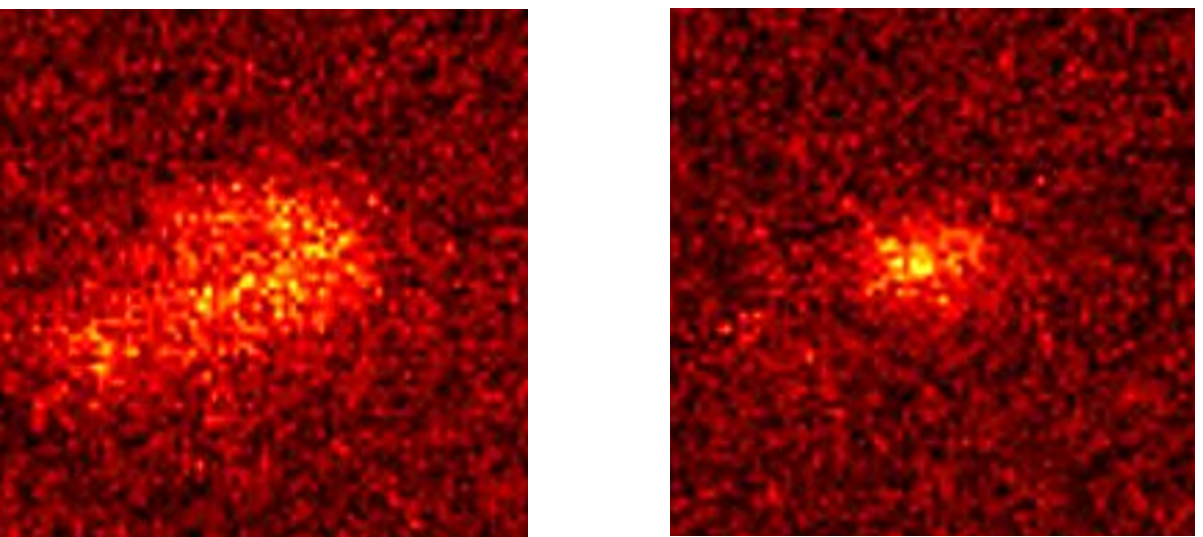}
\caption{(color online) X-ray source as retrieved by the EPiC at the Ti He$\alpha$ energy (left) and at  higher  energy ($>20\,\mathrm{keV}$) (right) in the RAL experiment. 
The corresponding energy intervals are also highlighted in Figure \ref{figure4} by the blue (left image) and purple (right image) vertical lines.
The overall size of the images shown is $67.5\times 67.5 \,\mathrm{\mu m^2}$.}
\label{figure5}
\end{center}
\end{figure}

\begin{figure}
\begin{center}
\includegraphics[width=\columnwidth]{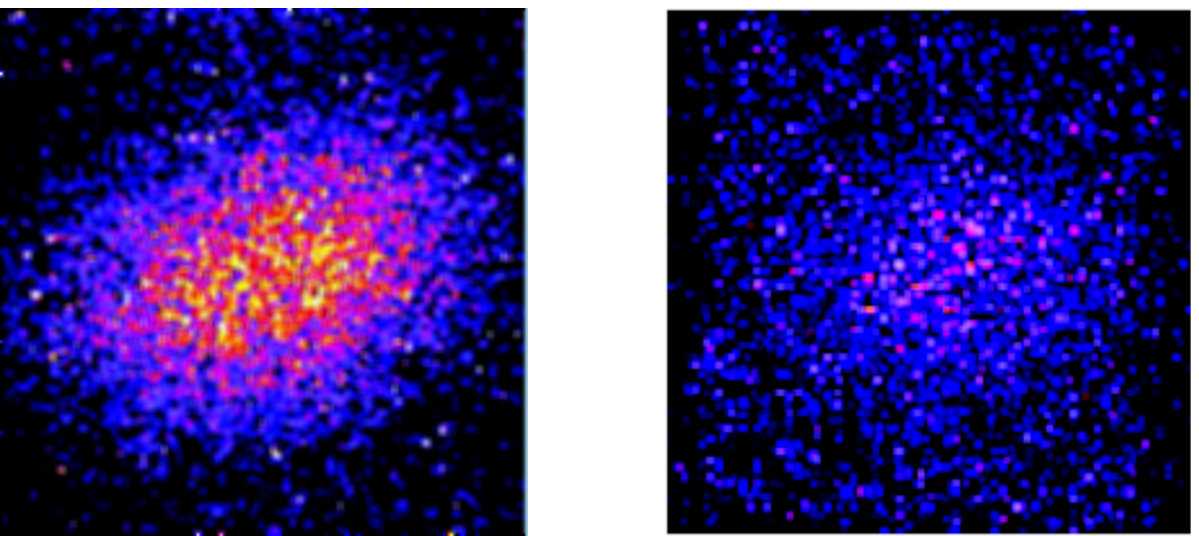}
\caption{(color online)  X-ray source as retrieved by the EPiC at the Ti K$\alpha$ energy (left) and at  higher  energy ($>10\,\mathrm{keV}$) (right) in the PALS experiment. The overall size of the images shown is $81\times 81 \,\mathrm{\mu m^2}$.  }
\label{figure6}
\end{center}
\end{figure}

Figure \ref{figure5} shows the EPiC images of the emitting plasma at the energy corresponding to the Ti He$\alpha$ line (left) and to the \textit{Bremsstrahlung} emission (right) at high photon energy ($>20\,\mathrm{keV}$), respectively.
In other words, only photons with energy  close to the He$\alpha$ energy were considered to recover the image on the left (and similarly, only photons with energy $>20\,\mathrm{keV}$ in the case on the right).
The slight asymmetry of the left image was attributed to a corresponding slight misalignment between the two laser beams incident on the two sides of the target.
The first conclusion one can draw from Figure \ref{figure5} is that the high energy continuum emission is spatially confined to a smaller region than the He$\alpha$ emission region, i.e. to the region of highest laser irradiance.\par

Figure \ref{figure6} shows similar data taken in  a different situation where continuum X-ray radiation is emitted from a region whose size is comparable to the line emission region.
The data were taken in an experiment carried out at the Prague Asterix Laser System (PALS) laboratory in Prague, devoted to study the role of the fast electrons in the shock ignition scenario.
Here the $400\,\mathrm{ps}$ duration, up to $\sim 300\,\mathrm{J}$ energy laser pulses were focused at an intensity up to $\sim 10^{16}\,\mathrm{W/cm^2}$ on thick targets.
Figure \ref{figure6} refers to a 2-layer Ti/Cu target and shows, in particular, the X-ray source imaged out at the Ti K$\alpha$ line and at higher ($>10\,\mathrm{keV}$) energy.
The two experiments demonstrate the capability of EPiC to provide a unique way to study the distribution of X-ray emission from hot and dense plasmas.\par

Finally, we want to briefly discuss here how the results provided by our diagnostic can be cross-checked with similar results obtained with other comparable, well established diagnostics.
We point out that this is not a straightforward task, as no other 2D
imaging diagnostic is available with a spectral discrimination
capability over such a broad energy range (spectrometers based on bent
crystals, for instance, allow 2D imaging over a very narrow spectral range when used in a 2D configuration \cite{Blasco_RSI2001}).
However, during the PALS experiment described above, we compared the image provided by the EPiC diagnostic integrated over the whole useful energy range to just one of the images through one of the pinholes acquired, in an \textit{ad hoc} laser shot, in a condition in which the CCD was not forced to operate in the single-photon mode (that is, no X-ray attenuators were used).
In such a configuration, the diagnostic is just equivalent to a conventional pinhole camera (or, better, to $N_{pinholes}$ independent pinhole cameras), so that the image provided by whatever pinhole should match the one retrieved using the EPiC scheme retaining all the collected photons (provided that the experimental conditions of the two shots as for the laser energy, target material and size, laser focusing, pre-plasma conditions and so on were the same).
Although being a rather indirect check, this allowed to rule out any possible effect due to the EPiC optical scheme potentially leading to image artifacts.
Furthermore, we point out that the plasma created in the PALS experiment was also spatially characterized by an X-ray interferometric diagnostic\cite{Nejdl_PoP2010}, whose results concerning the plasma size were fully consistent with the ones obtained using the EPiC scheme.

\section{Discussion and perspectives}
As we have shown, the EPiC diagnostic allows laser-plasmas to be imaged out in any X-ray photon energy interval, provided that a sufficient number of photons had been collected in that interval.
In detail, let us confine ourselves to the most common situation in which no ad-hoc X-ray attenuators are used to selectively attenuate the X-ray emission lines.
In this case  the smallest photon energy range needed to image out the X-ray source at a photon energy corresponding to a spectral line might be as small as the typical spectral resolution attainable with CCD detectors operating in the single-photon mode \cite{Fraser_NIMA1994}, that is of the order of a few percent.
In other words, a spectral region of a few tens of eV can be imaged out, around an X-ray emission line centered at a few keV or a few tens of keV.
Continuum (Bremsstrahlung) emission can also be imaged out at the same time; in this case, as we have shown above, a relatively larger energy range (typically of a few keV) is needed in order to have a sufficient number of photons to recover an image.
It is worth observing here that the ultimate spectral resolution achievable within the EPiC scheme is not comparable to the one of Bragg crystal spectrometers; nevertheless, it still allows a number of studies in different fields of ICF physics to be carried out.
Actually, it is important to point out that the EPiC diagnostic is mainly intended to observe a broad energy range, allowing to study, on a single-shot basis, both line and continuum emission.
This is a striking difference with respect to other kind of X-ray imaging spectrometers based on different principles.\par
A further feature of the single-shot EPiC is its capability to provide an absolute measurement of the X-ray flux, once the geometry had been fully known and the CCD detector Quantum Efficiency characterized \cite{Fourment_RSI2009,Maddox_RSI2008Cali,Howe_RSI2006}.\par
Finally, a few words about the available energy range.
As it is well known, the Quantum Efficiency of currently available CCD detectors swiftly drops down in the hard X-ray range, thus hindering the possibility of their use at high energy.
In particular, over the past few years, the use of single-photon, almost Fano-limited CCD spectroscopy has been reported up to a few tens of keV \cite{Labate_NIMA2008,Levato_NIMA2008,Fourment_RSI2009}.
Nevertheless, we observe that deep depletion CCD detectors are actually actively developed, allowing a more and more enhanced sensitivity at higher energy.
Moreover, as it can be easily realized, the conceptual scheme of the EPiC is not limited to the use of CCD cameras as detectors.
Indeed, any pixelized hard X/$\gamma$-ray detector with single-photon spectral capability can be used instead, such as, e.g., a CdTe based detector \cite{Chmeissani_IEEETNS2004}.
While this kind of detectors currently provides a limited number of pixels, a growing effort is ongoing to increase this figure.
This could open up new perspectives for an X/$\gamma$-ray diagnostic with imaging capability based on an EPiC scheme.

\section{Summary and conclusions}
A novel X-ray diagnostic, particularly suited for laser-plasma experiments in regimes relevant for the ICF, has been described.
As we have shown, the single-shot EPiC technique allows 2D images of hot and dense plasmas to be obtained at any X-ray photon energy, over a large energy domain, on a single-shot basis.
In particular, we have shown that our diagnostic is particularly useful when X-ray fluorescence has to be observed in order, e.g., to diagnose fast electron generation and transport.
The device (named Energy-encoded Pinhole Camera - EPiC) is based upon the use of an array of many pinholes
coupled to a large area CCD camera operating in the
single-photon mode.
The available X-ray spectral domain is only limited by the Quantum Efficiency
of the employed X-ray CCD camera, thus typically extending from a few keV up to
a few tens of keV.
Spectral 2D images of the emitting plasma can be obtained with spatial resolution comparable to the one of a conventional pinhole camera scheme.
We have reported results from two recent experiments where the EPiC was used. The two experiments demonstrate the capability of EPiC to provide a unique way to study the distribution of X-ray emission from laser-plasmas.
The unprecedented available spectral range of our X-ray imaging diagnostic provides a novel tool to characterize laser-plasma interaction regimes and  helps understanding complex processes like target heating and fast electron transport.

\begin{acknowledgments}
This research programme was partially supported by the project of European large scale infrastructure HiPER (``High Power laser Energy Research facility'').
We acknowledge support from the Italian Ministry of Education, University and Research through the PRIN project ``Fusione a confinamento inerziale via laser, con ignizione indotta da onde d'urto intense (shock ignition)''. The access to the RAL and PALS facilities was partially supported by the EU LaserLab II network.
We thank L. Antonelli, D. Batani, N. Booth, C.A. Cecchetti, Hui Chen, O. Ciricosta, G. Gregori, M. Kozlova, Bin Li,  M. Makita, D. Margarone, J. Mithen,  C. Murphy, M. Notley, A. Patria,  R. Pattathil, D. Riley, B. Rus, M. Sawitcka, N. Woolsey for their collaboration during the RAL and PALS experiments.

\end{acknowledgments}


\begin{thebibliography}{71}%
\makeatletter
\providecommand \@ifxundefined [1]{%
 \@ifx{#1\undefined}
}%
\providecommand \@ifnum [1]{%
 \ifnum #1\expandafter \@firstoftwo
 \else \expandafter \@secondoftwo
 \fi
}%
\providecommand \@ifx [1]{%
 \ifx #1\expandafter \@firstoftwo
 \else \expandafter \@secondoftwo
 \fi
}%
\providecommand \natexlab [1]{#1}%
\providecommand \enquote  [1]{``#1''}%
\providecommand \bibnamefont  [1]{#1}%
\providecommand \bibfnamefont [1]{#1}%
\providecommand \citenamefont [1]{#1}%
\providecommand \href@noop [0]{\@secondoftwo}%
\providecommand \href [0]{\begingroup \@sanitize@url \@href}%
\providecommand \@href[1]{\@@startlink{#1}\@@href}%
\providecommand \@@href[1]{\endgroup#1\@@endlink}%
\providecommand \@sanitize@url [0]{\catcode `\\12\catcode `\$12\catcode
  `\&12\catcode `\#12\catcode `\^12\catcode `\_12\catcode `\%12\relax}%
\providecommand \@@startlink[1]{}%
\providecommand \@@endlink[0]{}%
\providecommand \url  [0]{\begingroup\@sanitize@url \@url }%
\providecommand \@url [1]{\endgroup\@href {#1}{\urlprefix }}%
\providecommand \urlprefix  [0]{URL }%
\providecommand \Eprint [0]{\href }%
\providecommand \doibase [0]{http://dx.doi.org/}%
\providecommand \selectlanguage [0]{\@gobble}%
\providecommand \bibinfo  [0]{\@secondoftwo}%
\providecommand \bibfield  [0]{\@secondoftwo}%
\providecommand \translation [1]{[#1]}%
\providecommand \BibitemOpen [0]{}%
\providecommand \bibitemStop [0]{}%
\providecommand \bibitemNoStop [0]{.\EOS\space}%
\providecommand \EOS [0]{\spacefactor3000\relax}%
\providecommand \BibitemShut  [1]{\csname bibitem#1\endcsname}%
\let\auto@bib@innerbib\@empty
%</preamble>
\bibitem [{\citenamefont {Giulietti}\ and\ \citenamefont
  {Gizzi}(1998)}]{Giulietti_RNC1998}%
  \BibitemOpen
  \bibfield  {author} {\bibinfo {author} {\bibfnamefont {D.}~\bibnamefont
  {Giulietti}}\ and\ \bibinfo {author} {\bibfnamefont {L.~A.}\ \bibnamefont
  {Gizzi}},\ }\href@noop {} {\bibfield  {journal} {\bibinfo  {journal} {La
  Rivista del Nuovo Cimento}\ }\textbf {\bibinfo {volume} {21}},\ \bibinfo
  {pages} {1} (\bibinfo {year} {1998})}\BibitemShut {NoStop}%
\bibitem [{\citenamefont {Gizzi}(2011)}]{Gizzi_SSCP2011}%
  \BibitemOpen
  \bibfield  {author} {\bibinfo {author} {\bibfnamefont {L.~A.}\ \bibnamefont
  {Gizzi}},\ }in\ \href@noop {} {\emph {\bibinfo {booktitle} {Lectures on
  Ultrafast Intense Laser Science}}},\ \bibinfo {series} {Springer Series in
  Chemical Physics}, Vol.~\bibinfo {volume} {94},\ \bibinfo {editor} {edited
  by\ \bibinfo {editor} {\bibnamefont {Springer}}}\ (\bibinfo {year}
  {2011})\BibitemShut {NoStop}%
\bibitem [{\citenamefont {Schneider}\ \emph {et~al.}(2006)\citenamefont
  {Schneider}, \citenamefont {Holder}, \citenamefont {James}, \citenamefont
  {Bruns}, \citenamefont {Celeste}, \citenamefont {Compton}, \citenamefont
  {Costa}, \citenamefont {Ellis}, \citenamefont {Emig}, \citenamefont
  {Hargrove}, \citenamefont {Kalantar}, \citenamefont {MacGowan}, \citenamefont
  {Power}, \citenamefont {Sorce}, \citenamefont {Rekow}, \citenamefont
  {Widmann}, \citenamefont {Young}, \citenamefont {Young}, \citenamefont
  {Garcia}, \citenamefont {McKenney}, \citenamefont {Haugh}, \citenamefont
  {Goldin}, \citenamefont {MacNeil},\ and\ \citenamefont
  {Cone}}]{Schneider_RSI2006}%
  \BibitemOpen
  \bibfield  {author} {\bibinfo {author} {\bibfnamefont {M.~B.}\ \bibnamefont
  {Schneider}}, \bibinfo {author} {\bibfnamefont {J.~P.}\ \bibnamefont
  {Holder}}, \bibinfo {author} {\bibfnamefont {D.~L.}\ \bibnamefont {James}},
  \bibinfo {author} {\bibfnamefont {H.~C.}\ \bibnamefont {Bruns}}, \bibinfo
  {author} {\bibfnamefont {J.~R.}\ \bibnamefont {Celeste}}, \bibinfo {author}
  {\bibfnamefont {S.}~\bibnamefont {Compton}}, \bibinfo {author} {\bibfnamefont
  {R.~L.}\ \bibnamefont {Costa}}, \bibinfo {author} {\bibfnamefont {A.~D.}\
  \bibnamefont {Ellis}}, \bibinfo {author} {\bibfnamefont {J.~A.}\ \bibnamefont
  {Emig}}, \bibinfo {author} {\bibfnamefont {D.}~\bibnamefont {Hargrove}},
  \bibinfo {author} {\bibfnamefont {D.~H.}\ \bibnamefont {Kalantar}}, \bibinfo
  {author} {\bibfnamefont {B.~J.}\ \bibnamefont {MacGowan}}, \bibinfo {author}
  {\bibfnamefont {G.~D.}\ \bibnamefont {Power}}, \bibinfo {author}
  {\bibfnamefont {C.}~\bibnamefont {Sorce}}, \bibinfo {author} {\bibfnamefont
  {V.}~\bibnamefont {Rekow}}, \bibinfo {author} {\bibfnamefont
  {K.}~\bibnamefont {Widmann}}, \bibinfo {author} {\bibfnamefont {B.~K.}\
  \bibnamefont {Young}}, \bibinfo {author} {\bibfnamefont {P.~E.}\ \bibnamefont
  {Young}}, \bibinfo {author} {\bibfnamefont {O.~F.}\ \bibnamefont {Garcia}},
  \bibinfo {author} {\bibfnamefont {J.}~\bibnamefont {McKenney}}, \bibinfo
  {author} {\bibfnamefont {M.}~\bibnamefont {Haugh}}, \bibinfo {author}
  {\bibfnamefont {F.}~\bibnamefont {Goldin}}, \bibinfo {author} {\bibfnamefont
  {L.~P.}\ \bibnamefont {MacNeil}}, \ and\ \bibinfo {author} {\bibfnamefont
  {K.}~\bibnamefont {Cone}},\ }\href {\doibase 10.1063/1.2349748} {\bibfield
  {journal} {\bibinfo  {journal} {Review of Scientific Instruments}\ }\textbf
  {\bibinfo {volume} {77}},\ \bibinfo {eid} {10E321} (\bibinfo {year}
  {2006})}\BibitemShut {NoStop}%
\bibitem [{\citenamefont {Glenn}\ \emph {et~al.}(2010)\citenamefont {Glenn},
  \citenamefont {Koch}, \citenamefont {Bradley}, \citenamefont {Izumi},
  \citenamefont {Bell}, \citenamefont {Holder}, \citenamefont {Stone},
  \citenamefont {Prasad}, \citenamefont {MacKinnon}, \citenamefont {Springer},
  \citenamefont {Landen},\ and\ \citenamefont {Kyrala}}]{Glenn_RSI2010}%
  \BibitemOpen
  \bibfield  {author} {\bibinfo {author} {\bibfnamefont {S.}~\bibnamefont
  {Glenn}}, \bibinfo {author} {\bibfnamefont {J.}~\bibnamefont {Koch}},
  \bibinfo {author} {\bibfnamefont {D.~K.}\ \bibnamefont {Bradley}}, \bibinfo
  {author} {\bibfnamefont {N.}~\bibnamefont {Izumi}}, \bibinfo {author}
  {\bibfnamefont {P.}~\bibnamefont {Bell}}, \bibinfo {author} {\bibfnamefont
  {J.}~\bibnamefont {Holder}}, \bibinfo {author} {\bibfnamefont
  {G.}~\bibnamefont {Stone}}, \bibinfo {author} {\bibfnamefont
  {R.}~\bibnamefont {Prasad}}, \bibinfo {author} {\bibfnamefont
  {A.}~\bibnamefont {MacKinnon}}, \bibinfo {author} {\bibfnamefont
  {P.}~\bibnamefont {Springer}}, \bibinfo {author} {\bibfnamefont {O.~L.}\
  \bibnamefont {Landen}}, \ and\ \bibinfo {author} {\bibfnamefont
  {G.}~\bibnamefont {Kyrala}},\ }\href {\doibase 10.1063/1.3478897} {\bibfield
  {journal} {\bibinfo  {journal} {Review of Scientific Instruments}\ }\textbf
  {\bibinfo {volume} {81}},\ \bibinfo {eid} {10E539} (\bibinfo {year}
  {2010})}\BibitemShut {NoStop}%
\bibitem [{\citenamefont {Schneider}\ \emph {et~al.}(2010)\citenamefont
  {Schneider}, \citenamefont {Jones}, \citenamefont {Meezan}, \citenamefont
  {Milovich}, \citenamefont {Town}, \citenamefont {Alvarez}, \citenamefont
  {Beeler}, \citenamefont {Bradley}, \citenamefont {Celeste}, \citenamefont
  {Dixit}, \citenamefont {Edwards}, \citenamefont {Haugh}, \citenamefont
  {Kalantar}, \citenamefont {Kline}, \citenamefont {Kyrala}, \citenamefont
  {Landen}, \citenamefont {MacGowan}, \citenamefont {Michel}, \citenamefont
  {Moody}, \citenamefont {Oberhelman}, \citenamefont {Piston}, \citenamefont
  {Pivovaroff}, \citenamefont {Suter}, \citenamefont {Teruya}, \citenamefont
  {Thomas}, \citenamefont {Vernon}, \citenamefont {Warrick}, \citenamefont
  {Widmann}, \citenamefont {Wood},\ and\ \citenamefont
  {Young}}]{Schneider_RSI2010}%
  \BibitemOpen
  \bibfield  {author} {\bibinfo {author} {\bibfnamefont {M.~B.}\ \bibnamefont
  {Schneider}}, \bibinfo {author} {\bibfnamefont {O.~S.}\ \bibnamefont
  {Jones}}, \bibinfo {author} {\bibfnamefont {N.~B.}\ \bibnamefont {Meezan}},
  \bibinfo {author} {\bibfnamefont {J.~L.}\ \bibnamefont {Milovich}}, \bibinfo
  {author} {\bibfnamefont {R.~P.}\ \bibnamefont {Town}}, \bibinfo {author}
  {\bibfnamefont {S.~S.}\ \bibnamefont {Alvarez}}, \bibinfo {author}
  {\bibfnamefont {R.~G.}\ \bibnamefont {Beeler}}, \bibinfo {author}
  {\bibfnamefont {D.~K.}\ \bibnamefont {Bradley}}, \bibinfo {author}
  {\bibfnamefont {J.~R.}\ \bibnamefont {Celeste}}, \bibinfo {author}
  {\bibfnamefont {S.~N.}\ \bibnamefont {Dixit}}, \bibinfo {author}
  {\bibfnamefont {M.~J.}\ \bibnamefont {Edwards}}, \bibinfo {author}
  {\bibfnamefont {M.~J.}\ \bibnamefont {Haugh}}, \bibinfo {author}
  {\bibfnamefont {D.~H.}\ \bibnamefont {Kalantar}}, \bibinfo {author}
  {\bibfnamefont {J.~L.}\ \bibnamefont {Kline}}, \bibinfo {author}
  {\bibfnamefont {G.~A.}\ \bibnamefont {Kyrala}}, \bibinfo {author}
  {\bibfnamefont {O.~L.}\ \bibnamefont {Landen}}, \bibinfo {author}
  {\bibfnamefont {B.~J.}\ \bibnamefont {MacGowan}}, \bibinfo {author}
  {\bibfnamefont {P.}~\bibnamefont {Michel}}, \bibinfo {author} {\bibfnamefont
  {J.~D.}\ \bibnamefont {Moody}}, \bibinfo {author} {\bibfnamefont {S.~K.}\
  \bibnamefont {Oberhelman}}, \bibinfo {author} {\bibfnamefont {K.~W.}\
  \bibnamefont {Piston}}, \bibinfo {author} {\bibfnamefont {M.~J.}\
  \bibnamefont {Pivovaroff}}, \bibinfo {author} {\bibfnamefont {L.~J.}\
  \bibnamefont {Suter}}, \bibinfo {author} {\bibfnamefont {A.~T.}\ \bibnamefont
  {Teruya}}, \bibinfo {author} {\bibfnamefont {C.~A.}\ \bibnamefont {Thomas}},
  \bibinfo {author} {\bibfnamefont {S.~P.}\ \bibnamefont {Vernon}}, \bibinfo
  {author} {\bibfnamefont {A.~L.}\ \bibnamefont {Warrick}}, \bibinfo {author}
  {\bibfnamefont {K.}~\bibnamefont {Widmann}}, \bibinfo {author} {\bibfnamefont
  {R.~D.}\ \bibnamefont {Wood}}, \ and\ \bibinfo {author} {\bibfnamefont
  {B.~K.}\ \bibnamefont {Young}},\ }\href {\doibase 10.1063/1.3491316}
  {\bibfield  {journal} {\bibinfo  {journal} {Review of Scientific
  Instruments}\ }\textbf {\bibinfo {volume} {81}},\ \bibinfo {eid} {10E538}
  (\bibinfo {year} {2010})}\BibitemShut {NoStop}%
\bibitem [{\citenamefont {Kimbrough}\ \emph {et~al.}(2010)\citenamefont
  {Kimbrough}, \citenamefont {Bell}, \citenamefont {Bradley}, \citenamefont
  {Holder}, \citenamefont {Kalantar}, \citenamefont {MacPhee},\ and\
  \citenamefont {Telford}}]{Kimbrough_RSI2010}%
  \BibitemOpen
  \bibfield  {author} {\bibinfo {author} {\bibfnamefont {J.~R.}\ \bibnamefont
  {Kimbrough}}, \bibinfo {author} {\bibfnamefont {P.~M.}\ \bibnamefont {Bell}},
  \bibinfo {author} {\bibfnamefont {D.~K.}\ \bibnamefont {Bradley}}, \bibinfo
  {author} {\bibfnamefont {J.~P.}\ \bibnamefont {Holder}}, \bibinfo {author}
  {\bibfnamefont {D.~K.}\ \bibnamefont {Kalantar}}, \bibinfo {author}
  {\bibfnamefont {A.~G.}\ \bibnamefont {MacPhee}}, \ and\ \bibinfo {author}
  {\bibfnamefont {S.}~\bibnamefont {Telford}},\ }\href {\doibase
  10.1063/1.3496990} {\bibfield  {journal} {\bibinfo  {journal} {Review of
  Scientific Instruments}\ }\textbf {\bibinfo {volume} {81}},\ \bibinfo {eid}
  {10E530} (\bibinfo {year} {2010})}\BibitemShut {NoStop}%
\bibitem [{\citenamefont {Hammel}\ \emph {et~al.}(2011)\citenamefont {Hammel},
  \citenamefont {Scott}, \citenamefont {Regan}, \citenamefont {Cerjan},
  \citenamefont {Clark}, \citenamefont {Edwards}, \citenamefont {Epstein},
  \citenamefont {Glenzer}, \citenamefont {Haan}, \citenamefont {Izumi},
  \citenamefont {Koch}, \citenamefont {Kyrala}, \citenamefont {Landen},
  \citenamefont {Langer}, \citenamefont {Peterson}, \citenamefont {Smalyuk},
  \citenamefont {Suter},\ and\ \citenamefont {Wilson}}]{Hammel_PoP2011}%
  \BibitemOpen
  \bibfield  {author} {\bibinfo {author} {\bibfnamefont {B.~A.}\ \bibnamefont
  {Hammel}}, \bibinfo {author} {\bibfnamefont {H.~A.}\ \bibnamefont {Scott}},
  \bibinfo {author} {\bibfnamefont {S.~P.}\ \bibnamefont {Regan}}, \bibinfo
  {author} {\bibfnamefont {C.}~\bibnamefont {Cerjan}}, \bibinfo {author}
  {\bibfnamefont {D.~S.}\ \bibnamefont {Clark}}, \bibinfo {author}
  {\bibfnamefont {M.~J.}\ \bibnamefont {Edwards}}, \bibinfo {author}
  {\bibfnamefont {R.}~\bibnamefont {Epstein}}, \bibinfo {author} {\bibfnamefont
  {S.~H.}\ \bibnamefont {Glenzer}}, \bibinfo {author} {\bibfnamefont {S.~W.}\
  \bibnamefont {Haan}}, \bibinfo {author} {\bibfnamefont {N.}~\bibnamefont
  {Izumi}}, \bibinfo {author} {\bibfnamefont {J.~A.}\ \bibnamefont {Koch}},
  \bibinfo {author} {\bibfnamefont {G.~A.}\ \bibnamefont {Kyrala}}, \bibinfo
  {author} {\bibfnamefont {O.~L.}\ \bibnamefont {Landen}}, \bibinfo {author}
  {\bibfnamefont {S.~H.}\ \bibnamefont {Langer}}, \bibinfo {author}
  {\bibfnamefont {K.}~\bibnamefont {Peterson}}, \bibinfo {author}
  {\bibfnamefont {V.~A.}\ \bibnamefont {Smalyuk}}, \bibinfo {author}
  {\bibfnamefont {L.~J.}\ \bibnamefont {Suter}}, \ and\ \bibinfo {author}
  {\bibfnamefont {D.~C.}\ \bibnamefont {Wilson}},\ }\href {\doibase
  10.1063/1.3567520} {\bibfield  {journal} {\bibinfo  {journal} {Physics of
  Plasmas}\ }\textbf {\bibinfo {volume} {18}},\ \bibinfo {eid} {056310}
  (\bibinfo {year} {2011})}\BibitemShut {NoStop}%
\bibitem [{\citenamefont {Hansen}(2012)}]{Hansen_PoP2012}%
  \BibitemOpen
  \bibfield  {author} {\bibinfo {author} {\bibfnamefont {S.~B.}\ \bibnamefont
  {Hansen}},\ }\href@noop {} {\bibfield  {journal} {\bibinfo  {journal}
  {Physics of Plasmas}\ }\textbf {\bibinfo {volume} {19}},\ \bibinfo {pages}
  {056312} (\bibinfo {year} {2012})}\BibitemShut {NoStop}%
\bibitem [{\citenamefont {Tabak}\ \emph {et~al.}(1994)\citenamefont {Tabak},
  \citenamefont {Hammer}, \citenamefont {Glinsky}, \citenamefont {Kruer},
  \citenamefont {Wilks}, \citenamefont {Woodworth}, \citenamefont {Campbell},\
  and\ \citenamefont {Perry}}]{Tabak_PoP1994}%
  \BibitemOpen
  \bibfield  {author} {\bibinfo {author} {\bibfnamefont {M.}~\bibnamefont
  {Tabak}}, \bibinfo {author} {\bibfnamefont {J.}~\bibnamefont {Hammer}},
  \bibinfo {author} {\bibfnamefont {M.~E.}\ \bibnamefont {Glinsky}}, \bibinfo
  {author} {\bibfnamefont {W.~L.}\ \bibnamefont {Kruer}}, \bibinfo {author}
  {\bibfnamefont {S.~C.}\ \bibnamefont {Wilks}}, \bibinfo {author}
  {\bibfnamefont {J.}~\bibnamefont {Woodworth}}, \bibinfo {author}
  {\bibfnamefont {E.~M.}\ \bibnamefont {Campbell}}, \ and\ \bibinfo {author}
  {\bibfnamefont {M.~D.}\ \bibnamefont {Perry}},\ }\href@noop {} {\bibfield
  {journal} {\bibinfo  {journal} {Physics of Plasmas}\ }\textbf {\bibinfo
  {volume} {1}},\ \bibinfo {pages} {1626} (\bibinfo {year} {1994})}\BibitemShut
  {NoStop}%
\bibitem [{\citenamefont {Key}(2007)}]{Key_PoP2007}%
  \BibitemOpen
  \bibfield  {author} {\bibinfo {author} {\bibfnamefont {M.~H.}\ \bibnamefont
  {Key}},\ }\href@noop {} {\bibfield  {journal} {\bibinfo  {journal} {Physics
  of Plasmas}\ }\textbf {\bibinfo {volume} {14}},\ \bibinfo {pages} {055502}
  (\bibinfo {year} {2007})}\BibitemShut {NoStop}%
\bibitem [{\citenamefont {Norreys}\ \emph {et~al.}(2009)\citenamefont
  {Norreys}, \citenamefont {Scott}, \citenamefont {Lancaster}, \citenamefont
  {Green}, \citenamefont {Robinson}, \citenamefont {Sherlock}, \citenamefont
  {Evans}, \citenamefont {Haines}, \citenamefont {Kar}, \citenamefont {Zepf},
  \citenamefont {Key}, \citenamefont {King}, \citenamefont {Ma}, \citenamefont
  {Yabuuchi}, \citenamefont {Wei}, \citenamefont {Beg}, \citenamefont {Nilson},
  \citenamefont {Theobald}, \citenamefont {Stephens}, \citenamefont {Valente},
  \citenamefont {Davies}, \citenamefont {Takeda}, \citenamefont {Azechi},
  \citenamefont {Nakatsutsumi}, \citenamefont {Tanimoto}, \citenamefont
  {Kodama},\ and\ \citenamefont {Tanaka}}]{Norreys_NF2009}%
  \BibitemOpen
  \bibfield  {author} {\bibinfo {author} {\bibfnamefont {P.~A.}\ \bibnamefont
  {Norreys}}, \bibinfo {author} {\bibfnamefont {R.~H.~H.}\ \bibnamefont
  {Scott}}, \bibinfo {author} {\bibfnamefont {K.~L.}\ \bibnamefont
  {Lancaster}}, \bibinfo {author} {\bibfnamefont {J.~S.}\ \bibnamefont
  {Green}}, \bibinfo {author} {\bibfnamefont {A.~P.~L.}\ \bibnamefont
  {Robinson}}, \bibinfo {author} {\bibfnamefont {M.}~\bibnamefont {Sherlock}},
  \bibinfo {author} {\bibfnamefont {R.~G.}\ \bibnamefont {Evans}}, \bibinfo
  {author} {\bibfnamefont {M.~G.}\ \bibnamefont {Haines}}, \bibinfo {author}
  {\bibfnamefont {S.}~\bibnamefont {Kar}}, \bibinfo {author} {\bibfnamefont
  {M.}~\bibnamefont {Zepf}}, \bibinfo {author} {\bibfnamefont {M.~H.}\
  \bibnamefont {Key}}, \bibinfo {author} {\bibfnamefont {J.}~\bibnamefont
  {King}}, \bibinfo {author} {\bibfnamefont {T.}~\bibnamefont {Ma}}, \bibinfo
  {author} {\bibfnamefont {T.}~\bibnamefont {Yabuuchi}}, \bibinfo {author}
  {\bibfnamefont {M.~S.}\ \bibnamefont {Wei}}, \bibinfo {author} {\bibfnamefont
  {F.~N.}\ \bibnamefont {Beg}}, \bibinfo {author} {\bibfnamefont
  {P.}~\bibnamefont {Nilson}}, \bibinfo {author} {\bibfnamefont
  {W.}~\bibnamefont {Theobald}}, \bibinfo {author} {\bibfnamefont {R.~B.}\
  \bibnamefont {Stephens}}, \bibinfo {author} {\bibfnamefont {J.}~\bibnamefont
  {Valente}}, \bibinfo {author} {\bibfnamefont {J.~R.}\ \bibnamefont {Davies}},
  \bibinfo {author} {\bibfnamefont {K.}~\bibnamefont {Takeda}}, \bibinfo
  {author} {\bibfnamefont {H.}~\bibnamefont {Azechi}}, \bibinfo {author}
  {\bibfnamefont {N.}~\bibnamefont {Nakatsutsumi}}, \bibinfo {author}
  {\bibfnamefont {T.}~\bibnamefont {Tanimoto}}, \bibinfo {author}
  {\bibfnamefont {R.}~\bibnamefont {Kodama}}, \ and\ \bibinfo {author}
  {\bibfnamefont {K.~A.}\ \bibnamefont {Tanaka}},\ }\href@noop {} {\bibfield
  {journal} {\bibinfo  {journal} {Nuclear Fusion}\ }\textbf {\bibinfo {volume}
  {49}},\ \bibinfo {pages} {104023} (\bibinfo {year} {2009})}\BibitemShut
  {NoStop}%
\bibitem [{\citenamefont {Lancaster}\ \emph {et~al.}(2007)\citenamefont
  {Lancaster}, \citenamefont {Green}, \citenamefont {Hey}, \citenamefont
  {Akli}, \citenamefont {Davies}, \citenamefont {Clarke}, \citenamefont
  {Freeman}, \citenamefont {Habara}, \citenamefont {Key}, \citenamefont
  {Kodama}, \citenamefont {Krushelnick}, \citenamefont {Murphy}, \citenamefont
  {Nakatsutsumi}, \citenamefont {Simpson}, \citenamefont {Stephens},
  \citenamefont {Stoeckl}, \citenamefont {Yabuuchi}, \citenamefont {Zepf},\
  and\ \citenamefont {Norreys}}]{Lancaster_PRL2007}%
  \BibitemOpen
  \bibfield  {author} {\bibinfo {author} {\bibfnamefont {K.~L.}\ \bibnamefont
  {Lancaster}}, \bibinfo {author} {\bibfnamefont {J.~S.}\ \bibnamefont
  {Green}}, \bibinfo {author} {\bibfnamefont {D.~S.}\ \bibnamefont {Hey}},
  \bibinfo {author} {\bibfnamefont {K.~U.}\ \bibnamefont {Akli}}, \bibinfo
  {author} {\bibfnamefont {J.~R.}\ \bibnamefont {Davies}}, \bibinfo {author}
  {\bibfnamefont {R.~J.}\ \bibnamefont {Clarke}}, \bibinfo {author}
  {\bibfnamefont {R.~R.}\ \bibnamefont {Freeman}}, \bibinfo {author}
  {\bibfnamefont {H.}~\bibnamefont {Habara}}, \bibinfo {author} {\bibfnamefont
  {M.~H.}\ \bibnamefont {Key}}, \bibinfo {author} {\bibfnamefont
  {R.}~\bibnamefont {Kodama}}, \bibinfo {author} {\bibfnamefont
  {K.}~\bibnamefont {Krushelnick}}, \bibinfo {author} {\bibfnamefont {C.~D.}\
  \bibnamefont {Murphy}}, \bibinfo {author} {\bibfnamefont {M.}~\bibnamefont
  {Nakatsutsumi}}, \bibinfo {author} {\bibfnamefont {P.}~\bibnamefont
  {Simpson}}, \bibinfo {author} {\bibfnamefont {R.}~\bibnamefont {Stephens}},
  \bibinfo {author} {\bibfnamefont {C.}~\bibnamefont {Stoeckl}}, \bibinfo
  {author} {\bibfnamefont {T.}~\bibnamefont {Yabuuchi}}, \bibinfo {author}
  {\bibfnamefont {M.}~\bibnamefont {Zepf}}, \ and\ \bibinfo {author}
  {\bibfnamefont {P.~A.}\ \bibnamefont {Norreys}},\ }\href@noop {} {\bibfield
  {journal} {\bibinfo  {journal} {Physical Review Letters}\ }\textbf {\bibinfo
  {volume} {98}},\ \bibinfo {pages} {125002} (\bibinfo {year}
  {2007})}\BibitemShut {NoStop}%
\bibitem [{\citenamefont {Wei}\ \emph {et~al.}(2004)\citenamefont {Wei},
  \citenamefont {Beg}, \citenamefont {Clark}, \citenamefont {Dangor},
  \citenamefont {Evans}, \citenamefont {Gopal}, \citenamefont {Ledingham},
  \citenamefont {McKenna}, \citenamefont {Norreys}, \citenamefont {Tatarakis},
  \citenamefont {Zepf},\ and\ \citenamefont {Krushelnick}}]{Wei_PRE2004}%
  \BibitemOpen
  \bibfield  {author} {\bibinfo {author} {\bibfnamefont {M.~S.}\ \bibnamefont
  {Wei}}, \bibinfo {author} {\bibfnamefont {F.~N.}\ \bibnamefont {Beg}},
  \bibinfo {author} {\bibfnamefont {E.~L.}\ \bibnamefont {Clark}}, \bibinfo
  {author} {\bibfnamefont {A.~E.}\ \bibnamefont {Dangor}}, \bibinfo {author}
  {\bibfnamefont {R.~G.}\ \bibnamefont {Evans}}, \bibinfo {author}
  {\bibfnamefont {A.}~\bibnamefont {Gopal}}, \bibinfo {author} {\bibfnamefont
  {K.~W.~D.}\ \bibnamefont {Ledingham}}, \bibinfo {author} {\bibfnamefont
  {P.}~\bibnamefont {McKenna}}, \bibinfo {author} {\bibfnamefont {P.~A.}\
  \bibnamefont {Norreys}}, \bibinfo {author} {\bibfnamefont {M.}~\bibnamefont
  {Tatarakis}}, \bibinfo {author} {\bibfnamefont {M.}~\bibnamefont {Zepf}}, \
  and\ \bibinfo {author} {\bibfnamefont {K.}~\bibnamefont {Krushelnick}},\
  }\href {\doibase 10.1103/PhysRevE.70.056412} {\bibfield  {journal} {\bibinfo
  {journal} {Phys. Rev. E}\ }\textbf {\bibinfo {volume} {70}},\ \bibinfo
  {pages} {056412} (\bibinfo {year} {2004})}\BibitemShut {NoStop}%
\bibitem [{\citenamefont {Labate}\ \emph
  {et~al.}(2007{\natexlab{a}})\citenamefont {Labate}, \citenamefont
  {Galimberti}, \citenamefont {Giulietti}, \citenamefont {Giulietti},
  \citenamefont {K\"oster}, \citenamefont {Tomassini},\ and\ \citenamefont
  {Gizzi}}]{Labate_APB2007}%
  \BibitemOpen
  \bibfield  {author} {\bibinfo {author} {\bibfnamefont {L.}~\bibnamefont
  {Labate}}, \bibinfo {author} {\bibfnamefont {M.}~\bibnamefont {Galimberti}},
  \bibinfo {author} {\bibfnamefont {A.}~\bibnamefont {Giulietti}}, \bibinfo
  {author} {\bibfnamefont {D.}~\bibnamefont {Giulietti}}, \bibinfo {author}
  {\bibfnamefont {P.}~\bibnamefont {K\"oster}}, \bibinfo {author}
  {\bibfnamefont {P.}~\bibnamefont {Tomassini}}, \ and\ \bibinfo {author}
  {\bibfnamefont {L.~A.}\ \bibnamefont {Gizzi}},\ }\href@noop {} {\bibfield
  {journal} {\bibinfo  {journal} {Applied Physics B}\ }\textbf {\bibinfo
  {volume} {86}},\ \bibinfo {pages} {229} (\bibinfo {year}
  {2007}{\natexlab{a}})}\BibitemShut {NoStop}%
\bibitem [{\citenamefont {Norreys}\ \emph {et~al.}(2006)\citenamefont
  {Norreys}, \citenamefont {Green}, \citenamefont {Davies}, \citenamefont
  {Tatarakis}, \citenamefont {Clark}, \citenamefont {Beg}, \citenamefont
  {Dangor}, \citenamefont {Lancaster}, \citenamefont {Wei}, \citenamefont
  {Zepf},\ and\ \citenamefont {Krushelnick}}]{Norreys_PPCF2006}%
  \BibitemOpen
  \bibfield  {author} {\bibinfo {author} {\bibfnamefont {P.~A.}\ \bibnamefont
  {Norreys}}, \bibinfo {author} {\bibfnamefont {J.~S.}\ \bibnamefont {Green}},
  \bibinfo {author} {\bibfnamefont {J.~R.}\ \bibnamefont {Davies}}, \bibinfo
  {author} {\bibfnamefont {M.}~\bibnamefont {Tatarakis}}, \bibinfo {author}
  {\bibfnamefont {E.~L.}\ \bibnamefont {Clark}}, \bibinfo {author}
  {\bibfnamefont {F.~N.}\ \bibnamefont {Beg}}, \bibinfo {author} {\bibfnamefont
  {A.~E.}\ \bibnamefont {Dangor}}, \bibinfo {author} {\bibfnamefont {K.~L.}\
  \bibnamefont {Lancaster}}, \bibinfo {author} {\bibfnamefont {M.~S.}\
  \bibnamefont {Wei}}, \bibinfo {author} {\bibfnamefont {M.}~\bibnamefont
  {Zepf}}, \ and\ \bibinfo {author} {\bibfnamefont {K.}~\bibnamefont
  {Krushelnick}},\ }\href {http://stacks.iop.org/0741-3335/48/i=2/a=L01}
  {\bibfield  {journal} {\bibinfo  {journal} {Plasma Physics and Controlled
  Fusion}\ }\textbf {\bibinfo {volume} {48}},\ \bibinfo {pages} {L11} (\bibinfo
  {year} {2006})}\BibitemShut {NoStop}%
\bibitem [{\citenamefont {Manclossi}\ \emph {et~al.}(2006)\citenamefont
  {Manclossi}, \citenamefont {Santos}, \citenamefont {Batani}, \citenamefont
  {Faure}, \citenamefont {Debayle}, \citenamefont {Tikhonchuk},\ and\
  \citenamefont {Malka}}]{Manclossi_PRL2006}%
  \BibitemOpen
  \bibfield  {author} {\bibinfo {author} {\bibfnamefont {M.}~\bibnamefont
  {Manclossi}}, \bibinfo {author} {\bibfnamefont {J.~J.}\ \bibnamefont
  {Santos}}, \bibinfo {author} {\bibfnamefont {D.}~\bibnamefont {Batani}},
  \bibinfo {author} {\bibfnamefont {J.}~\bibnamefont {Faure}}, \bibinfo
  {author} {\bibfnamefont {A.}~\bibnamefont {Debayle}}, \bibinfo {author}
  {\bibfnamefont {V.~T.}\ \bibnamefont {Tikhonchuk}}, \ and\ \bibinfo {author}
  {\bibfnamefont {V.}~\bibnamefont {Malka}},\ }\href {\doibase
  10.1103/PhysRevLett.96.125002} {\bibfield  {journal} {\bibinfo  {journal}
  {Phys. Rev. Lett.}\ }\textbf {\bibinfo {volume} {96}},\ \bibinfo {pages}
  {125002} (\bibinfo {year} {2006})}\BibitemShut {NoStop}%
\bibitem [{\citenamefont {Santos}\ \emph {et~al.}(2002)\citenamefont {Santos},
  \citenamefont {Amiranoff}, \citenamefont {Baton}, \citenamefont {Gremillet},
  \citenamefont {Koenig}, \citenamefont {Martinolli}, \citenamefont {{Rabec Le
  Gloahec}}, \citenamefont {Rousseaux}, \citenamefont {Batani}, \citenamefont
  {Bernardinello}, \citenamefont {Greison},\ and\ \citenamefont
  {Hall}}]{Santos_PRL2002}%
  \BibitemOpen
  \bibfield  {author} {\bibinfo {author} {\bibfnamefont {J.}~\bibnamefont
  {Santos}, \bibfnamefont {J}}, \bibinfo {author} {\bibfnamefont
  {F.}~\bibnamefont {Amiranoff}}, \bibinfo {author} {\bibfnamefont {S.~D.}\
  \bibnamefont {Baton}}, \bibinfo {author} {\bibfnamefont {L.}~\bibnamefont
  {Gremillet}}, \bibinfo {author} {\bibfnamefont {M.}~\bibnamefont {Koenig}},
  \bibinfo {author} {\bibfnamefont {E.}~\bibnamefont {Martinolli}}, \bibinfo
  {author} {\bibfnamefont {M.}~\bibnamefont {{Rabec Le Gloahec}}}, \bibinfo
  {author} {\bibfnamefont {C.}~\bibnamefont {Rousseaux}}, \bibinfo {author}
  {\bibfnamefont {D.}~\bibnamefont {Batani}}, \bibinfo {author} {\bibfnamefont
  {A.}~\bibnamefont {Bernardinello}}, \bibinfo {author} {\bibfnamefont
  {G.}~\bibnamefont {Greison}}, \ and\ \bibinfo {author} {\bibfnamefont
  {T.}~\bibnamefont {Hall}},\ }\href@noop {} {\bibfield  {journal} {\bibinfo
  {journal} {Physical Review Letters}\ }\textbf {\bibinfo {volume} {89}},\
  \bibinfo {pages} {025001} (\bibinfo {year} {2002})}\BibitemShut {NoStop}%
\bibitem [{\citenamefont {P\'erez}\ \emph {et~al.}(2011)\citenamefont
  {P\'erez}, \citenamefont {Debayle}, \citenamefont {Honrubia}, \citenamefont
  {Koenig}, \citenamefont {Batani}, \citenamefont {Baton}, \citenamefont {Beg},
  \citenamefont {Benedetti}, \citenamefont {Brambrink}, \citenamefont {Chawla},
  \citenamefont {Dorchies}, \citenamefont {Forment}, \citenamefont
  {Galimberti}, \citenamefont {Gizzi}, \citenamefont {Gremillet}, \citenamefont
  {Heathcote}, \citenamefont {Higginson}, \citenamefont {Hulin}, \citenamefont
  {Jafer}, \citenamefont {Koester}, \citenamefont {Labate}, \citenamefont
  {Lancaster}, \citenamefont {MacKinnon}, \citenamefont {MacPhee},
  \citenamefont {Nazarov}, \citenamefont {Nicolai}, \citenamefont {Pasley},
  \citenamefont {Ramis}, \citenamefont {Richetta}, \citenamefont {Santos},
  \citenamefont {Sgattoni}, \citenamefont {Spindloe}, \citenamefont {Vauzour},
  \citenamefont {Vinci},\ and\ \citenamefont {Volpe}}]{Perez_PRL2011}%
  \BibitemOpen
  \bibfield  {author} {\bibinfo {author} {\bibfnamefont {F.}~\bibnamefont
  {P\'erez}}, \bibinfo {author} {\bibfnamefont {A.}~\bibnamefont {Debayle}},
  \bibinfo {author} {\bibfnamefont {J.}~\bibnamefont {Honrubia}}, \bibinfo
  {author} {\bibfnamefont {M.}~\bibnamefont {Koenig}}, \bibinfo {author}
  {\bibfnamefont {D.}~\bibnamefont {Batani}}, \bibinfo {author} {\bibfnamefont
  {S.~D.}\ \bibnamefont {Baton}}, \bibinfo {author} {\bibfnamefont {F.~N.}\
  \bibnamefont {Beg}}, \bibinfo {author} {\bibfnamefont {C.}~\bibnamefont
  {Benedetti}}, \bibinfo {author} {\bibfnamefont {E.}~\bibnamefont
  {Brambrink}}, \bibinfo {author} {\bibfnamefont {S.}~\bibnamefont {Chawla}},
  \bibinfo {author} {\bibfnamefont {F.}~\bibnamefont {Dorchies}}, \bibinfo
  {author} {\bibfnamefont {C.}~\bibnamefont {Forment}}, \bibinfo {author}
  {\bibfnamefont {M.}~\bibnamefont {Galimberti}}, \bibinfo {author}
  {\bibfnamefont {L.~A.}\ \bibnamefont {Gizzi}}, \bibinfo {author}
  {\bibfnamefont {L.}~\bibnamefont {Gremillet}}, \bibinfo {author}
  {\bibfnamefont {R.}~\bibnamefont {Heathcote}}, \bibinfo {author}
  {\bibfnamefont {D.~P.}\ \bibnamefont {Higginson}}, \bibinfo {author}
  {\bibfnamefont {S.}~\bibnamefont {Hulin}}, \bibinfo {author} {\bibfnamefont
  {R.}~\bibnamefont {Jafer}}, \bibinfo {author} {\bibfnamefont
  {P.}~\bibnamefont {Koester}}, \bibinfo {author} {\bibfnamefont
  {L.}~\bibnamefont {Labate}}, \bibinfo {author} {\bibfnamefont {K.~L.}\
  \bibnamefont {Lancaster}}, \bibinfo {author} {\bibfnamefont {A.~J.}\
  \bibnamefont {MacKinnon}}, \bibinfo {author} {\bibfnamefont {A.~G.}\
  \bibnamefont {MacPhee}}, \bibinfo {author} {\bibfnamefont {W.}~\bibnamefont
  {Nazarov}}, \bibinfo {author} {\bibfnamefont {P.}~\bibnamefont {Nicolai}},
  \bibinfo {author} {\bibfnamefont {J.}~\bibnamefont {Pasley}}, \bibinfo
  {author} {\bibfnamefont {R.}~\bibnamefont {Ramis}}, \bibinfo {author}
  {\bibfnamefont {M.}~\bibnamefont {Richetta}}, \bibinfo {author}
  {\bibfnamefont {J.~J.}\ \bibnamefont {Santos}}, \bibinfo {author}
  {\bibfnamefont {A.}~\bibnamefont {Sgattoni}}, \bibinfo {author}
  {\bibfnamefont {C.}~\bibnamefont {Spindloe}}, \bibinfo {author}
  {\bibfnamefont {B.}~\bibnamefont {Vauzour}}, \bibinfo {author} {\bibfnamefont
  {T.}~\bibnamefont {Vinci}}, \ and\ \bibinfo {author} {\bibfnamefont
  {L.}~\bibnamefont {Volpe}},\ }\href@noop {} {\bibfield  {journal} {\bibinfo
  {journal} {Physical Review Letters}\ }\textbf {\bibinfo {volume} {107}},\
  \bibinfo {pages} {065004} (\bibinfo {year} {2011})}\BibitemShut {NoStop}%
\bibitem [{\citenamefont {Vauzour}\ \emph {et~al.}(2011)\citenamefont
  {Vauzour}, \citenamefont {P\'erez}, \citenamefont {Volpe}, \citenamefont
  {Lancaster}, \citenamefont {{Nicola\"\i}}, \citenamefont {Batani},
  \citenamefont {Baton}, \citenamefont {Beg}, \citenamefont {Benedetti},
  \citenamefont {Brambrink}, \citenamefont {Chawla}, \citenamefont {Dorchies},
  \citenamefont {Fourment}, \citenamefont {Galimberti}, \citenamefont {Gizzi},
  \citenamefont {Heathcote}, \citenamefont {Higginson}, \citenamefont {Hulin},
  \citenamefont {Jafer}, \citenamefont {K\"oster}, \citenamefont {Labate},
  \citenamefont {MacKinnon}, \citenamefont {MacPhee}, \citenamefont {Nazarov},
  \citenamefont {Pasley}, \citenamefont {Regan}, \citenamefont {Ribeyre},
  \citenamefont {Richetta}, \citenamefont {Schurtz}, \citenamefont {Sgattoni},\
  and\ \citenamefont {Santos}}]{Vauzour_PoP2011}%
  \BibitemOpen
  \bibfield  {author} {\bibinfo {author} {\bibfnamefont {B.}~\bibnamefont
  {Vauzour}}, \bibinfo {author} {\bibfnamefont {F.}~\bibnamefont {P\'erez}},
  \bibinfo {author} {\bibfnamefont {L.}~\bibnamefont {Volpe}}, \bibinfo
  {author} {\bibfnamefont {K.}~\bibnamefont {Lancaster}}, \bibinfo {author}
  {\bibfnamefont {P.}~\bibnamefont {{Nicola\"\i}}}, \bibinfo {author}
  {\bibfnamefont {D.}~\bibnamefont {Batani}}, \bibinfo {author} {\bibfnamefont
  {S.~D.}\ \bibnamefont {Baton}}, \bibinfo {author} {\bibfnamefont {F.~N.}\
  \bibnamefont {Beg}}, \bibinfo {author} {\bibfnamefont {C.}~\bibnamefont
  {Benedetti}}, \bibinfo {author} {\bibfnamefont {E.}~\bibnamefont
  {Brambrink}}, \bibinfo {author} {\bibfnamefont {S.}~\bibnamefont {Chawla}},
  \bibinfo {author} {\bibfnamefont {F.}~\bibnamefont {Dorchies}}, \bibinfo
  {author} {\bibfnamefont {C.}~\bibnamefont {Fourment}}, \bibinfo {author}
  {\bibfnamefont {M.}~\bibnamefont {Galimberti}}, \bibinfo {author}
  {\bibfnamefont {L.~A.}\ \bibnamefont {Gizzi}}, \bibinfo {author}
  {\bibfnamefont {R.}~\bibnamefont {Heathcote}}, \bibinfo {author}
  {\bibfnamefont {D.~P.}\ \bibnamefont {Higginson}}, \bibinfo {author}
  {\bibfnamefont {D.}~\bibnamefont {Hulin}}, \bibinfo {author} {\bibfnamefont
  {R.}~\bibnamefont {Jafer}}, \bibinfo {author} {\bibfnamefont
  {P.}~\bibnamefont {K\"oster}}, \bibinfo {author} {\bibfnamefont
  {L.}~\bibnamefont {Labate}}, \bibinfo {author} {\bibfnamefont {A.~J.}\
  \bibnamefont {MacKinnon}}, \bibinfo {author} {\bibfnamefont {A.~G.}\
  \bibnamefont {MacPhee}}, \bibinfo {author} {\bibfnamefont {W.}~\bibnamefont
  {Nazarov}}, \bibinfo {author} {\bibfnamefont {J.}~\bibnamefont {Pasley}},
  \bibinfo {author} {\bibfnamefont {C.}~\bibnamefont {Regan}}, \bibinfo
  {author} {\bibfnamefont {X.}~\bibnamefont {Ribeyre}}, \bibinfo {author}
  {\bibfnamefont {M.}~\bibnamefont {Richetta}}, \bibinfo {author}
  {\bibfnamefont {G.}~\bibnamefont {Schurtz}}, \bibinfo {author} {\bibfnamefont
  {A.}~\bibnamefont {Sgattoni}}, \ and\ \bibinfo {author} {\bibfnamefont
  {J.~J.}\ \bibnamefont {Santos}},\ }\href@noop {} {\bibfield  {journal}
  {\bibinfo  {journal} {Physics of Plasmas}\ }\textbf {\bibinfo {volume}
  {18}},\ \bibinfo {pages} {043108} (\bibinfo {year} {2011})}\BibitemShut
  {NoStop}%
\bibitem [{\citenamefont {Martinolli}\ \emph {et~al.}(2006)\citenamefont
  {Martinolli}, \citenamefont {Koenig}, \citenamefont {Baton}, \citenamefont
  {Santos}, \citenamefont {Amiranoff}, \citenamefont {Batani}, \citenamefont
  {Perelli-Cippo}, \citenamefont {Scianitti}, \citenamefont {Gremillet},
  \citenamefont {M{\'e}lizzi}, \citenamefont {Decoster}, \citenamefont
  {Rousseaux}, \citenamefont {Hall}, \citenamefont {Key}, \citenamefont
  {Snavely}, \citenamefont {MacKinnon}, \citenamefont {Freeman}, \citenamefont
  {King}, \citenamefont {Stephens}, \citenamefont {Neely},\ and\ \citenamefont
  {Clarke}}]{Martinolli_PRE2006}%
  \BibitemOpen
  \bibfield  {author} {\bibinfo {author} {\bibfnamefont {E.}~\bibnamefont
  {Martinolli}}, \bibinfo {author} {\bibfnamefont {M.}~\bibnamefont {Koenig}},
  \bibinfo {author} {\bibfnamefont {S.~D.}\ \bibnamefont {Baton}}, \bibinfo
  {author} {\bibfnamefont {J.~J.}\ \bibnamefont {Santos}}, \bibinfo {author}
  {\bibfnamefont {F.}~\bibnamefont {Amiranoff}}, \bibinfo {author}
  {\bibfnamefont {D.}~\bibnamefont {Batani}}, \bibinfo {author} {\bibfnamefont
  {E.}~\bibnamefont {Perelli-Cippo}}, \bibinfo {author} {\bibfnamefont
  {F.}~\bibnamefont {Scianitti}}, \bibinfo {author} {\bibfnamefont
  {L.}~\bibnamefont {Gremillet}}, \bibinfo {author} {\bibfnamefont
  {R.}~\bibnamefont {M{\'e}lizzi}}, \bibinfo {author} {\bibfnamefont
  {A.}~\bibnamefont {Decoster}}, \bibinfo {author} {\bibfnamefont
  {C.}~\bibnamefont {Rousseaux}}, \bibinfo {author} {\bibfnamefont {T.~A.}\
  \bibnamefont {Hall}}, \bibinfo {author} {\bibfnamefont {M.~H.}\ \bibnamefont
  {Key}}, \bibinfo {author} {\bibfnamefont {R.}~\bibnamefont {Snavely}},
  \bibinfo {author} {\bibfnamefont {A.~J.}\ \bibnamefont {MacKinnon}}, \bibinfo
  {author} {\bibfnamefont {R.~R.}\ \bibnamefont {Freeman}}, \bibinfo {author}
  {\bibfnamefont {J.~A.}\ \bibnamefont {King}}, \bibinfo {author}
  {\bibfnamefont {R.}~\bibnamefont {Stephens}}, \bibinfo {author}
  {\bibfnamefont {D.}~\bibnamefont {Neely}}, \ and\ \bibinfo {author}
  {\bibfnamefont {R.~J.}\ \bibnamefont {Clarke}},\ }\href@noop {} {\bibfield
  {journal} {\bibinfo  {journal} {Physical Review E}\ }\textbf {\bibinfo
  {volume} {73}},\ \bibinfo {pages} {046402} (\bibinfo {year}
  {2006})}\BibitemShut {NoStop}%
\bibitem [{\citenamefont {Theobald}\ \emph {et~al.}(2006)\citenamefont
  {Theobald}, \citenamefont {Akli}, \citenamefont {Clarke}, \citenamefont
  {Delettrez}, \citenamefont {Freeman}, \citenamefont {Glenzer}, \citenamefont
  {Green}, \citenamefont {Gregori}, \citenamefont {Heathcote}, \citenamefont
  {Izumi}, \citenamefont {King}, \citenamefont {Koch}, \citenamefont {Kuba},
  \citenamefont {Lancaster}, \citenamefont {MacKinnon}, \citenamefont {Key},
  \citenamefont {Mileham}, \citenamefont {Myatt}, \citenamefont {Neely},
  \citenamefont {Norreys}, \citenamefont {Park}, \citenamefont {Pasley},
  \citenamefont {Patel}, \citenamefont {Regan}, \citenamefont {Sawada},
  \citenamefont {Shepherd}, \citenamefont {Snavely}, \citenamefont {Stephens},
  \citenamefont {Stoeckl}, \citenamefont {Storm}, \citenamefont {Zhang},\ and\
  \citenamefont {Sangster}}]{Theobald_PoP2006}%
  \BibitemOpen
  \bibfield  {author} {\bibinfo {author} {\bibfnamefont {W.}~\bibnamefont
  {Theobald}}, \bibinfo {author} {\bibfnamefont {K.}~\bibnamefont {Akli}},
  \bibinfo {author} {\bibfnamefont {R.}~\bibnamefont {Clarke}}, \bibinfo
  {author} {\bibfnamefont {J.~A.}\ \bibnamefont {Delettrez}}, \bibinfo {author}
  {\bibfnamefont {R.~R.}\ \bibnamefont {Freeman}}, \bibinfo {author}
  {\bibfnamefont {S.}~\bibnamefont {Glenzer}}, \bibinfo {author} {\bibfnamefont
  {J.}~\bibnamefont {Green}}, \bibinfo {author} {\bibfnamefont
  {G.}~\bibnamefont {Gregori}}, \bibinfo {author} {\bibfnamefont
  {R.}~\bibnamefont {Heathcote}}, \bibinfo {author} {\bibfnamefont
  {N.}~\bibnamefont {Izumi}}, \bibinfo {author} {\bibfnamefont {J.~A.}\
  \bibnamefont {King}}, \bibinfo {author} {\bibfnamefont {J.~A.}\ \bibnamefont
  {Koch}}, \bibinfo {author} {\bibfnamefont {J.}~\bibnamefont {Kuba}}, \bibinfo
  {author} {\bibfnamefont {K.}~\bibnamefont {Lancaster}}, \bibinfo {author}
  {\bibfnamefont {A.~J.}\ \bibnamefont {MacKinnon}}, \bibinfo {author}
  {\bibfnamefont {M.}~\bibnamefont {Key}}, \bibinfo {author} {\bibfnamefont
  {C.}~\bibnamefont {Mileham}}, \bibinfo {author} {\bibfnamefont
  {J.}~\bibnamefont {Myatt}}, \bibinfo {author} {\bibfnamefont
  {D.}~\bibnamefont {Neely}}, \bibinfo {author} {\bibfnamefont {P.~A.}\
  \bibnamefont {Norreys}}, \bibinfo {author} {\bibfnamefont {H.-S.}\
  \bibnamefont {Park}}, \bibinfo {author} {\bibfnamefont {J.}~\bibnamefont
  {Pasley}}, \bibinfo {author} {\bibfnamefont {P.}~\bibnamefont {Patel}},
  \bibinfo {author} {\bibfnamefont {S.~P.}\ \bibnamefont {Regan}}, \bibinfo
  {author} {\bibfnamefont {H.}~\bibnamefont {Sawada}}, \bibinfo {author}
  {\bibfnamefont {R.}~\bibnamefont {Shepherd}}, \bibinfo {author}
  {\bibfnamefont {R.}~\bibnamefont {Snavely}}, \bibinfo {author} {\bibfnamefont
  {R.~B.}\ \bibnamefont {Stephens}}, \bibinfo {author} {\bibfnamefont
  {C.}~\bibnamefont {Stoeckl}}, \bibinfo {author} {\bibfnamefont
  {M.}~\bibnamefont {Storm}}, \bibinfo {author} {\bibfnamefont
  {B.}~\bibnamefont {Zhang}}, \ and\ \bibinfo {author} {\bibfnamefont {T.~C.}\
  \bibnamefont {Sangster}},\ }\href {\doibase 10.1063/1.2188912} {\bibfield
  {journal} {\bibinfo  {journal} {Physics of Plasmas}\ }\textbf {\bibinfo
  {volume} {13}},\ \bibinfo {eid} {043102} (\bibinfo {year}
  {2006})}\BibitemShut {NoStop}%
\bibitem [{\citenamefont {Gizzi}\ \emph {et~al.}(2007)\citenamefont {Gizzi},
  \citenamefont {Giulietti}, \citenamefont {Giulietti}, \citenamefont
  {K\"oster}, \citenamefont {Labate}, \citenamefont {Levato}, \citenamefont
  {Zamponi}, \citenamefont {L\"ubcke}, \citenamefont {K\"ampfer}, \citenamefont
  {Uschmann}, \citenamefont {F\"orster}, \citenamefont {Antonicci},\ and\
  \citenamefont {Batani}}]{Gizzi_PPCF2007}%
  \BibitemOpen
  \bibfield  {author} {\bibinfo {author} {\bibfnamefont {L.~A.}\ \bibnamefont
  {Gizzi}}, \bibinfo {author} {\bibfnamefont {A.}~\bibnamefont {Giulietti}},
  \bibinfo {author} {\bibfnamefont {D.}~\bibnamefont {Giulietti}}, \bibinfo
  {author} {\bibfnamefont {P.}~\bibnamefont {K\"oster}}, \bibinfo {author}
  {\bibfnamefont {L.}~\bibnamefont {Labate}}, \bibinfo {author} {\bibfnamefont
  {T.}~\bibnamefont {Levato}}, \bibinfo {author} {\bibfnamefont
  {F.}~\bibnamefont {Zamponi}}, \bibinfo {author} {\bibfnamefont
  {A.}~\bibnamefont {L\"ubcke}}, \bibinfo {author} {\bibfnamefont
  {T.}~\bibnamefont {K\"ampfer}}, \bibinfo {author} {\bibfnamefont
  {I.}~\bibnamefont {Uschmann}}, \bibinfo {author} {\bibfnamefont
  {E.}~\bibnamefont {F\"orster}}, \bibinfo {author} {\bibfnamefont
  {A.}~\bibnamefont {Antonicci}}, \ and\ \bibinfo {author} {\bibfnamefont
  {D.}~\bibnamefont {Batani}},\ }\href@noop {} {\bibfield  {journal} {\bibinfo
  {journal} {Plasma Physics and Controlled Fusion}\ }\textbf {\bibinfo {volume}
  {49}},\ \bibinfo {pages} {B221} (\bibinfo {year} {2007})}\BibitemShut
  {NoStop}%
\bibitem [{\citenamefont {Chen}\ \emph {et~al.}(2009)\citenamefont {Chen},
  \citenamefont {Patel}, \citenamefont {Hey}, \citenamefont {Mackinnon},
  \citenamefont {Key}, \citenamefont {Akli}, \citenamefont {Bartal},
  \citenamefont {Beg}, \citenamefont {Chawla}, \citenamefont {Chen},
  \citenamefont {Freeman}, \citenamefont {Higginson}, \citenamefont {Link},
  \citenamefont {Ma}, \citenamefont {MacPhee}, \citenamefont {Stephens},
  \citenamefont {{Van Woerkom}}, \citenamefont {Westover},\ and\ \citenamefont
  {Porkolab}}]{Chen_PoP2009Brem}%
  \BibitemOpen
  \bibfield  {author} {\bibinfo {author} {\bibfnamefont {C.~D.}\ \bibnamefont
  {Chen}}, \bibinfo {author} {\bibfnamefont {P.~K.}\ \bibnamefont {Patel}},
  \bibinfo {author} {\bibfnamefont {D.~S.}\ \bibnamefont {Hey}}, \bibinfo
  {author} {\bibfnamefont {A.~J.}\ \bibnamefont {Mackinnon}}, \bibinfo {author}
  {\bibfnamefont {M.~H.}\ \bibnamefont {Key}}, \bibinfo {author} {\bibfnamefont
  {K.~U.}\ \bibnamefont {Akli}}, \bibinfo {author} {\bibfnamefont
  {T.}~\bibnamefont {Bartal}}, \bibinfo {author} {\bibfnamefont {F.~N.}\
  \bibnamefont {Beg}}, \bibinfo {author} {\bibfnamefont {S.}~\bibnamefont
  {Chawla}}, \bibinfo {author} {\bibfnamefont {H.}~\bibnamefont {Chen}},
  \bibinfo {author} {\bibfnamefont {R.~R.}\ \bibnamefont {Freeman}}, \bibinfo
  {author} {\bibfnamefont {D.~P.}\ \bibnamefont {Higginson}}, \bibinfo {author}
  {\bibfnamefont {A.}~\bibnamefont {Link}}, \bibinfo {author} {\bibfnamefont
  {T.~Y.}\ \bibnamefont {Ma}}, \bibinfo {author} {\bibfnamefont {A.~G.}\
  \bibnamefont {MacPhee}}, \bibinfo {author} {\bibfnamefont {R.~B.}\
  \bibnamefont {Stephens}}, \bibinfo {author} {\bibfnamefont {L.~D.}\
  \bibnamefont {{Van Woerkom}}}, \bibinfo {author} {\bibfnamefont
  {B.}~\bibnamefont {Westover}}, \ and\ \bibinfo {author} {\bibfnamefont
  {M.}~\bibnamefont {Porkolab}},\ }\href@noop {} {\bibfield  {journal}
  {\bibinfo  {journal} {Physics of Plasmas}\ }\textbf {\bibinfo {volume}
  {16}},\ \bibinfo {pages} {082705} (\bibinfo {year} {2009})}\BibitemShut
  {NoStop}%
\bibitem [{\citenamefont {Scott}\ \emph {et~al.}(2012)\citenamefont {Scott},
  \citenamefont {Perez}, \citenamefont {Santos}, \citenamefont {Ridgers},
  \citenamefont {Davies}, \citenamefont {Lancaster}, \citenamefont {Baton},
  \citenamefont {Nicolai}, \citenamefont {Trines}, \citenamefont {Bell},
  \citenamefont {Hulin}, \citenamefont {Tzoufras}, \citenamefont {Rose},\ and\
  \citenamefont {Norreys}}]{Scott_PoP2012}%
  \BibitemOpen
  \bibfield  {author} {\bibinfo {author} {\bibfnamefont {R.~H.~H.}\
  \bibnamefont {Scott}}, \bibinfo {author} {\bibfnamefont {F.}~\bibnamefont
  {Perez}}, \bibinfo {author} {\bibfnamefont {J.~J.}\ \bibnamefont {Santos}},
  \bibinfo {author} {\bibfnamefont {C.~P.}\ \bibnamefont {Ridgers}}, \bibinfo
  {author} {\bibfnamefont {J.~R.}\ \bibnamefont {Davies}}, \bibinfo {author}
  {\bibfnamefont {K.~L.}\ \bibnamefont {Lancaster}}, \bibinfo {author}
  {\bibfnamefont {S.~D.}\ \bibnamefont {Baton}}, \bibinfo {author}
  {\bibfnamefont {P.}~\bibnamefont {Nicolai}}, \bibinfo {author} {\bibfnamefont
  {R.~M.~G.~M.}\ \bibnamefont {Trines}}, \bibinfo {author} {\bibfnamefont
  {A.~R.}\ \bibnamefont {Bell}}, \bibinfo {author} {\bibfnamefont
  {S.}~\bibnamefont {Hulin}}, \bibinfo {author} {\bibfnamefont
  {M.}~\bibnamefont {Tzoufras}}, \bibinfo {author} {\bibfnamefont {S.~J.}\
  \bibnamefont {Rose}}, \ and\ \bibinfo {author} {\bibfnamefont {P.~A.}\
  \bibnamefont {Norreys}},\ }\href@noop {} {\bibfield  {journal} {\bibinfo
  {journal} {Physics of Plasmas}\ }\textbf {\bibinfo {volume} {19}},\ \bibinfo
  {pages} {053104} (\bibinfo {year} {2012})}\BibitemShut {NoStop}%
\bibitem [{\citenamefont {Yasuike}\ \emph {et~al.}(2001)\citenamefont
  {Yasuike}, \citenamefont {Key}, \citenamefont {Hatchett}, \citenamefont
  {Snavely},\ and\ \citenamefont {Wharton}}]{Yasuike_RSI2001}%
  \BibitemOpen
  \bibfield  {author} {\bibinfo {author} {\bibfnamefont {K.}~\bibnamefont
  {Yasuike}}, \bibinfo {author} {\bibfnamefont {M.~H.}\ \bibnamefont {Key}},
  \bibinfo {author} {\bibfnamefont {S.~P.}\ \bibnamefont {Hatchett}}, \bibinfo
  {author} {\bibfnamefont {R.~A.}\ \bibnamefont {Snavely}}, \ and\ \bibinfo
  {author} {\bibfnamefont {K.~B.}\ \bibnamefont {Wharton}},\ }\href@noop {}
  {\bibfield  {journal} {\bibinfo  {journal} {Review of Scientific
  Instruments}\ }\textbf {\bibinfo {volume} {72}},\ \bibinfo {pages} {1236}
  (\bibinfo {year} {2001})}\BibitemShut {NoStop}%
\bibitem [{\citenamefont {Nilson}\ \emph {et~al.}(2008)\citenamefont {Nilson},
  \citenamefont {Theobald}, \citenamefont {Myatt}, \citenamefont {Stoeckl},
  \citenamefont {Storm}, \citenamefont {Gotchev}, \citenamefont {Zuegel},
  \citenamefont {Betti}, \citenamefont {Meyerhofer},\ and\ \citenamefont
  {Sangster}}]{Nilson_PoP2008}%
  \BibitemOpen
  \bibfield  {author} {\bibinfo {author} {\bibfnamefont {P.~M.}\ \bibnamefont
  {Nilson}}, \bibinfo {author} {\bibfnamefont {W.}~\bibnamefont {Theobald}},
  \bibinfo {author} {\bibfnamefont {J.}~\bibnamefont {Myatt}}, \bibinfo
  {author} {\bibfnamefont {C.}~\bibnamefont {Stoeckl}}, \bibinfo {author}
  {\bibfnamefont {M.}~\bibnamefont {Storm}}, \bibinfo {author} {\bibfnamefont
  {O.~V.}\ \bibnamefont {Gotchev}}, \bibinfo {author} {\bibfnamefont {J.~D.}\
  \bibnamefont {Zuegel}}, \bibinfo {author} {\bibfnamefont {R.}~\bibnamefont
  {Betti}}, \bibinfo {author} {\bibfnamefont {D.~D.}\ \bibnamefont
  {Meyerhofer}}, \ and\ \bibinfo {author} {\bibfnamefont {T.~C.}\ \bibnamefont
  {Sangster}},\ }\href {\doibase 10.1063/1.2889449} {\bibfield  {journal}
  {\bibinfo  {journal} {Physics of Plasmas}\ }\textbf {\bibinfo {volume}
  {15}},\ \bibinfo {eid} {056308} (\bibinfo {year} {2008})}\BibitemShut
  {NoStop}%
\bibitem [{\citenamefont {Tanimoto}\ \emph {et~al.}(2009)\citenamefont
  {Tanimoto}, \citenamefont {Habara}, \citenamefont {Kodama}, \citenamefont
  {Nakatsutsumi}, \citenamefont {Tanaka}, \citenamefont {Lancaster},
  \citenamefont {Green}, \citenamefont {Scott}, \citenamefont {Sherlock},
  \citenamefont {Norreys}, \citenamefont {Evans}, \citenamefont {Haines},
  \citenamefont {Kar}, \citenamefont {Zepf}, \citenamefont {King},
  \citenamefont {Ma}, \citenamefont {Wei}, \citenamefont {Yabuuchi},
  \citenamefont {Beg}, \citenamefont {Key}, \citenamefont {Nilson},
  \citenamefont {Stephens}, \citenamefont {Azechi}, \citenamefont {Nagai},
  \citenamefont {Norimatsu}, \citenamefont {Takeda}, \citenamefont {Valente},\
  and\ \citenamefont {Davies}}]{Tanimoto_PoP2009}%
  \BibitemOpen
  \bibfield  {author} {\bibinfo {author} {\bibfnamefont {T.}~\bibnamefont
  {Tanimoto}}, \bibinfo {author} {\bibfnamefont {H.}~\bibnamefont {Habara}},
  \bibinfo {author} {\bibfnamefont {R.}~\bibnamefont {Kodama}}, \bibinfo
  {author} {\bibfnamefont {M.}~\bibnamefont {Nakatsutsumi}}, \bibinfo {author}
  {\bibfnamefont {K.~A.}\ \bibnamefont {Tanaka}}, \bibinfo {author}
  {\bibfnamefont {K.~L.}\ \bibnamefont {Lancaster}}, \bibinfo {author}
  {\bibfnamefont {J.~S.}\ \bibnamefont {Green}}, \bibinfo {author}
  {\bibfnamefont {R.~H.~H.}\ \bibnamefont {Scott}}, \bibinfo {author}
  {\bibfnamefont {M.}~\bibnamefont {Sherlock}}, \bibinfo {author}
  {\bibfnamefont {P.~A.}\ \bibnamefont {Norreys}}, \bibinfo {author}
  {\bibfnamefont {R.~G.}\ \bibnamefont {Evans}}, \bibinfo {author}
  {\bibfnamefont {M.~G.}\ \bibnamefont {Haines}}, \bibinfo {author}
  {\bibfnamefont {S.}~\bibnamefont {Kar}}, \bibinfo {author} {\bibfnamefont
  {M.}~\bibnamefont {Zepf}}, \bibinfo {author} {\bibfnamefont {J.}~\bibnamefont
  {King}}, \bibinfo {author} {\bibfnamefont {T.}~\bibnamefont {Ma}}, \bibinfo
  {author} {\bibfnamefont {M.~S.}\ \bibnamefont {Wei}}, \bibinfo {author}
  {\bibfnamefont {T.}~\bibnamefont {Yabuuchi}}, \bibinfo {author}
  {\bibfnamefont {F.~N.}\ \bibnamefont {Beg}}, \bibinfo {author} {\bibfnamefont
  {M.~H.}\ \bibnamefont {Key}}, \bibinfo {author} {\bibfnamefont
  {P.}~\bibnamefont {Nilson}}, \bibinfo {author} {\bibfnamefont {R.~B.}\
  \bibnamefont {Stephens}}, \bibinfo {author} {\bibfnamefont {H.}~\bibnamefont
  {Azechi}}, \bibinfo {author} {\bibfnamefont {K.}~\bibnamefont {Nagai}},
  \bibinfo {author} {\bibfnamefont {T.}~\bibnamefont {Norimatsu}}, \bibinfo
  {author} {\bibfnamefont {K.}~\bibnamefont {Takeda}}, \bibinfo {author}
  {\bibfnamefont {J.}~\bibnamefont {Valente}}, \ and\ \bibinfo {author}
  {\bibfnamefont {J.~R.}\ \bibnamefont {Davies}},\ }\href {\doibase
  10.1063/1.3155086} {\bibfield  {journal} {\bibinfo  {journal} {Physics of
  Plasmas}\ }\textbf {\bibinfo {volume} {16}},\ \bibinfo {eid} {062703}
  (\bibinfo {year} {2009})}\BibitemShut {NoStop}%
\bibitem [{\citenamefont {Stephens}\ \emph {et~al.}(2004)\citenamefont
  {Stephens}, \citenamefont {Snavely}, \citenamefont {Aglitskiy}, \citenamefont
  {Amiranoff}, \citenamefont {Andersen}, \citenamefont {Batani}, \citenamefont
  {Baton}, \citenamefont {Cowan}, \citenamefont {Freeman}, \citenamefont
  {Hall}, \citenamefont {Hatchett}, \citenamefont {Hill}, \citenamefont {Key},
  \citenamefont {King}, \citenamefont {Koch}, \citenamefont {Koenig},
  \citenamefont {MacKinnon}, \citenamefont {Lancaster}, \citenamefont
  {Martinolli}, \citenamefont {Norreys}, \citenamefont {Perelli-Cippo},
  \citenamefont {Rabec Le~Gloahec}, \citenamefont {Rousseaux}, \citenamefont
  {Santos},\ and\ \citenamefont {Scianitti}}]{Stephens_PRE2004}%
  \BibitemOpen
  \bibfield  {author} {\bibinfo {author} {\bibfnamefont {R.~B.}\ \bibnamefont
  {Stephens}}, \bibinfo {author} {\bibfnamefont {R.~A.}\ \bibnamefont
  {Snavely}}, \bibinfo {author} {\bibfnamefont {Y.}~\bibnamefont {Aglitskiy}},
  \bibinfo {author} {\bibfnamefont {F.}~\bibnamefont {Amiranoff}}, \bibinfo
  {author} {\bibfnamefont {C.}~\bibnamefont {Andersen}}, \bibinfo {author}
  {\bibfnamefont {D.}~\bibnamefont {Batani}}, \bibinfo {author} {\bibfnamefont
  {S.~D.}\ \bibnamefont {Baton}}, \bibinfo {author} {\bibfnamefont
  {T.}~\bibnamefont {Cowan}}, \bibinfo {author} {\bibfnamefont {R.~R.}\
  \bibnamefont {Freeman}}, \bibinfo {author} {\bibfnamefont {T.}~\bibnamefont
  {Hall}}, \bibinfo {author} {\bibfnamefont {S.~P.}\ \bibnamefont {Hatchett}},
  \bibinfo {author} {\bibfnamefont {J.~M.}\ \bibnamefont {Hill}}, \bibinfo
  {author} {\bibfnamefont {M.~H.}\ \bibnamefont {Key}}, \bibinfo {author}
  {\bibfnamefont {J.~A.}\ \bibnamefont {King}}, \bibinfo {author}
  {\bibfnamefont {J.~A.}\ \bibnamefont {Koch}}, \bibinfo {author}
  {\bibfnamefont {M.}~\bibnamefont {Koenig}}, \bibinfo {author} {\bibfnamefont
  {A.~J.}\ \bibnamefont {MacKinnon}}, \bibinfo {author} {\bibfnamefont {K.~L.}\
  \bibnamefont {Lancaster}}, \bibinfo {author} {\bibfnamefont {E.}~\bibnamefont
  {Martinolli}}, \bibinfo {author} {\bibfnamefont {P.}~\bibnamefont {Norreys}},
  \bibinfo {author} {\bibfnamefont {E.}~\bibnamefont {Perelli-Cippo}}, \bibinfo
  {author} {\bibfnamefont {M.}~\bibnamefont {Rabec Le~Gloahec}}, \bibinfo
  {author} {\bibfnamefont {C.}~\bibnamefont {Rousseaux}}, \bibinfo {author}
  {\bibfnamefont {J.~J.}\ \bibnamefont {Santos}}, \ and\ \bibinfo {author}
  {\bibfnamefont {F.}~\bibnamefont {Scianitti}},\ }\href@noop {} {\bibfield
  {journal} {\bibinfo  {journal} {Physical Review E}\ }\textbf {\bibinfo
  {volume} {69}},\ \bibinfo {pages} {066414} (\bibinfo {year}
  {2004})}\BibitemShut {NoStop}%
\bibitem [{\citenamefont {Booth}\ \emph {et~al.}(2011)\citenamefont {Booth},
  \citenamefont {Clarke}, \citenamefont {Doria}, \citenamefont {Gizzi},
  \citenamefont {Gregori}, \citenamefont {Hake}, \citenamefont {Koester},
  \citenamefont {Labate}, \citenamefont {Levato}, \citenamefont {Li},
  \citenamefont {Makita}, \citenamefont {Mancini}, \citenamefont {Pasley},
  \citenamefont {Rajeev}, \citenamefont {Riley}, \citenamefont {Robinson},
  \citenamefont {Wagenaars}, \citenamefont {Waugh},\ and\ \citenamefont
  {Woolsey}}]{Booth_NIMA2011}%
  \BibitemOpen
  \bibfield  {author} {\bibinfo {author} {\bibfnamefont {N.}~\bibnamefont
  {Booth}}, \bibinfo {author} {\bibfnamefont {R.~J.}\ \bibnamefont {Clarke}},
  \bibinfo {author} {\bibfnamefont {D.}~\bibnamefont {Doria}}, \bibinfo
  {author} {\bibfnamefont {L.~A.}\ \bibnamefont {Gizzi}}, \bibinfo {author}
  {\bibfnamefont {G.}~\bibnamefont {Gregori}}, \bibinfo {author} {\bibfnamefont
  {P.}~\bibnamefont {Hake}}, \bibinfo {author} {\bibfnamefont {P.}~\bibnamefont
  {Koester}}, \bibinfo {author} {\bibfnamefont {L.}~\bibnamefont {Labate}},
  \bibinfo {author} {\bibfnamefont {T.}~\bibnamefont {Levato}}, \bibinfo
  {author} {\bibfnamefont {B.}~\bibnamefont {Li}}, \bibinfo {author}
  {\bibfnamefont {M.}~\bibnamefont {Makita}}, \bibinfo {author} {\bibfnamefont
  {R.~C.}\ \bibnamefont {Mancini}}, \bibinfo {author} {\bibfnamefont
  {J.}~\bibnamefont {Pasley}}, \bibinfo {author} {\bibfnamefont {P.~P.}\
  \bibnamefont {Rajeev}}, \bibinfo {author} {\bibfnamefont {D.}~\bibnamefont
  {Riley}}, \bibinfo {author} {\bibfnamefont {A.~P.~L.}\ \bibnamefont
  {Robinson}}, \bibinfo {author} {\bibfnamefont {E.}~\bibnamefont {Wagenaars}},
  \bibinfo {author} {\bibfnamefont {J.~N.}\ \bibnamefont {Waugh}}, \ and\
  \bibinfo {author} {\bibfnamefont {N.~C.}\ \bibnamefont {Woolsey}},\
  }\href@noop {} {\bibfield  {journal} {\bibinfo  {journal} {Nuclear
  Instruments and Methods in Physics Research A}\ }\textbf {\bibinfo {volume}
  {653}},\ \bibinfo {pages} {137} (\bibinfo {year} {2011})}\BibitemShut
  {NoStop}%
\bibitem [{\citenamefont {Chen}\ \emph {et~al.}(2008)\citenamefont {Chen},
  \citenamefont {King}, \citenamefont {Key}, \citenamefont {Akli},
  \citenamefont {Beg}, \citenamefont {Chen}, \citenamefont {Freeman},
  \citenamefont {Link}, \citenamefont {Mackinnon}, \citenamefont {MacPhee},
  \citenamefont {Patel}, \citenamefont {Porkolab}, \citenamefont {Stephens},\
  and\ \citenamefont {{Van Woerkom}}}]{Chen_RSI2008}%
  \BibitemOpen
  \bibfield  {author} {\bibinfo {author} {\bibfnamefont {C.~D.}\ \bibnamefont
  {Chen}}, \bibinfo {author} {\bibfnamefont {J.~A.}\ \bibnamefont {King}},
  \bibinfo {author} {\bibfnamefont {M.~H.}\ \bibnamefont {Key}}, \bibinfo
  {author} {\bibfnamefont {K.~U.}\ \bibnamefont {Akli}}, \bibinfo {author}
  {\bibfnamefont {F.~N.}\ \bibnamefont {Beg}}, \bibinfo {author} {\bibfnamefont
  {H.}~\bibnamefont {Chen}}, \bibinfo {author} {\bibfnamefont {R.~R.}\
  \bibnamefont {Freeman}}, \bibinfo {author} {\bibfnamefont {A.}~\bibnamefont
  {Link}}, \bibinfo {author} {\bibfnamefont {A.~J.}\ \bibnamefont {Mackinnon}},
  \bibinfo {author} {\bibfnamefont {A.~G.}\ \bibnamefont {MacPhee}}, \bibinfo
  {author} {\bibfnamefont {P.~K.}\ \bibnamefont {Patel}}, \bibinfo {author}
  {\bibfnamefont {M.}~\bibnamefont {Porkolab}}, \bibinfo {author}
  {\bibfnamefont {R.~B.}\ \bibnamefont {Stephens}}, \ and\ \bibinfo {author}
  {\bibfnamefont {L.~D.}\ \bibnamefont {{Van Woerkom}}},\ }\href@noop {}
  {\bibfield  {journal} {\bibinfo  {journal} {Review of Scientific
  Instruments}\ }\textbf {\bibinfo {volume} {79}},\ \bibinfo {pages} {10E305}
  (\bibinfo {year} {2008})}\BibitemShut {NoStop}%
\bibitem [{\citenamefont {McDonald}\ \emph {et~al.}(2004)\citenamefont
  {McDonald}, \citenamefont {Kauffman}, \citenamefont {Celeste}, \citenamefont
  {Rhodes}, \citenamefont {Lee}, \citenamefont {Suter}, \citenamefont {Lee},
  \citenamefont {Foster},\ and\ \citenamefont {Slark}}]{McDonald_RSI2004}%
  \BibitemOpen
  \bibfield  {author} {\bibinfo {author} {\bibfnamefont {J.~W.}\ \bibnamefont
  {McDonald}}, \bibinfo {author} {\bibfnamefont {R.~L.}\ \bibnamefont
  {Kauffman}}, \bibinfo {author} {\bibfnamefont {J.~R.}\ \bibnamefont
  {Celeste}}, \bibinfo {author} {\bibfnamefont {M.~A.}\ \bibnamefont {Rhodes}},
  \bibinfo {author} {\bibfnamefont {F.~D.}\ \bibnamefont {Lee}}, \bibinfo
  {author} {\bibfnamefont {L.~J.}\ \bibnamefont {Suter}}, \bibinfo {author}
  {\bibfnamefont {A.~P.}\ \bibnamefont {Lee}}, \bibinfo {author} {\bibfnamefont
  {J.~M.}\ \bibnamefont {Foster}}, \ and\ \bibinfo {author} {\bibfnamefont
  {G.}~\bibnamefont {Slark}},\ }\href {\doibase 10.1063/1.1788871} {\bibfield
  {journal} {\bibinfo  {journal} {Review of Scientific Instruments}\ }\textbf
  {\bibinfo {volume} {75}},\ \bibinfo {pages} {3753} (\bibinfo {year}
  {2004})}\BibitemShut {NoStop}%
\bibitem [{\citenamefont {Dewald}\ \emph {et~al.}(2010)\citenamefont {Dewald},
  \citenamefont {Thomas}, \citenamefont {Hunter}, \citenamefont {Divol},
  \citenamefont {Meezan}, \citenamefont {Glenzer}, \citenamefont {Suter},
  \citenamefont {Bond}, \citenamefont {Kline}, \citenamefont {Celeste},
  \citenamefont {Bradley}, \citenamefont {Bell}, \citenamefont {Kauffman},
  \citenamefont {Kilkenny},\ and\ \citenamefont {Landen}}]{Dewald_RSI2010}%
  \BibitemOpen
  \bibfield  {author} {\bibinfo {author} {\bibfnamefont {E.~L.}\ \bibnamefont
  {Dewald}}, \bibinfo {author} {\bibfnamefont {C.}~\bibnamefont {Thomas}},
  \bibinfo {author} {\bibfnamefont {S.}~\bibnamefont {Hunter}}, \bibinfo
  {author} {\bibfnamefont {L.}~\bibnamefont {Divol}}, \bibinfo {author}
  {\bibfnamefont {N.}~\bibnamefont {Meezan}}, \bibinfo {author} {\bibfnamefont
  {S.~H.}\ \bibnamefont {Glenzer}}, \bibinfo {author} {\bibfnamefont {L.~J.}\
  \bibnamefont {Suter}}, \bibinfo {author} {\bibfnamefont {E.}~\bibnamefont
  {Bond}}, \bibinfo {author} {\bibfnamefont {J.~L.}\ \bibnamefont {Kline}},
  \bibinfo {author} {\bibfnamefont {J.}~\bibnamefont {Celeste}}, \bibinfo
  {author} {\bibfnamefont {D.}~\bibnamefont {Bradley}}, \bibinfo {author}
  {\bibfnamefont {P.}~\bibnamefont {Bell}}, \bibinfo {author} {\bibfnamefont
  {R.~L.}\ \bibnamefont {Kauffman}}, \bibinfo {author} {\bibfnamefont
  {J.}~\bibnamefont {Kilkenny}}, \ and\ \bibinfo {author} {\bibfnamefont
  {O.~L.}\ \bibnamefont {Landen}},\ }\href {\doibase 10.1063/1.3478683}
  {\bibfield  {journal} {\bibinfo  {journal} {Review of Scientific
  Instruments}\ }\textbf {\bibinfo {volume} {81}},\ \bibinfo {eid} {10D938}
  (\bibinfo {year} {2010})}\BibitemShut {NoStop}%
\bibitem [{\citenamefont {Dewald}\ \emph {et~al.}(2006)\citenamefont {Dewald},
  \citenamefont {Landen}, \citenamefont {Suter}, \citenamefont {Schein},
  \citenamefont {Holder}, \citenamefont {Campbell}, \citenamefont {Glenzer},
  \citenamefont {McDonald}, \citenamefont {Niemann}, \citenamefont {Mackinnon},
  \citenamefont {Schneider}, \citenamefont {Haynam}, \citenamefont {Hinkel},\
  and\ \citenamefont {Hammel}}]{Dewald_PoP2006}%
  \BibitemOpen
  \bibfield  {author} {\bibinfo {author} {\bibfnamefont {E.~L.}\ \bibnamefont
  {Dewald}}, \bibinfo {author} {\bibfnamefont {O.~L.}\ \bibnamefont {Landen}},
  \bibinfo {author} {\bibfnamefont {L.~J.}\ \bibnamefont {Suter}}, \bibinfo
  {author} {\bibfnamefont {J.}~\bibnamefont {Schein}}, \bibinfo {author}
  {\bibfnamefont {J.}~\bibnamefont {Holder}}, \bibinfo {author} {\bibfnamefont
  {K.}~\bibnamefont {Campbell}}, \bibinfo {author} {\bibfnamefont {S.~H.}\
  \bibnamefont {Glenzer}}, \bibinfo {author} {\bibfnamefont {J.~W.}\
  \bibnamefont {McDonald}}, \bibinfo {author} {\bibfnamefont {C.}~\bibnamefont
  {Niemann}}, \bibinfo {author} {\bibfnamefont {A.~J.}\ \bibnamefont
  {Mackinnon}}, \bibinfo {author} {\bibfnamefont {M.~S.}\ \bibnamefont
  {Schneider}}, \bibinfo {author} {\bibfnamefont {C.}~\bibnamefont {Haynam}},
  \bibinfo {author} {\bibfnamefont {D.}~\bibnamefont {Hinkel}}, \ and\ \bibinfo
  {author} {\bibfnamefont {B.~A.}\ \bibnamefont {Hammel}},\ }\href {\doibase
  10.1063/1.2178783} {\bibfield  {journal} {\bibinfo  {journal} {Physics of
  Plasmas}\ }\textbf {\bibinfo {volume} {13}},\ \bibinfo {eid} {056315}
  (\bibinfo {year} {2006})}\BibitemShut {NoStop}%
\bibitem [{\citenamefont {Labate}\ \emph {et~al.}(2005)\citenamefont {Labate},
  \citenamefont {Cecchetti}, \citenamefont {Galimberti}, \citenamefont
  {Giulietti}, \citenamefont {Giulietti},\ and\ \citenamefont
  {Gizzi}}]{Labate_PoP2005}%
  \BibitemOpen
  \bibfield  {author} {\bibinfo {author} {\bibfnamefont {L.}~\bibnamefont
  {Labate}}, \bibinfo {author} {\bibfnamefont {C.~A.}\ \bibnamefont
  {Cecchetti}}, \bibinfo {author} {\bibfnamefont {M.}~\bibnamefont
  {Galimberti}}, \bibinfo {author} {\bibfnamefont {A.}~\bibnamefont
  {Giulietti}}, \bibinfo {author} {\bibfnamefont {D.}~\bibnamefont
  {Giulietti}}, \ and\ \bibinfo {author} {\bibfnamefont {L.~A.}\ \bibnamefont
  {Gizzi}},\ }\href@noop {} {\bibfield  {journal} {\bibinfo  {journal} {Physics
  of Plasmas}\ }\textbf {\bibinfo {volume} {12}},\ \bibinfo {pages} {083101}
  (\bibinfo {year} {2005})}\BibitemShut {NoStop}%
\bibitem [{\citenamefont {Tommasini}\ \emph {et~al.}(2006)\citenamefont
  {Tommasini}, \citenamefont {Koch}, \citenamefont {Izumi}, \citenamefont
  {Welser}, \citenamefont {Mancini}, \citenamefont {Delettrez}, \citenamefont
  {Regan},\ and\ \citenamefont {Smalyuk}}]{Tommasini_RSI2006}%
  \BibitemOpen
  \bibfield  {author} {\bibinfo {author} {\bibfnamefont {R.}~\bibnamefont
  {Tommasini}}, \bibinfo {author} {\bibfnamefont {J.~A.}\ \bibnamefont {Koch}},
  \bibinfo {author} {\bibfnamefont {N.}~\bibnamefont {Izumi}}, \bibinfo
  {author} {\bibfnamefont {L.~A.}\ \bibnamefont {Welser}}, \bibinfo {author}
  {\bibfnamefont {R.~C.}\ \bibnamefont {Mancini}}, \bibinfo {author}
  {\bibfnamefont {J.}~\bibnamefont {Delettrez}}, \bibinfo {author}
  {\bibfnamefont {S.}~\bibnamefont {Regan}}, \ and\ \bibinfo {author}
  {\bibfnamefont {V.}~\bibnamefont {Smalyuk}},\ }\href@noop {} {\bibfield
  {journal} {\bibinfo  {journal} {Review of Scientific Instruments}\ }\textbf
  {\bibinfo {volume} {77}},\ \bibinfo {pages} {10E303} (\bibinfo {year}
  {2006})}\BibitemShut {NoStop}%
\bibitem [{\citenamefont {King}\ \emph {et~al.}(2005)\citenamefont {King},
  \citenamefont {Akli}, \citenamefont {Snavely}, \citenamefont {Zhang},
  \citenamefont {Key}, \citenamefont {Chen}, \citenamefont {Chen},
  \citenamefont {Hatchett}, \citenamefont {Koch}, \citenamefont {MacKinnon},
  \citenamefont {Patel}, \citenamefont {Phillips}, \citenamefont {Town},
  \citenamefont {Freeman}, \citenamefont {Borghesi}, \citenamefont {Romagnani},
  \citenamefont {Zepf}, \citenamefont {Cowan}, \citenamefont {Stephens},
  \citenamefont {Lancaster}, \citenamefont {Murphy}, \citenamefont {Norreys},\
  and\ \citenamefont {Stoeckl}}]{King_RSI2005}%
  \BibitemOpen
  \bibfield  {author} {\bibinfo {author} {\bibfnamefont {J.~A.}\ \bibnamefont
  {King}}, \bibinfo {author} {\bibfnamefont {K.}~\bibnamefont {Akli}}, \bibinfo
  {author} {\bibfnamefont {R.~A.}\ \bibnamefont {Snavely}}, \bibinfo {author}
  {\bibfnamefont {B.}~\bibnamefont {Zhang}}, \bibinfo {author} {\bibfnamefont
  {M.~H.}\ \bibnamefont {Key}}, \bibinfo {author} {\bibfnamefont {C.~D.}\
  \bibnamefont {Chen}}, \bibinfo {author} {\bibfnamefont {M.}~\bibnamefont
  {Chen}}, \bibinfo {author} {\bibfnamefont {S.~P.}\ \bibnamefont {Hatchett}},
  \bibinfo {author} {\bibfnamefont {J.~A.}\ \bibnamefont {Koch}}, \bibinfo
  {author} {\bibfnamefont {A.~J.}\ \bibnamefont {MacKinnon}}, \bibinfo {author}
  {\bibfnamefont {P.~K.}\ \bibnamefont {Patel}}, \bibinfo {author}
  {\bibfnamefont {T.}~\bibnamefont {Phillips}}, \bibinfo {author}
  {\bibfnamefont {R.~P.~J.}\ \bibnamefont {Town}}, \bibinfo {author}
  {\bibfnamefont {R.~R.}\ \bibnamefont {Freeman}}, \bibinfo {author}
  {\bibfnamefont {M.}~\bibnamefont {Borghesi}}, \bibinfo {author}
  {\bibfnamefont {L.}~\bibnamefont {Romagnani}}, \bibinfo {author}
  {\bibfnamefont {M.}~\bibnamefont {Zepf}}, \bibinfo {author} {\bibfnamefont
  {T.}~\bibnamefont {Cowan}}, \bibinfo {author} {\bibfnamefont
  {R.}~\bibnamefont {Stephens}}, \bibinfo {author} {\bibfnamefont {K.~L.}\
  \bibnamefont {Lancaster}}, \bibinfo {author} {\bibfnamefont {C.~D.}\
  \bibnamefont {Murphy}}, \bibinfo {author} {\bibfnamefont {P.}~\bibnamefont
  {Norreys}}, \ and\ \bibinfo {author} {\bibfnamefont {C.}~\bibnamefont
  {Stoeckl}},\ }\href@noop {} {\bibfield  {journal} {\bibinfo  {journal}
  {Review of Scientific Instruments}\ }\textbf {\bibinfo {volume} {76}},\
  \bibinfo {pages} {076102} (\bibinfo {year} {2005})}\BibitemShut {NoStop}%
\bibitem [{\citenamefont {Norreys}\ \emph {et~al.}(2004)\citenamefont
  {Norreys}, \citenamefont {Lancaster}, \citenamefont {Murphy}, \citenamefont
  {Habara}, \citenamefont {Karsch}, \citenamefont {Clarke}, \citenamefont
  {Collier}, \citenamefont {Heathcote}, \citenamefont {Hemandez-Gomez},
  \citenamefont {Hawkes}, \citenamefont {Neely}, \citenamefont {Hutchinson},
  \citenamefont {Evans}, \citenamefont {Borghesi}, \citenamefont {Romagnani},
  \citenamefont {Zepf}, \citenamefont {Akli}, \citenamefont {King},
  \citenamefont {Zhang}, \citenamefont {Freeman}, \citenamefont {MacKinnon},
  \citenamefont {Hatchett}, \citenamefont {Patel}, \citenamefont {Snavely},
  \citenamefont {Key}, \citenamefont {Nikroo}, \citenamefont {Stephens},
  \citenamefont {Stoeckl}, \citenamefont {Tanaka}, \citenamefont {Norimatsu},
  \citenamefont {Toyama},\ and\ \citenamefont {Kodama}}]{Norreys_PoP2004}%
  \BibitemOpen
  \bibfield  {author} {\bibinfo {author} {\bibfnamefont {P.~A.}\ \bibnamefont
  {Norreys}}, \bibinfo {author} {\bibfnamefont {K.~L.}\ \bibnamefont
  {Lancaster}}, \bibinfo {author} {\bibfnamefont {C.~D.}\ \bibnamefont
  {Murphy}}, \bibinfo {author} {\bibfnamefont {H.}~\bibnamefont {Habara}},
  \bibinfo {author} {\bibfnamefont {S.}~\bibnamefont {Karsch}}, \bibinfo
  {author} {\bibfnamefont {R.~J.}\ \bibnamefont {Clarke}}, \bibinfo {author}
  {\bibfnamefont {J.}~\bibnamefont {Collier}}, \bibinfo {author} {\bibfnamefont
  {R.}~\bibnamefont {Heathcote}}, \bibinfo {author} {\bibfnamefont
  {C.}~\bibnamefont {Hemandez-Gomez}}, \bibinfo {author} {\bibfnamefont
  {S.}~\bibnamefont {Hawkes}}, \bibinfo {author} {\bibfnamefont
  {D.}~\bibnamefont {Neely}}, \bibinfo {author} {\bibfnamefont {M.~H.~R.}\
  \bibnamefont {Hutchinson}}, \bibinfo {author} {\bibfnamefont {R.~G.}\
  \bibnamefont {Evans}}, \bibinfo {author} {\bibfnamefont {M.}~\bibnamefont
  {Borghesi}}, \bibinfo {author} {\bibfnamefont {L.}~\bibnamefont {Romagnani}},
  \bibinfo {author} {\bibfnamefont {M.}~\bibnamefont {Zepf}}, \bibinfo {author}
  {\bibfnamefont {K.}~\bibnamefont {Akli}}, \bibinfo {author} {\bibfnamefont
  {J.~A.}\ \bibnamefont {King}}, \bibinfo {author} {\bibfnamefont
  {B.}~\bibnamefont {Zhang}}, \bibinfo {author} {\bibfnamefont {R.~R.}\
  \bibnamefont {Freeman}}, \bibinfo {author} {\bibfnamefont {A.~J.}\
  \bibnamefont {MacKinnon}}, \bibinfo {author} {\bibfnamefont {S.~P.}\
  \bibnamefont {Hatchett}}, \bibinfo {author} {\bibfnamefont {P.}~\bibnamefont
  {Patel}}, \bibinfo {author} {\bibfnamefont {R.}~\bibnamefont {Snavely}},
  \bibinfo {author} {\bibfnamefont {M.~H.}\ \bibnamefont {Key}}, \bibinfo
  {author} {\bibfnamefont {A.}~\bibnamefont {Nikroo}}, \bibinfo {author}
  {\bibfnamefont {R.}~\bibnamefont {Stephens}}, \bibinfo {author}
  {\bibfnamefont {C.}~\bibnamefont {Stoeckl}}, \bibinfo {author} {\bibfnamefont
  {K.~A.}\ \bibnamefont {Tanaka}}, \bibinfo {author} {\bibfnamefont
  {T.}~\bibnamefont {Norimatsu}}, \bibinfo {author} {\bibfnamefont
  {Y.}~\bibnamefont {Toyama}}, \ and\ \bibinfo {author} {\bibfnamefont
  {R.}~\bibnamefont {Kodama}},\ }\href {\doibase 10.1063/1.1688790} {\bibfield
  {journal} {\bibinfo  {journal} {Physics of Plasmas}\ }\textbf {\bibinfo
  {volume} {11}},\ \bibinfo {pages} {2746} (\bibinfo {year}
  {2004})}\BibitemShut {NoStop}%
\bibitem [{\citenamefont {Zhang}\ \emph {et~al.}(2012)\citenamefont {Zhang},
  \citenamefont {Nishimura}, \citenamefont {Namimoto}, \citenamefont {Fujioka},
  \citenamefont {Arikawa}, \citenamefont {Nishikino}, \citenamefont {Kawachi},
  \citenamefont {Sagisaka}, \citenamefont {Hosoda}, \citenamefont {Orimo},
  \citenamefont {Ogura}, \citenamefont {Pirozhkov}, \citenamefont {Yogo},
  \citenamefont {Okano}, \citenamefont {Kiriyama}, \citenamefont {Kondo},
  \citenamefont {Ohshima},\ and\ \citenamefont {Azechi}}]{Zhang_RSI2012}%
  \BibitemOpen
  \bibfield  {author} {\bibinfo {author} {\bibfnamefont {Z.}~\bibnamefont
  {Zhang}}, \bibinfo {author} {\bibfnamefont {H.}~\bibnamefont {Nishimura}},
  \bibinfo {author} {\bibfnamefont {T.}~\bibnamefont {Namimoto}}, \bibinfo
  {author} {\bibfnamefont {S.}~\bibnamefont {Fujioka}}, \bibinfo {author}
  {\bibfnamefont {Y.}~\bibnamefont {Arikawa}}, \bibinfo {author} {\bibfnamefont
  {M.}~\bibnamefont {Nishikino}}, \bibinfo {author} {\bibfnamefont
  {T.}~\bibnamefont {Kawachi}}, \bibinfo {author} {\bibfnamefont
  {A.}~\bibnamefont {Sagisaka}}, \bibinfo {author} {\bibfnamefont
  {H.}~\bibnamefont {Hosoda}}, \bibinfo {author} {\bibfnamefont
  {S.}~\bibnamefont {Orimo}}, \bibinfo {author} {\bibfnamefont
  {K.}~\bibnamefont {Ogura}}, \bibinfo {author} {\bibfnamefont
  {A.}~\bibnamefont {Pirozhkov}}, \bibinfo {author} {\bibfnamefont
  {A.}~\bibnamefont {Yogo}}, \bibinfo {author} {\bibfnamefont {Y.}~\bibnamefont
  {Okano}}, \bibinfo {author} {\bibfnamefont {H.}~\bibnamefont {Kiriyama}},
  \bibinfo {author} {\bibfnamefont {K.}~\bibnamefont {Kondo}}, \bibinfo
  {author} {\bibfnamefont {S.}~\bibnamefont {Ohshima}}, \ and\ \bibinfo
  {author} {\bibfnamefont {H.}~\bibnamefont {Azechi}},\ }\href {\doibase
  10.1063/1.4717677} {\bibfield  {journal} {\bibinfo  {journal} {Review of
  Scientific Instruments}\ }\textbf {\bibinfo {volume} {83}},\ \bibinfo {eid}
  {053502} (\bibinfo {year} {2012})}\BibitemShut {NoStop}%
\bibitem [{\citenamefont {Park}\ \emph {et~al.}(2010)\citenamefont {Park},
  \citenamefont {Dewald}, \citenamefont {Glenzer}, \citenamefont {Kalantar},
  \citenamefont {Kilkenny}, \citenamefont {MacGowan}, \citenamefont {Maddox},
  \citenamefont {Milovich}, \citenamefont {Prasad}, \citenamefont {Remington},
  \citenamefont {Robey},\ and\ \citenamefont {Thomas}}]{Park_RSI2010}%
  \BibitemOpen
  \bibfield  {author} {\bibinfo {author} {\bibfnamefont {H.-S.}\ \bibnamefont
  {Park}}, \bibinfo {author} {\bibfnamefont {E.~D.}\ \bibnamefont {Dewald}},
  \bibinfo {author} {\bibfnamefont {S.}~\bibnamefont {Glenzer}}, \bibinfo
  {author} {\bibfnamefont {D.~H.}\ \bibnamefont {Kalantar}}, \bibinfo {author}
  {\bibfnamefont {J.~D.}\ \bibnamefont {Kilkenny}}, \bibinfo {author}
  {\bibfnamefont {B.~J.}\ \bibnamefont {MacGowan}}, \bibinfo {author}
  {\bibfnamefont {B.~R.}\ \bibnamefont {Maddox}}, \bibinfo {author}
  {\bibfnamefont {J.~L.}\ \bibnamefont {Milovich}}, \bibinfo {author}
  {\bibfnamefont {R.~R.}\ \bibnamefont {Prasad}}, \bibinfo {author}
  {\bibfnamefont {B.~A.}\ \bibnamefont {Remington}}, \bibinfo {author}
  {\bibfnamefont {H.~F.}\ \bibnamefont {Robey}}, \ and\ \bibinfo {author}
  {\bibfnamefont {C.~A.}\ \bibnamefont {Thomas}},\ }\href {\doibase
  10.1063/1.3478682} {\bibfield  {journal} {\bibinfo  {journal} {Review of
  Scientific Instruments}\ }\textbf {\bibinfo {volume} {81}},\ \bibinfo {eid}
  {10E519} (\bibinfo {year} {2010})}\BibitemShut {NoStop}%
\bibitem [{\citenamefont {Dirksm{\"o}ller}\ \emph {et~al.}(1995)\citenamefont
  {Dirksm{\"o}ller}, \citenamefont {Rancu}, \citenamefont {Uschmann},
  \citenamefont {Renaudin}, \citenamefont {Chenais-Popovics}, \citenamefont
  {Gauthier},\ and\ \citenamefont {F{\"o}rster}}]{Dirksmoller_OC1995}%
  \BibitemOpen
  \bibfield  {author} {\bibinfo {author} {\bibfnamefont {M.}~\bibnamefont
  {Dirksm{\"o}ller}}, \bibinfo {author} {\bibfnamefont {O.}~\bibnamefont
  {Rancu}}, \bibinfo {author} {\bibfnamefont {I.}~\bibnamefont {Uschmann}},
  \bibinfo {author} {\bibfnamefont {P.}~\bibnamefont {Renaudin}}, \bibinfo
  {author} {\bibfnamefont {C.}~\bibnamefont {Chenais-Popovics}}, \bibinfo
  {author} {\bibfnamefont {J.~C.}\ \bibnamefont {Gauthier}}, \ and\ \bibinfo
  {author} {\bibfnamefont {E.}~\bibnamefont {F{\"o}rster}},\ }\href@noop {}
  {\bibfield  {journal} {\bibinfo  {journal} {Optics Communications}\ }\textbf
  {\bibinfo {volume} {118}},\ \bibinfo {pages} {379} (\bibinfo {year}
  {1995})}\BibitemShut {NoStop}%
\bibitem [{\citenamefont {Talukder}, \citenamefont {Bose},\ and\ \citenamefont
  {Takamura}(2008)}]{Talukder_IJMS2008}%
  \BibitemOpen
  \bibfield  {author} {\bibinfo {author} {\bibfnamefont {M.~R.}\ \bibnamefont
  {Talukder}}, \bibinfo {author} {\bibfnamefont {S.}~\bibnamefont {Bose}}, \
  and\ \bibinfo {author} {\bibfnamefont {S.}~\bibnamefont {Takamura}},\
  }\href@noop {} {\bibfield  {journal} {\bibinfo  {journal} {International
  Journal of Mass Spectrometry}\ }\textbf {\bibinfo {volume} {269}},\ \bibinfo
  {pages} {118} (\bibinfo {year} {2008})}\BibitemShut {NoStop}%
\bibitem [{\citenamefont {Batani}(2002)}]{Batani_LPB2002}%
  \BibitemOpen
  \bibfield  {author} {\bibinfo {author} {\bibfnamefont {D.}~\bibnamefont
  {Batani}},\ }\href@noop {} {\bibfield  {journal} {\bibinfo  {journal} {Laser
  and Particle Beams}\ }\textbf {\bibinfo {volume} {20}},\ \bibinfo {pages}
  {321} (\bibinfo {year} {2002})}\BibitemShut {NoStop}%
\bibitem [{\citenamefont {Perez}\ \emph {et~al.}(2010)\citenamefont {Perez},
  \citenamefont {Baton}, \citenamefont {Koenig}, \citenamefont {Chen},
  \citenamefont {Hey}, \citenamefont {Key}, \citenamefont {{Le Pape}},
  \citenamefont {Ma}, \citenamefont {McLean}, \citenamefont {MacPhee},
  \citenamefont {Patel}, \citenamefont {Ping}, \citenamefont {Beg},
  \citenamefont {Higginson}, \citenamefont {Murphy}, \citenamefont {Sawada},
  \citenamefont {Westover}, \citenamefont {Yabuuchi}, \citenamefont {Akli},
  \citenamefont {Giraldez}, \citenamefont {{Hoppe Jr.}}, \citenamefont
  {Shearer}, \citenamefont {Stephens}, \citenamefont {Gremillet}, \citenamefont
  {Lefebvre}, \citenamefont {Freeman}, \citenamefont {Kemp}, \citenamefont
  {Krygier}, \citenamefont {{Van Woerkom}}, \citenamefont {Fedosejevs},
  \citenamefont {Friesen}, \citenamefont {Tsui},\ and\ \citenamefont
  {Tumbull}}]{Perez_PoP2010}%
  \BibitemOpen
  \bibfield  {author} {\bibinfo {author} {\bibfnamefont {F.}~\bibnamefont
  {Perez}}, \bibinfo {author} {\bibfnamefont {S.~D.}\ \bibnamefont {Baton}},
  \bibinfo {author} {\bibfnamefont {M.}~\bibnamefont {Koenig}}, \bibinfo
  {author} {\bibfnamefont {C.~D.}\ \bibnamefont {Chen}}, \bibinfo {author}
  {\bibfnamefont {D.}~\bibnamefont {Hey}}, \bibinfo {author} {\bibfnamefont
  {M.~H.}\ \bibnamefont {Key}}, \bibinfo {author} {\bibfnamefont
  {S.}~\bibnamefont {{Le Pape}}}, \bibinfo {author} {\bibfnamefont
  {T.}~\bibnamefont {Ma}}, \bibinfo {author} {\bibfnamefont {H.~S.}\
  \bibnamefont {McLean}}, \bibinfo {author} {\bibfnamefont {A.~G.}\
  \bibnamefont {MacPhee}}, \bibinfo {author} {\bibfnamefont {P.~K.}\
  \bibnamefont {Patel}}, \bibinfo {author} {\bibfnamefont {Y.}~\bibnamefont
  {Ping}}, \bibinfo {author} {\bibfnamefont {F.~N.}\ \bibnamefont {Beg}},
  \bibinfo {author} {\bibfnamefont {D.~P.}\ \bibnamefont {Higginson}}, \bibinfo
  {author} {\bibfnamefont {C.~W.}\ \bibnamefont {Murphy}}, \bibinfo {author}
  {\bibfnamefont {H.}~\bibnamefont {Sawada}}, \bibinfo {author} {\bibfnamefont
  {B.}~\bibnamefont {Westover}}, \bibinfo {author} {\bibfnamefont
  {T.}~\bibnamefont {Yabuuchi}}, \bibinfo {author} {\bibfnamefont {K.~U.}\
  \bibnamefont {Akli}}, \bibinfo {author} {\bibfnamefont {E.}~\bibnamefont
  {Giraldez}}, \bibinfo {author} {\bibfnamefont {M.}~\bibnamefont {{Hoppe
  Jr.}}}, \bibinfo {author} {\bibfnamefont {C.}~\bibnamefont {Shearer}},
  \bibinfo {author} {\bibfnamefont {R.~B.}\ \bibnamefont {Stephens}}, \bibinfo
  {author} {\bibfnamefont {L.}~\bibnamefont {Gremillet}}, \bibinfo {author}
  {\bibfnamefont {E.}~\bibnamefont {Lefebvre}}, \bibinfo {author}
  {\bibfnamefont {R.~R.}\ \bibnamefont {Freeman}}, \bibinfo {author}
  {\bibfnamefont {G.~E.}\ \bibnamefont {Kemp}}, \bibinfo {author}
  {\bibfnamefont {A.~G.}\ \bibnamefont {Krygier}}, \bibinfo {author}
  {\bibfnamefont {L.~D.}\ \bibnamefont {{Van Woerkom}}}, \bibinfo {author}
  {\bibfnamefont {R.}~\bibnamefont {Fedosejevs}}, \bibinfo {author}
  {\bibfnamefont {R.~H.}\ \bibnamefont {Friesen}}, \bibinfo {author}
  {\bibfnamefont {Y.~Y.}\ \bibnamefont {Tsui}}, \ and\ \bibinfo {author}
  {\bibfnamefont {D.}~\bibnamefont {Tumbull}},\ }\href@noop {} {\bibfield
  {journal} {\bibinfo  {journal} {Physics of Plasmas}\ }\textbf {\bibinfo
  {volume} {17}},\ \bibinfo {pages} {113106} (\bibinfo {year}
  {2010})}\BibitemShut {NoStop}%
\bibitem [{\citenamefont {Akli}\ \emph {et~al.}(2007)\citenamefont {Akli},
  \citenamefont {Key}, \citenamefont {Chung}, \citenamefont {Hansen},
  \citenamefont {Freeman}, \citenamefont {Chen}, \citenamefont {Gregori},
  \citenamefont {Hatchett}, \citenamefont {Hey}, \citenamefont {Izumi},
  \citenamefont {King}, \citenamefont {Kuba}, \citenamefont {Norreys},
  \citenamefont {MacKinnon}, \citenamefont {Murphy}, \citenamefont {Snavely},
  \citenamefont {Stephens}, \citenamefont {Stoeckl}, \citenamefont {Theobald},\
  and\ \citenamefont {Zhang}}]{Akli_PoP2007}%
  \BibitemOpen
  \bibfield  {author} {\bibinfo {author} {\bibfnamefont {K.~U.}\ \bibnamefont
  {Akli}}, \bibinfo {author} {\bibfnamefont {M.~H.}\ \bibnamefont {Key}},
  \bibinfo {author} {\bibfnamefont {H.~K.}\ \bibnamefont {Chung}}, \bibinfo
  {author} {\bibfnamefont {S.~B.}\ \bibnamefont {Hansen}}, \bibinfo {author}
  {\bibfnamefont {R.~R.}\ \bibnamefont {Freeman}}, \bibinfo {author}
  {\bibfnamefont {M.~H.}\ \bibnamefont {Chen}}, \bibinfo {author}
  {\bibfnamefont {G.}~\bibnamefont {Gregori}}, \bibinfo {author} {\bibfnamefont
  {S.}~\bibnamefont {Hatchett}}, \bibinfo {author} {\bibfnamefont
  {D.}~\bibnamefont {Hey}}, \bibinfo {author} {\bibfnamefont {N.}~\bibnamefont
  {Izumi}}, \bibinfo {author} {\bibfnamefont {J.}~\bibnamefont {King}},
  \bibinfo {author} {\bibfnamefont {J.}~\bibnamefont {Kuba}}, \bibinfo {author}
  {\bibfnamefont {P.}~\bibnamefont {Norreys}}, \bibinfo {author} {\bibfnamefont
  {A.~J.}\ \bibnamefont {MacKinnon}}, \bibinfo {author} {\bibfnamefont {C.~D.}\
  \bibnamefont {Murphy}}, \bibinfo {author} {\bibfnamefont {R.}~\bibnamefont
  {Snavely}}, \bibinfo {author} {\bibfnamefont {R.~B.}\ \bibnamefont
  {Stephens}}, \bibinfo {author} {\bibfnamefont {C.}~\bibnamefont {Stoeckl}},
  \bibinfo {author} {\bibfnamefont {W.}~\bibnamefont {Theobald}}, \ and\
  \bibinfo {author} {\bibfnamefont {B.}~\bibnamefont {Zhang}},\ }\href@noop {}
  {\bibfield  {journal} {\bibinfo  {journal} {Physics of Plasmas}\ }\textbf
  {\bibinfo {volume} {14}},\ \bibinfo {pages} {023102} (\bibinfo {year}
  {2007})}\BibitemShut {NoStop}%
\bibitem [{\citenamefont {Betti}\ \emph {et~al.}(2007)\citenamefont {Betti},
  \citenamefont {Zhou}, \citenamefont {Anderson}, \citenamefont {Perkins},
  \citenamefont {Theobald},\ and\ \citenamefont {Solodov}}]{Betti_PRL2007}%
  \BibitemOpen
  \bibfield  {author} {\bibinfo {author} {\bibfnamefont {R.}~\bibnamefont
  {Betti}}, \bibinfo {author} {\bibfnamefont {C.~D.}\ \bibnamefont {Zhou}},
  \bibinfo {author} {\bibfnamefont {K.~S.}\ \bibnamefont {Anderson}}, \bibinfo
  {author} {\bibfnamefont {L.~J.}\ \bibnamefont {Perkins}}, \bibinfo {author}
  {\bibfnamefont {W.}~\bibnamefont {Theobald}}, \ and\ \bibinfo {author}
  {\bibfnamefont {A.~A.}\ \bibnamefont {Solodov}},\ }\href@noop {} {\bibfield
  {journal} {\bibinfo  {journal} {Physical Reveiw Letters}\ }\textbf {\bibinfo
  {volume} {98}},\ \bibinfo {pages} {155001} (\bibinfo {year}
  {2007})}\BibitemShut {NoStop}%
\bibitem [{\citenamefont {Atzeni}, \citenamefont {Schiavi},\ and\ \citenamefont
  {Marocchino}(2011)}]{Atzeni_PPCF2011}%
  \BibitemOpen
  \bibfield  {author} {\bibinfo {author} {\bibfnamefont {S.}~\bibnamefont
  {Atzeni}}, \bibinfo {author} {\bibfnamefont {A.}~\bibnamefont {Schiavi}}, \
  and\ \bibinfo {author} {\bibfnamefont {A.}~\bibnamefont {Marocchino}},\
  }\href@noop {} {\bibfield  {journal} {\bibinfo  {journal} {Plasma Physics and
  Controlled Fusion}\ }\textbf {\bibinfo {volume} {53}},\ \bibinfo {pages}
  {035010} (\bibinfo {year} {2011})}\BibitemShut {NoStop}%
\bibitem [{\citenamefont {Bell}\ and\ \citenamefont
  {Tzoufras}(2011)}]{Bell_PPCF2011}%
  \BibitemOpen
  \bibfield  {author} {\bibinfo {author} {\bibfnamefont {A.~R.}\ \bibnamefont
  {Bell}}\ and\ \bibinfo {author} {\bibfnamefont {M.}~\bibnamefont
  {Tzoufras}},\ }\href {http://stacks.iop.org/0741-3335/53/i=4/a=045010}
  {\bibfield  {journal} {\bibinfo  {journal} {Plasma Physics and Controlled
  Fusion}\ }\textbf {\bibinfo {volume} {53}},\ \bibinfo {pages} {045010}
  (\bibinfo {year} {2011})}\BibitemShut {NoStop}%
\bibitem [{\citenamefont {Baton}\ \emph {et~al.}(2012)\citenamefont {Baton},
  \citenamefont {Koenig}, \citenamefont {Brambrink}, \citenamefont
  {Schlenvoigt}, \citenamefont {Rousseaux}, \citenamefont {Debras},
  \citenamefont {Laffite}, \citenamefont {Loiseau}, \citenamefont {Philippe},
  \citenamefont {Ribeyre},\ and\ \citenamefont {Schurtz}}]{Baton_PRL2012}%
  \BibitemOpen
  \bibfield  {author} {\bibinfo {author} {\bibfnamefont {S.~D.}\ \bibnamefont
  {Baton}}, \bibinfo {author} {\bibfnamefont {M.}~\bibnamefont {Koenig}},
  \bibinfo {author} {\bibfnamefont {E.}~\bibnamefont {Brambrink}}, \bibinfo
  {author} {\bibfnamefont {H.-P.}\ \bibnamefont {Schlenvoigt}}, \bibinfo
  {author} {\bibfnamefont {C.}~\bibnamefont {Rousseaux}}, \bibinfo {author}
  {\bibfnamefont {G.}~\bibnamefont {Debras}}, \bibinfo {author} {\bibfnamefont
  {S.}~\bibnamefont {Laffite}}, \bibinfo {author} {\bibfnamefont
  {P.}~\bibnamefont {Loiseau}}, \bibinfo {author} {\bibfnamefont
  {F.}~\bibnamefont {Philippe}}, \bibinfo {author} {\bibfnamefont
  {X.}~\bibnamefont {Ribeyre}}, \ and\ \bibinfo {author} {\bibfnamefont
  {G.}~\bibnamefont {Schurtz}},\ }\href@noop {} {\bibfield  {journal} {\bibinfo
   {journal} {Physical Review Letters}\ }\textbf {\bibinfo {volume} {108}},\
  \bibinfo {pages} {195002} (\bibinfo {year} {2012})}\BibitemShut {NoStop}%
\bibitem [{\citenamefont {Nejdl}\ \emph {et~al.}(2011)\citenamefont {Nejdl},
  \citenamefont {Kozlov{\'a}}, \citenamefont {Sawicka}, \citenamefont
  {Margarone}, \citenamefont {Krus}, \citenamefont {Prokupek}, \citenamefont
  {Rus}, \citenamefont {Batani}, \citenamefont {Antonelli}, \citenamefont
  {Patria}, \citenamefont {Ciricosta}, \citenamefont {Cecchetti}, \citenamefont
  {Koester}, \citenamefont {Labate}, \citenamefont {Giulietti}, \citenamefont
  {Gizzi}, \citenamefont {Moretti}, \citenamefont {Richetta}, \citenamefont
  {Schurtz}, \citenamefont {Ribeyre}, \citenamefont {Lafon}, \citenamefont
  {Spindloe},\ and\ \citenamefont {{O'Dell}}}]{Nejdl_ProcSPIE2011}%
  \BibitemOpen
  \bibfield  {author} {\bibinfo {author} {\bibfnamefont {J.}~\bibnamefont
  {Nejdl}}, \bibinfo {author} {\bibfnamefont {M.}~\bibnamefont {Kozlov{\'a}}},
  \bibinfo {author} {\bibfnamefont {M.}~\bibnamefont {Sawicka}}, \bibinfo
  {author} {\bibfnamefont {D.}~\bibnamefont {Margarone}}, \bibinfo {author}
  {\bibfnamefont {M.}~\bibnamefont {Krus}}, \bibinfo {author} {\bibfnamefont
  {J.}~\bibnamefont {Prokupek}}, \bibinfo {author} {\bibfnamefont
  {B.}~\bibnamefont {Rus}}, \bibinfo {author} {\bibfnamefont {D.}~\bibnamefont
  {Batani}}, \bibinfo {author} {\bibfnamefont {L.}~\bibnamefont {Antonelli}},
  \bibinfo {author} {\bibfnamefont {A.}~\bibnamefont {Patria}}, \bibinfo
  {author} {\bibfnamefont {O.}~\bibnamefont {Ciricosta}}, \bibinfo {author}
  {\bibfnamefont {C.}~\bibnamefont {Cecchetti}}, \bibinfo {author}
  {\bibfnamefont {P.}~\bibnamefont {Koester}}, \bibinfo {author} {\bibfnamefont
  {L.}~\bibnamefont {Labate}}, \bibinfo {author} {\bibfnamefont
  {A.}~\bibnamefont {Giulietti}}, \bibinfo {author} {\bibfnamefont {L.~A.}\
  \bibnamefont {Gizzi}}, \bibinfo {author} {\bibfnamefont {A.}~\bibnamefont
  {Moretti}}, \bibinfo {author} {\bibfnamefont {M.}~\bibnamefont {Richetta}},
  \bibinfo {author} {\bibfnamefont {G.}~\bibnamefont {Schurtz}}, \bibinfo
  {author} {\bibfnamefont {X.}~\bibnamefont {Ribeyre}}, \bibinfo {author}
  {\bibfnamefont {M.}~\bibnamefont {Lafon}}, \bibinfo {author} {\bibfnamefont
  {C.}~\bibnamefont {Spindloe}}, \ and\ \bibinfo {author} {\bibfnamefont
  {T.}~\bibnamefont {{O'Dell}}},\ }in\ \href@noop {} {\emph {\bibinfo
  {booktitle} {Proceedings of SPIE}}},\ Vol.\ \bibinfo {volume} {8080}\
  (\bibinfo {year} {2011})\BibitemShut {NoStop}%
\bibitem [{\citenamefont {Gizzi}\ \emph {et~al.}(1996)\citenamefont {Gizzi},
  \citenamefont {Giulietti}, \citenamefont {Giulietti}, \citenamefont
  {Audebert}, \citenamefont {Bastiani}, \citenamefont {Geindre},\ and\
  \citenamefont {Mysyrowicz}}]{Gizzi_PRL1996}%
  \BibitemOpen
  \bibfield  {author} {\bibinfo {author} {\bibfnamefont {L.~A.}\ \bibnamefont
  {Gizzi}}, \bibinfo {author} {\bibfnamefont {A.}~\bibnamefont {Giulietti}},
  \bibinfo {author} {\bibfnamefont {D.}~\bibnamefont {Giulietti}}, \bibinfo
  {author} {\bibfnamefont {P.}~\bibnamefont {Audebert}}, \bibinfo {author}
  {\bibfnamefont {S.}~\bibnamefont {Bastiani}}, \bibinfo {author}
  {\bibfnamefont {J.-P.}\ \bibnamefont {Geindre}}, \ and\ \bibinfo {author}
  {\bibfnamefont {A.}~\bibnamefont {Mysyrowicz}},\ }\href@noop {} {\bibfield
  {journal} {\bibinfo  {journal} {Physical Review Letters}\ }\textbf {\bibinfo
  {volume} {76}},\ \bibinfo {pages} {2278} (\bibinfo {year}
  {1996})}\BibitemShut {NoStop}%
\bibitem [{\citenamefont {Labate}\ \emph
  {et~al.}(2007{\natexlab{b}})\citenamefont {Labate}, \citenamefont
  {Giulietti}, \citenamefont {Giulietti}, \citenamefont {K\"oster},
  \citenamefont {Levato}, \citenamefont {Gizzi}, \citenamefont {Zamponi},
  \citenamefont {L\"ubcke}, \citenamefont {K\"ampfer}, \citenamefont
  {Uschmann},\ and\ \citenamefont {F\"orster}}]{Labate_RSI2007}%
  \BibitemOpen
  \bibfield  {author} {\bibinfo {author} {\bibfnamefont {L.}~\bibnamefont
  {Labate}}, \bibinfo {author} {\bibfnamefont {A.}~\bibnamefont {Giulietti}},
  \bibinfo {author} {\bibfnamefont {D.}~\bibnamefont {Giulietti}}, \bibinfo
  {author} {\bibfnamefont {P.}~\bibnamefont {K\"oster}}, \bibinfo {author}
  {\bibfnamefont {T.}~\bibnamefont {Levato}}, \bibinfo {author} {\bibfnamefont
  {L.~A.}\ \bibnamefont {Gizzi}}, \bibinfo {author} {\bibfnamefont
  {F.}~\bibnamefont {Zamponi}}, \bibinfo {author} {\bibfnamefont
  {A.}~\bibnamefont {L\"ubcke}}, \bibinfo {author} {\bibfnamefont
  {T.}~\bibnamefont {K\"ampfer}}, \bibinfo {author} {\bibfnamefont
  {I.}~\bibnamefont {Uschmann}}, \ and\ \bibinfo {author} {\bibfnamefont
  {E.}~\bibnamefont {F\"orster}},\ }\href@noop {} {\bibfield  {journal}
  {\bibinfo  {journal} {Review of Scientific Instruments}\ }\textbf {\bibinfo
  {volume} {78}},\ \bibinfo {pages} {103506} (\bibinfo {year}
  {2007}{\natexlab{b}})}\BibitemShut {NoStop}%
\bibitem [{\citenamefont {Zamponi}\ \emph {et~al.}(2010)\citenamefont
  {Zamponi}, \citenamefont {L\"ubcke}, \citenamefont {K\"ampfer}, \citenamefont
  {Uschmann}, \citenamefont {F\"orster}, \citenamefont {Robinson},
  \citenamefont {Giulietti}, \citenamefont {K\"oster}, \citenamefont {Labate},
  \citenamefont {Levato},\ and\ \citenamefont {Gizzi}}]{Zamponi_PRL2010}%
  \BibitemOpen
  \bibfield  {author} {\bibinfo {author} {\bibfnamefont {F.}~\bibnamefont
  {Zamponi}}, \bibinfo {author} {\bibfnamefont {A.}~\bibnamefont {L\"ubcke}},
  \bibinfo {author} {\bibfnamefont {T.}~\bibnamefont {K\"ampfer}}, \bibinfo
  {author} {\bibfnamefont {I.}~\bibnamefont {Uschmann}}, \bibinfo {author}
  {\bibfnamefont {E.}~\bibnamefont {F\"orster}}, \bibinfo {author}
  {\bibfnamefont {A.~P.~L.}\ \bibnamefont {Robinson}}, \bibinfo {author}
  {\bibfnamefont {A.}~\bibnamefont {Giulietti}}, \bibinfo {author}
  {\bibfnamefont {P.}~\bibnamefont {K\"oster}}, \bibinfo {author}
  {\bibfnamefont {L.}~\bibnamefont {Labate}}, \bibinfo {author} {\bibfnamefont
  {T.}~\bibnamefont {Levato}}, \ and\ \bibinfo {author} {\bibfnamefont {L.~A.}\
  \bibnamefont {Gizzi}},\ }\href@noop {} {\bibfield  {journal} {\bibinfo
  {journal} {Physical Review Letters}\ }\textbf {\bibinfo {volume} {105}},\
  \bibinfo {pages} {085001} (\bibinfo {year} {2010})}\BibitemShut {NoStop}%
\bibitem [{\citenamefont {Park}\ \emph {et~al.}(2004)\citenamefont {Park},
  \citenamefont {Izumi}, \citenamefont {Key}, \citenamefont {Koch},
  \citenamefont {Landen}, \citenamefont {Patel}, \citenamefont {Phillips},\
  and\ \citenamefont {Zhang}}]{Park_RSI2004}%
  \BibitemOpen
  \bibfield  {author} {\bibinfo {author} {\bibfnamefont {H.-S.}\ \bibnamefont
  {Park}}, \bibinfo {author} {\bibfnamefont {N.}~\bibnamefont {Izumi}},
  \bibinfo {author} {\bibfnamefont {M.~H.}\ \bibnamefont {Key}}, \bibinfo
  {author} {\bibfnamefont {J.~A.}\ \bibnamefont {Koch}}, \bibinfo {author}
  {\bibfnamefont {O.~L.}\ \bibnamefont {Landen}}, \bibinfo {author}
  {\bibfnamefont {P.~K.}\ \bibnamefont {Patel}}, \bibinfo {author}
  {\bibfnamefont {T.~W.}\ \bibnamefont {Phillips}}, \ and\ \bibinfo {author}
  {\bibfnamefont {B.~B.}\ \bibnamefont {Zhang}},\ }\href@noop {} {\bibfield
  {journal} {\bibinfo  {journal} {Review of Scientific Instruments}\ }\textbf
  {\bibinfo {volume} {75}},\ \bibinfo {pages} {4048} (\bibinfo {year}
  {2004})}\BibitemShut {NoStop}%
\bibitem [{\citenamefont {Labate}\ \emph {et~al.}(2002)\citenamefont {Labate},
  \citenamefont {Galimberti}, \citenamefont {Giulietti}, \citenamefont
  {Giulietti}, \citenamefont {Gizzi}, \citenamefont {Tomassini},\ and\
  \citenamefont {{Di Cocco}}}]{Labate_NIMA2002}%
  \BibitemOpen
  \bibfield  {author} {\bibinfo {author} {\bibfnamefont {L.}~\bibnamefont
  {Labate}}, \bibinfo {author} {\bibfnamefont {M.}~\bibnamefont {Galimberti}},
  \bibinfo {author} {\bibfnamefont {A.}~\bibnamefont {Giulietti}}, \bibinfo
  {author} {\bibfnamefont {D.}~\bibnamefont {Giulietti}}, \bibinfo {author}
  {\bibfnamefont {L.}~\bibnamefont {Gizzi}}, \bibinfo {author} {\bibfnamefont
  {P.}~\bibnamefont {Tomassini}}, \ and\ \bibinfo {author} {\bibfnamefont
  {G.}~\bibnamefont {{Di Cocco}}},\ }\href@noop {} {\bibfield  {journal}
  {\bibinfo  {journal} {Nuclear Instruments and Methods in Physics Research A}\
  }\textbf {\bibinfo {volume} {495}},\ \bibinfo {pages} {148} (\bibinfo {year}
  {2002})}\BibitemShut {NoStop}%
\bibitem [{\citenamefont {Fourment}\ \emph {et~al.}(2009)\citenamefont
  {Fourment}, \citenamefont {Arazam}, \citenamefont {Bonte}, \citenamefont
  {Caillaud}, \citenamefont {Descamps}, \citenamefont {Dorchies}, \citenamefont
  {Harmand}, \citenamefont {Hulin}, \citenamefont {Petit},\ and\ \citenamefont
  {Santos}}]{Fourment_RSI2009}%
  \BibitemOpen
  \bibfield  {author} {\bibinfo {author} {\bibfnamefont {C.}~\bibnamefont
  {Fourment}}, \bibinfo {author} {\bibfnamefont {N.}~\bibnamefont {Arazam}},
  \bibinfo {author} {\bibfnamefont {C.}~\bibnamefont {Bonte}}, \bibinfo
  {author} {\bibfnamefont {T.}~\bibnamefont {Caillaud}}, \bibinfo {author}
  {\bibfnamefont {D.}~\bibnamefont {Descamps}}, \bibinfo {author}
  {\bibfnamefont {F.}~\bibnamefont {Dorchies}}, \bibinfo {author}
  {\bibfnamefont {M.}~\bibnamefont {Harmand}}, \bibinfo {author} {\bibfnamefont
  {S.}~\bibnamefont {Hulin}}, \bibinfo {author} {\bibfnamefont
  {S.}~\bibnamefont {Petit}}, \ and\ \bibinfo {author} {\bibfnamefont {J.~J.}\
  \bibnamefont {Santos}},\ }\href@noop {} {\bibfield  {journal} {\bibinfo
  {journal} {Review of Scientific Instruments}\ ,\ \bibinfo {pages} {083505}}
  (\bibinfo {year} {2009})}\BibitemShut {NoStop}%
\bibitem [{\citenamefont {Maddox}\ \emph {et~al.}(2008)\citenamefont {Maddox},
  \citenamefont {Park}, \citenamefont {Remington},\ and\ \citenamefont
  {McKernan}}]{Maddox_RSI2008Cali}%
  \BibitemOpen
  \bibfield  {author} {\bibinfo {author} {\bibfnamefont {B.~R.}\ \bibnamefont
  {Maddox}}, \bibinfo {author} {\bibfnamefont {H.~S.}\ \bibnamefont {Park}},
  \bibinfo {author} {\bibfnamefont {B.~A.}\ \bibnamefont {Remington}}, \ and\
  \bibinfo {author} {\bibfnamefont {M.}~\bibnamefont {McKernan}},\ }\href@noop
  {} {\bibfield  {journal} {\bibinfo  {journal} {Review of Scientific
  Instruments}\ }\textbf {\bibinfo {volume} {79}},\ \bibinfo {pages} {10E924}
  (\bibinfo {year} {2008})}\BibitemShut {NoStop}%
\bibitem [{\citenamefont {Levato}\ \emph {et~al.}(2010)\citenamefont {Levato},
  \citenamefont {Labate}, \citenamefont {Pathak}, \citenamefont {Cecchetti},
  \citenamefont {K\"oster}, \citenamefont {{Di Fabrizio}}, \citenamefont
  {Delogu}, \citenamefont {Giulietti}, \citenamefont {Giulietti},\ and\
  \citenamefont {Gizzi}}]{Levato_NIMA2010}%
  \BibitemOpen
  \bibfield  {author} {\bibinfo {author} {\bibfnamefont {T.}~\bibnamefont
  {Levato}}, \bibinfo {author} {\bibfnamefont {L.}~\bibnamefont {Labate}},
  \bibinfo {author} {\bibfnamefont {N.~C.}\ \bibnamefont {Pathak}}, \bibinfo
  {author} {\bibfnamefont {C.~A.}\ \bibnamefont {Cecchetti}}, \bibinfo {author}
  {\bibfnamefont {P.}~\bibnamefont {K\"oster}}, \bibinfo {author}
  {\bibfnamefont {E.}~\bibnamefont {{Di Fabrizio}}}, \bibinfo {author}
  {\bibfnamefont {P.}~\bibnamefont {Delogu}}, \bibinfo {author} {\bibfnamefont
  {A.}~\bibnamefont {Giulietti}}, \bibinfo {author} {\bibfnamefont
  {D.}~\bibnamefont {Giulietti}}, \ and\ \bibinfo {author} {\bibfnamefont
  {L.~A.}\ \bibnamefont {Gizzi}},\ }\href@noop {} {\bibfield  {journal}
  {\bibinfo  {journal} {Nuclear Instruments and Methods in Physics Research A}\
  }\textbf {\bibinfo {volume} {623}},\ \bibinfo {pages} {842} (\bibinfo {year}
  {2010})}\BibitemShut {NoStop}%
\bibitem [{\citenamefont {Labate}\ \emph {et~al.}(2009)\citenamefont {Labate},
  \citenamefont {Förster}, \citenamefont {Giulietti}, \citenamefont
  {Giulietti}, \citenamefont {Höfer}, \citenamefont {Kämpfer}, \citenamefont
  {Köster}, \citenamefont {Kozlova}, \citenamefont {Levato}, \citenamefont
  {Lötzsch}, \citenamefont {Lübcke}, \citenamefont {Mocek}, \citenamefont
  {Polan}, \citenamefont {Rus}, \citenamefont {Uschmann}, \citenamefont
  {Zamponi},\ and\ \citenamefont {Gizzi}}]{Labate_LPB2009}%
  \BibitemOpen
  \bibfield  {author} {\bibinfo {author} {\bibfnamefont {L.}~\bibnamefont
  {Labate}}, \bibinfo {author} {\bibfnamefont {E.}~\bibnamefont {Förster}},
  \bibinfo {author} {\bibfnamefont {A.}~\bibnamefont {Giulietti}}, \bibinfo
  {author} {\bibfnamefont {D.}~\bibnamefont {Giulietti}}, \bibinfo {author}
  {\bibfnamefont {S.}~\bibnamefont {Höfer}}, \bibinfo {author} {\bibfnamefont
  {T.}~\bibnamefont {Kämpfer}}, \bibinfo {author} {\bibfnamefont
  {P.}~\bibnamefont {Köster}}, \bibinfo {author} {\bibfnamefont
  {M.}~\bibnamefont {Kozlova}}, \bibinfo {author} {\bibfnamefont
  {T.}~\bibnamefont {Levato}}, \bibinfo {author} {\bibfnamefont
  {R.}~\bibnamefont {Lötzsch}}, \bibinfo {author} {\bibfnamefont
  {A.}~\bibnamefont {Lübcke}}, \bibinfo {author} {\bibfnamefont
  {T.}~\bibnamefont {Mocek}}, \bibinfo {author} {\bibfnamefont
  {J.}~\bibnamefont {Polan}}, \bibinfo {author} {\bibfnamefont
  {B.}~\bibnamefont {Rus}}, \bibinfo {author} {\bibfnamefont {I.}~\bibnamefont
  {Uschmann}}, \bibinfo {author} {\bibfnamefont {F.}~\bibnamefont {Zamponi}}, \
  and\ \bibinfo {author} {\bibfnamefont {L.}~\bibnamefont {Gizzi}},\
  }\href@noop {} {\bibfield  {journal} {\bibinfo  {journal} {Laser and Particle
  Beams}\ }\textbf {\bibinfo {volume} {27}},\ \bibinfo {pages} {643} (\bibinfo
  {year} {2009})}\BibitemShut {NoStop}%
\bibitem [{\citenamefont {Knoll}(1989)}]{KnollRadiationDetectionBook_1989}%
  \BibitemOpen
  \bibfield  {author} {\bibinfo {author} {\bibfnamefont {G.~F.}\ \bibnamefont
  {Knoll}},\ }\href@noop {} {\emph {\bibinfo {title} {Radiation detection and
  measurement}}},\ \bibinfo {edition} {2nd}\ ed.\ (\bibinfo  {publisher} {John
  Wiley \& Sons},\ \bibinfo {address} {New York},\ \bibinfo {year}
  {1989})\BibitemShut {NoStop}%
\bibitem [{\citenamefont {Hopkinson}(1983)}]{Hopkinson_NIMA1983}%
  \BibitemOpen
  \bibfield  {author} {\bibinfo {author} {\bibfnamefont {G.~R.}\ \bibnamefont
  {Hopkinson}},\ }\href@noop {} {\bibfield  {journal} {\bibinfo  {journal}
  {Nuclear Instruments and Methods in Physics Research A}\ }\textbf {\bibinfo
  {volume} {216}},\ \bibinfo {pages} {423} (\bibinfo {year}
  {1983})}\BibitemShut {NoStop}%
\bibitem [{\citenamefont {Prigozhin}\ \emph {et~al.}(2003)\citenamefont
  {Prigozhin}, \citenamefont {Butler}, \citenamefont {Kissel},\ and\
  \citenamefont {Ricker}}]{Prigozhin_IEEETED2003}%
  \BibitemOpen
  \bibfield  {author} {\bibinfo {author} {\bibfnamefont {G.}~\bibnamefont
  {Prigozhin}}, \bibinfo {author} {\bibfnamefont {N.~R.}\ \bibnamefont
  {Butler}}, \bibinfo {author} {\bibfnamefont {S.~E.}\ \bibnamefont {Kissel}},
  \ and\ \bibinfo {author} {\bibfnamefont {G.~R.}\ \bibnamefont {Ricker}},\
  }\href@noop {} {\bibfield  {journal} {\bibinfo  {journal} {IEEE Transaction
  on Electron Devices}\ }\textbf {\bibinfo {volume} {50}},\ \bibinfo {pages}
  {246} (\bibinfo {year} {2003})}\BibitemShut {NoStop}%
\bibitem [{\citenamefont {Pavlov}\ and\ \citenamefont
  {Nousek}(1999)}]{Pavlov_NIMA1999}%
  \BibitemOpen
  \bibfield  {author} {\bibinfo {author} {\bibfnamefont {G.~G.}\ \bibnamefont
  {Pavlov}}\ and\ \bibinfo {author} {\bibfnamefont {J.~A.}\ \bibnamefont
  {Nousek}},\ }\href@noop {} {\bibfield  {journal} {\bibinfo  {journal}
  {Nuclear Instruments and Methods in Physics Research A}\ }\textbf {\bibinfo
  {volume} {428}},\ \bibinfo {pages} {348} (\bibinfo {year}
  {1999})}\BibitemShut {NoStop}%
\bibitem [{\citenamefont {Howe}\ \emph {et~al.}(2006)\citenamefont {Howe},
  \citenamefont {Chambers}, \citenamefont {Courtois}, \citenamefont {Forster},
  \citenamefont {Gregory}, \citenamefont {Hall}, \citenamefont {Renner},
  \citenamefont {Uschmann},\ and\ \citenamefont {Woolsey}}]{Howe_RSI2006}%
  \BibitemOpen
  \bibfield  {author} {\bibinfo {author} {\bibfnamefont {J.}~\bibnamefont
  {Howe}}, \bibinfo {author} {\bibfnamefont {D.~M.}\ \bibnamefont {Chambers}},
  \bibinfo {author} {\bibfnamefont {C.}~\bibnamefont {Courtois}}, \bibinfo
  {author} {\bibfnamefont {E.}~\bibnamefont {Forster}}, \bibinfo {author}
  {\bibfnamefont {C.~D.}\ \bibnamefont {Gregory}}, \bibinfo {author}
  {\bibfnamefont {I.~M.}\ \bibnamefont {Hall}}, \bibinfo {author}
  {\bibfnamefont {O.}~\bibnamefont {Renner}}, \bibinfo {author} {\bibfnamefont
  {I.}~\bibnamefont {Uschmann}}, \ and\ \bibinfo {author} {\bibfnamefont
  {N.~C.}\ \bibnamefont {Woolsey}},\ }\href {\doibase 10.1063/1.2166515}
  {\bibfield  {journal} {\bibinfo  {journal} {Review of Scientific
  Instruments}\ }\textbf {\bibinfo {volume} {77}},\ \bibinfo {eid} {036105}
  (\bibinfo {year} {2006})}\BibitemShut {NoStop}%
\bibitem [{\citenamefont {Labate}\ \emph {et~al.}(2008)\citenamefont {Labate},
  \citenamefont {Levato}, \citenamefont {Galimberti}, \citenamefont
  {Giulietti}, \citenamefont {Giulietti}, \citenamefont {Sanna}, \citenamefont
  {Traino}, \citenamefont {Lazzeri},\ and\ \citenamefont
  {Gizzi}}]{Labate_NIMA2008}%
  \BibitemOpen
  \bibfield  {author} {\bibinfo {author} {\bibfnamefont {L.}~\bibnamefont
  {Labate}}, \bibinfo {author} {\bibfnamefont {T.}~\bibnamefont {Levato}},
  \bibinfo {author} {\bibfnamefont {M.}~\bibnamefont {Galimberti}}, \bibinfo
  {author} {\bibfnamefont {A.}~\bibnamefont {Giulietti}}, \bibinfo {author}
  {\bibfnamefont {D.}~\bibnamefont {Giulietti}}, \bibinfo {author}
  {\bibfnamefont {M.}~\bibnamefont {Sanna}}, \bibinfo {author} {\bibfnamefont
  {C.}~\bibnamefont {Traino}}, \bibinfo {author} {\bibfnamefont
  {M.}~\bibnamefont {Lazzeri}}, \ and\ \bibinfo {author} {\bibfnamefont
  {L.~A.}\ \bibnamefont {Gizzi}},\ }\href@noop {} {\bibfield  {journal}
  {\bibinfo  {journal} {Nuclear Instruments and Methods in Physics Research A}\
  }\textbf {\bibinfo {volume} {594}},\ \bibinfo {pages} {278} (\bibinfo {year}
  {2008})}\BibitemShut {NoStop}%
\bibitem [{\citenamefont {K\"oster}\ \emph {et~al.}(2009)\citenamefont
  {K\"oster}, \citenamefont {Akli}, \citenamefont {Batani}, \citenamefont
  {Baton}, \citenamefont {Evans}, \citenamefont {Giulietti}, \citenamefont
  {Giulietti}, \citenamefont {Gizzi}, \citenamefont {Green}, \citenamefont
  {Koenig}, \citenamefont {Labate}, \citenamefont {Morace}, \citenamefont
  {Norreys}, \citenamefont {Perez}, \citenamefont {Waugh}, \citenamefont
  {Woolsey},\ and\ \citenamefont {Lancaster}}]{Koester_PPCF2009}%
  \BibitemOpen
  \bibfield  {author} {\bibinfo {author} {\bibfnamefont {P.}~\bibnamefont
  {K\"oster}}, \bibinfo {author} {\bibfnamefont {K.}~\bibnamefont {Akli}},
  \bibinfo {author} {\bibfnamefont {D.}~\bibnamefont {Batani}}, \bibinfo
  {author} {\bibfnamefont {S.}~\bibnamefont {Baton}}, \bibinfo {author}
  {\bibfnamefont {R.~G.}\ \bibnamefont {Evans}}, \bibinfo {author}
  {\bibfnamefont {A.}~\bibnamefont {Giulietti}}, \bibinfo {author}
  {\bibfnamefont {D.}~\bibnamefont {Giulietti}}, \bibinfo {author}
  {\bibfnamefont {L.~A.}\ \bibnamefont {Gizzi}}, \bibinfo {author}
  {\bibfnamefont {J.~S.}\ \bibnamefont {Green}}, \bibinfo {author}
  {\bibfnamefont {M.}~\bibnamefont {Koenig}}, \bibinfo {author} {\bibfnamefont
  {L.}~\bibnamefont {Labate}}, \bibinfo {author} {\bibfnamefont
  {A.}~\bibnamefont {Morace}}, \bibinfo {author} {\bibfnamefont
  {P.}~\bibnamefont {Norreys}}, \bibinfo {author} {\bibfnamefont
  {F.}~\bibnamefont {Perez}}, \bibinfo {author} {\bibfnamefont
  {J.}~\bibnamefont {Waugh}}, \bibinfo {author} {\bibfnamefont
  {N.}~\bibnamefont {Woolsey}}, \ and\ \bibinfo {author} {\bibfnamefont
  {K.~L.}\ \bibnamefont {Lancaster}},\ }\href@noop {} {\bibfield  {journal}
  {\bibinfo  {journal} {Plasma Physics and Controlled Fusion}\ }\textbf
  {\bibinfo {volume} {51}},\ \bibinfo {pages} {014007} (\bibinfo {year}
  {2009})}\BibitemShut {NoStop}%
\bibitem [{\citenamefont {Gizzi}\ \emph {et~al.}(2010)\citenamefont {Gizzi},
  \citenamefont {K\"oster}, \citenamefont {Labate},\ and\ \citenamefont
  {Levato}}]{Gizzi_NIMA2010}%
  \BibitemOpen
  \bibfield  {author} {\bibinfo {author} {\bibfnamefont {L.~A.}\ \bibnamefont
  {Gizzi}}, \bibinfo {author} {\bibfnamefont {P.}~\bibnamefont {K\"oster}},
  \bibinfo {author} {\bibfnamefont {L.}~\bibnamefont {Labate}}, \ and\ \bibinfo
  {author} {\bibfnamefont {T.}~\bibnamefont {Levato}},\ }\href@noop {}
  {\bibfield  {journal} {\bibinfo  {journal} {Nuclear Instruments and Methods
  in Physics Research A}\ }\textbf {\bibinfo {volume} {623}},\ \bibinfo {pages}
  {836} (\bibinfo {year} {2010})}\BibitemShut {NoStop}%
\bibitem [{\citenamefont {Blasco}\ \emph {et~al.}(2001)\citenamefont {Blasco},
  \citenamefont {Stenz}, \citenamefont {Salin}, \citenamefont {Faenov},
  \citenamefont {Magunov}, \citenamefont {Pikuz},\ and\ \citenamefont
  {Skobelev}}]{Blasco_RSI2001}%
  \BibitemOpen
  \bibfield  {author} {\bibinfo {author} {\bibfnamefont {F.}~\bibnamefont
  {Blasco}}, \bibinfo {author} {\bibfnamefont {C.}~\bibnamefont {Stenz}},
  \bibinfo {author} {\bibfnamefont {F.}~\bibnamefont {Salin}}, \bibinfo
  {author} {\bibfnamefont {A.~Y.}\ \bibnamefont {Faenov}}, \bibinfo {author}
  {\bibfnamefont {A.~I.}\ \bibnamefont {Magunov}}, \bibinfo {author}
  {\bibfnamefont {T.~A.}\ \bibnamefont {Pikuz}}, \ and\ \bibinfo {author}
  {\bibfnamefont {I.~Y.}\ \bibnamefont {Skobelev}},\ }\href@noop {} {\bibfield
  {journal} {\bibinfo  {journal} {Review of Scientific Instruments}\ }\textbf
  {\bibinfo {volume} {72}},\ \bibinfo {pages} {1956} (\bibinfo {year}
  {2001})}\BibitemShut {NoStop}%
\bibitem [{\citenamefont {Nejdl}\ \emph {et~al.}(2010)\citenamefont {Nejdl},
  \citenamefont {Kozlova}, \citenamefont {Mocek},\ and\ \citenamefont
  {Rus}}]{Nejdl_PoP2010}%
  \BibitemOpen
  \bibfield  {author} {\bibinfo {author} {\bibfnamefont {J.}~\bibnamefont
  {Nejdl}}, \bibinfo {author} {\bibfnamefont {M.}~\bibnamefont {Kozlova}},
  \bibinfo {author} {\bibfnamefont {T.}~\bibnamefont {Mocek}}, \ and\ \bibinfo
  {author} {\bibfnamefont {B.}~\bibnamefont {Rus}},\ }\href {\doibase
  10.1063/1.3525570} {\bibfield  {journal} {\bibinfo  {journal} {Physics of
  Plasmas}\ }\textbf {\bibinfo {volume} {17}},\ \bibinfo {eid} {122705}
  (\bibinfo {year} {2010})}\BibitemShut {NoStop}%
\bibitem [{\citenamefont {Fraser}\ \emph {et~al.}(1994)\citenamefont {Fraser},
  \citenamefont {Abbey}, \citenamefont {Holland}, \citenamefont {McCarthy},
  \citenamefont {Owens},\ and\ \citenamefont {Wells}}]{Fraser_NIMA1994}%
  \BibitemOpen
  \bibfield  {author} {\bibinfo {author} {\bibfnamefont {G.}~\bibnamefont
  {Fraser}}, \bibinfo {author} {\bibfnamefont {A.}~\bibnamefont {Abbey}},
  \bibinfo {author} {\bibfnamefont {A.}~\bibnamefont {Holland}}, \bibinfo
  {author} {\bibfnamefont {K.}~\bibnamefont {McCarthy}}, \bibinfo {author}
  {\bibfnamefont {A.}~\bibnamefont {Owens}}, \ and\ \bibinfo {author}
  {\bibfnamefont {A.}~\bibnamefont {Wells}},\ }\href {\doibase
  10.1016/0168-9002(94)91185-1} {\bibfield  {journal} {\bibinfo  {journal}
  {Nuclear Instruments and Methods in Physics Research A}\ }\textbf {\bibinfo
  {volume} {350}},\ \bibinfo {pages} {368 } (\bibinfo {year}
  {1994})}\BibitemShut {NoStop}%
\bibitem [{\citenamefont {Levato}\ \emph {et~al.}(2008)\citenamefont {Levato},
  \citenamefont {Labate}, \citenamefont {Galimberti}, \citenamefont
  {Giulietti}, \citenamefont {Giulietti},\ and\ \citenamefont
  {Gizzi}}]{Levato_NIMA2008}%
  \BibitemOpen
  \bibfield  {author} {\bibinfo {author} {\bibfnamefont {T.}~\bibnamefont
  {Levato}}, \bibinfo {author} {\bibfnamefont {L.}~\bibnamefont {Labate}},
  \bibinfo {author} {\bibfnamefont {M.}~\bibnamefont {Galimberti}}, \bibinfo
  {author} {\bibfnamefont {A.}~\bibnamefont {Giulietti}}, \bibinfo {author}
  {\bibfnamefont {D.}~\bibnamefont {Giulietti}}, \ and\ \bibinfo {author}
  {\bibfnamefont {L.~A.}\ \bibnamefont {Gizzi}},\ }\href@noop {} {\bibfield
  {journal} {\bibinfo  {journal} {Nuclear Instruments and Methods in Physics
  Research A}\ }\textbf {\bibinfo {volume} {592}},\ \bibinfo {pages} {346}
  (\bibinfo {year} {2008})}\BibitemShut {NoStop}%
\bibitem [{\citenamefont {Chmeissani}\ \emph {et~al.}(2004)\citenamefont
  {Chmeissani}, \citenamefont {Frojdh}, \citenamefont {Gal}, \citenamefont
  {Llopart}, \citenamefont {Ludwig}, \citenamefont {Maiorino}, \citenamefont
  {Manach}, \citenamefont {Mettivier}, \citenamefont {Montesi}, \citenamefont
  {Ponchut},\ and\ \citenamefont {Russo}}]{Chmeissani_IEEETNS2004}%
  \BibitemOpen
  \bibfield  {author} {\bibinfo {author} {\bibfnamefont {M.}~\bibnamefont
  {Chmeissani}}, \bibinfo {author} {\bibfnamefont {C.}~\bibnamefont {Frojdh}},
  \bibinfo {author} {\bibfnamefont {O.}~\bibnamefont {Gal}}, \bibinfo {author}
  {\bibfnamefont {X.}~\bibnamefont {Llopart}}, \bibinfo {author} {\bibfnamefont
  {J.}~\bibnamefont {Ludwig}}, \bibinfo {author} {\bibfnamefont
  {M.}~\bibnamefont {Maiorino}}, \bibinfo {author} {\bibfnamefont
  {E.}~\bibnamefont {Manach}}, \bibinfo {author} {\bibfnamefont
  {G.}~\bibnamefont {Mettivier}}, \bibinfo {author} {\bibfnamefont {M.~C.}\
  \bibnamefont {Montesi}}, \bibinfo {author} {\bibfnamefont {C.}~\bibnamefont
  {Ponchut}}, \ and\ \bibinfo {author} {\bibfnamefont {P.}~\bibnamefont
  {Russo}},\ }\href@noop {} {\bibfield  {journal} {\bibinfo  {journal} {IEEE
  Transactions on Nuclear Science}\ }\textbf {\bibinfo {volume} {51}},\
  \bibinfo {pages} {2379} (\bibinfo {year} {2004})}\BibitemShut {NoStop}%
\end{thebibliography}
\end{document}